\documentclass[fleqn,usenatbib]{mnras}

\usepackage{newtxtext}

\usepackage[T1]{fontenc}
\usepackage{xspace}
\newcommand{\source}{Aql~X-1~}
\newcommand{\fluxcgs}{erg~s$^{-1}$~cm$^{-2}$}
\newcommand{\flencecgs}{erg~cm$^{-2}$}
\newcommand{\nicer}{\emph{NICER}\xspace}
\newcommand{\fa}{$f_a$~}

\DeclareRobustCommand{\VAN}[3]{#2}
\let\VANthebibliography\thebibliography
\def\thebibliography{\DeclareRobustCommand{\VAN}[3]{##3}\VANthebibliography}

\usepackage{graphicx}	
\usepackage{amsmath}	
\usepackage{amssymb}	
\usepackage[normalem]{ulem}

\title[X-ray Bursts from \source]{A NICER look at thermonuclear X-ray bursts from \source}

\author[G\"uver, T. et al.]{
Tolga G\"uver,$^{1,2}$\thanks{E-mail: tolga.guver@istanbul.edu.tr}
Tu\u{g}ba Boztepe,$^{3}$ 
D.R. Ballantyne, $^{4}$
Z. Funda Bostanc\i,$^{1,2}$
Peter Bult,$^{5,6}$\newauthor
Gaurava K. Jaisawal,$^{7}$ 
Ersin G\"o\u{g}\"u\c{s},$^{8}$
{Tod E. Strohmayer},$^{9}$
Diego Altamirano,$^{10}$
Sebastien Guillot,$^{11}$\newauthor
Deepto Chakrabarty$^{12}$
\vspace{0.2cm}\\
$^{1}$ Istanbul University, Science Faculty, Department of Astronomy and Space Sciences, Beyaz\i t, 34119, \.Istanbul, Turkey \\
$^{2}$ Istanbul University Observatory Research and Application Center, Istanbul University 34119, \.Istanbul Turkey\\
$^{3}$ Istanbul University, Graduate School of Sciences, Department of Astronomy and Space Sciences, Beyaz\i t, 34119, \.Istanbul, Turkey \\
$^4$ Center for Relativistic Astrophysics, School of Physics, Georgia Institute of Technology, 837 State Street, Atlanta, GA 30332-0430, USA\\
$^{5}$ Department of Astronomy, University of Maryland, College Park, MD 20742, USA \\
$^{6}$ Astrophysics Science Division, NASA Goddard Space Flight Center, Greenbelt, MD 20771, USA\\
$^7$ National Space Institute, Technical University of Denmark, Elektrovej 327-328, DK-2800 Lyngby, Denmark\\
$^{8}$ Faculty of Engineering and Natural Sciences, Sabanc\i~University, Orhanl\i-Tuzla 34956, \.Istanbul, Turkey \\
$^{9}$ Astrophysics Science Division and Joint Space-Science Institute, NASA's Goddard Space Flight Center, Greenbelt, MD 20771, USA \\
$^{10}$ School of Physics and Astronomy, University of Southampton, Southampton, SO17 1BJ, UK \\
$^{11}$ IRAP, UPS-OMP, CNRS, CNES, 9 avenue du Colonel Roche, BP 44346, F-31028 Toulouse Cedex 4, France \\
$^{12}$ MIT Kavli Institute for Astrophysics and Space Research, Massachusetts Institute of Technology, Cambridge, MA 02139, USA 
}

\date{Accepted XXX. Received YYY; in original form ZZZ}

\pubyear{2021}

\begin{document}
\label{firstpage}
\pagerange{\pageref{firstpage}--\pageref{lastpage}}
\maketitle
\begin{abstract}

We present spectral and temporal properties of all the thermonuclear X-ray bursts observed from \source by the Neutron Star Interior and Composition Explorer (\nicer) between 2017 July and 2021 April. This is the first systematic investigation of a large sample of type I X-ray bursts from \source with improved sensitivity at low energies. We detect 22 X-ray bursts including two short recurrence burst events in which the separation was only 451 s and 496~s. We perform time resolved spectroscopy of the bursts using the fixed and scaled background (\fa method) approaches. We show that the use of a scaling factor to the pre-burst emission is the statistically preferred model in about 68\% of all the spectra compared to the fixed background approach. Typically the \fa values are clustered around 1--3, but can reach up to 11 in a burst where photospheric radius expansion is observed. Such \fa values indicate a very significant increase in the pre-burst emission especially at around the peak flux moments of the bursts. We show that the use of the \fa factor alters the best fit spectral parameters of the burst emission. Finally, we employed a reflection model instead of scaling the pre-burst emission. We show that reflection models also do fit the spectra and improve the goodness of the fits. In all cases we see that the disc is highly ionized by the burst emission and the fraction of the reprocessed emission to the incident burst flux is typically clustered around 20\%.
\end{abstract}




\section{Introduction}

Thermonuclear~(type I) X-ray bursts are caused by the unstable ignition of H/He deposited on the surface of a neutron star in a low-mass X-ray binary \citep[e.g.][]{1986ASIC..167.....T, 1993SSRv...62..223L, 1998ASIC..515..419B, 2006csxs.book..113S, 2021ASSL..461..209G}. When the gas is accreted from a low mass companion~($\lesssim$ 1 M$_\odot$) it accumulates on the surface of the neutron star where it is hydrostatically compressed and heated, the temperature and pressure increases and eventually a thermonuclear explosion can be triggered \citep{1981ApJ...247..267F, 2003ApJ...599..419N}.
These events are observed in X-rays as sudden increases in intensity reaching up to a factor of ten to a hundred times the persistent level \citep{2003astro.ph..1544S,Galloway2008}. 
Typically, the rise times of bursts can be within 1--10 seconds, while the bursts themselves can last from a few tens of seconds to several minutes. The total energy output of a burst can be $\approx10^{39}-10^{40}$~ergs \citep[e.g.][]{2020ApJS..249...32G}. The intensity produced in thermonuclear bursts can have an effect on the surrounding accretion flow \citep{2018SSRv..214...15D}, which consists of an accretion disc and a hot electron corona, allowing the bursts to affect the dynamics of the accretion process.

High time and moderate energy resolution observations of thermonuclear X-ray bursts could only be systematically performed with the launch of the Rossi X-Ray Timing Explorer~(RXTE) Proportional Counter Array (PCA) in the 2.5--25.0 keV band \citep[see e.g.][]{1997ApJ...485L..83T,Galloway2008, 2013ApJ...777L...9C, 2014A&A...562A..16I, 2019ApJS..245...19B,2020ApJS..249...32G}. Generally, using a single blackbody component fits X-ray burst spectra well in addition to a constant persistent component \citep{Galloway2008}, especially during the cooling tails of bursts \citep{2012ApJ...747...76G}. The persistent component defines the X-ray emission from accretion processes, as measured before and assumed not to change during the burst. Considering a large sample of observations with RXTE/PCA, burst spectra are found to have deviations from this simple blackbody model. \cite{2013ApJ...772...94W, 2015ApJ...801...60W} characterized the deviation from a pure blackbody model using a constant factor to scale the pre-burst emission, that is allowed to vary through a given burst. The so-called {\em \fa method} basically allows the pre-burst emission to vary during the burst to obtain a better fit to the observed data. It was found that allowing the pre-burst emission to vary significantly improves the fits, especially during the bursts where photospheric radius expansion (PRE) is observed and most obviously at around the peak flux times. This variation in the normalization of the pre-burst spectrum is attributed to an increase in the mass accretion rate likely due to Poynting-Robertson drag \citep{1937MNRAS..97..423R,1989ApJ...346..844W,1992ApJ...385..642W}. This interpretation is supported by recent simulations \citep{2018ApJ...867L..28F,2020NatAs...4..541F}. However, the fact that the \fa parameter is only a multiplication factor to the persistent emission and does not allow for a change in the spectral shape prevents the method from helping understand what really varies within the pre-burst emission.

Systematic studies of the effects of X-ray bursts to their environment have mostly been limited to observations of superbursts ($ \gtrsim$ 1000 s), which happen rarely \citep{2014ApJ...789..121K,2004ApJ...602L.105B}. These bursts are thought to be powered by the unstable combustion of deeper layers containing helium and carbon, instead of the hydrogen and/or helium expected from a normal type I X-ray burst \citep{2017symm.conf..121I}. The fact that superbursts are much longer allows for higher integration times for X-ray spectroscopic studies, which allows collecting higher quality spectra than a normal typical burst. Detailed spectral analyses of superbursts revealed the presence of a reflection component of the burst emission from the surface of the neutron star off of the accretion disc \citep{2004ApJ...602L.105B}. The reflection spectrum depends on the composition of the reflective material \citep[see for 4U~1820\textminus30 ][]{2004MNRAS.351...57B} as well as the properties of the burst.

\nicer~combines the sensitivity in the soft X-rays (0.2--12~keV) with large effective area at around 1~keV \citep{2012SPIE.8443E..13G,2016SPIE.9905E..1HG,2017AAS...22930903G} together with good energy resolution, which provides an exciting opportunity to study  burst - disc interactions in a much more systematic manner. Taking advantage of the soft X-ray sensitivity of \nicer, \cite{2018ApJ...855L...4K,2018ApJ...856L..37K} reported the detection of a strong soft excess in bursts from \source and 4U~1820\textminus30. In both cases, application of the \fa factor was shown to improve the goodness of the fits to the data with \fa values reaching up to 2.5 and 10 for \source and 4U~1820\textminus30, respectively. In agreement with \cite{2013ApJ...772...94W}, \cite{2018ApJ...855L...4K,2018ApJ...856L..37K} found that the existence and strength of PRE appears to drive the significance of the soft excess. Similarly, for Swift~J1858.6$-$081 \citep{2020MNRAS.499..793B} and for MAXI J1807$+$132 \citep{2021MNRAS.501..261A} application of the \fa method improved the fits, although in some cases adding a second blackbody component also improved the fits \citep{2020MNRAS.499..793B} and may even be preferred as for SAX~J1808.4$-$3658  \citep{2019ApJ...885L...1B}. On the other hand, for XTE~J1739$-$285 \citep{2021ApJ...907...79B} and 4U~1608$-$52 \citep{2021ApJ...910...37G} no statistically significant deviation from a blackbody model has been reported. In both cases the hydrogen column density in the line of sight is significantly larger, which may be affecting the detection of the soft excess.

As a follow-up to these studies, we here present a spectral and temporal analysis of all the bursts observed with \nicer from \source between July 2017 and April 2021. Our goal is to  report the X-ray bursts detected with \nicer and systematically study the soft excess that has been reported to be observed from this source before \citep{2018ApJ...855L...4K}. \source is one of the ideal sources for such studies given the relatively low hydrogen column density in the line of sight and the fact that the source frequently shows outbursts. \source was discovered in 1965 by \citet{1967Sci...156..374F}. The source is transient, showing frequent, roughly one per year outbursts \citep{2014MNRAS.439.2717G}. The system has an orbital period of 18.9~hr \citep{1991A&A...251L..11C}. Intermittent pulsations \citep{2008ApJ...674L..41C} and burst oscillations \citep{1998ApJ...495L...9Z,Galloway2008,2019ApJS..245...19B,2020ApJS..249...32G} at around 549~Hz have been observed, indicating a fast rotating neutron star in the system. Utilizing phase resolved near-IR spectroscopic observations, the spectral type of the companion is estimated as $K4$ and the distance and orbital inclination of the system was found to be  6$\pm2$~kpc and between $36^\circ-47^\circ$, respectively \citep{2017MNRAS.464L..41M}. Also,  the X-ray reflection modelling of the persistent emission gives an inclination of $i$ = $(36 \pm2)^\circ $ \citep{2016ApJ...819L..29K,2017ApJ...847..135L}. In good agreement with these results, the distance of this source is calculated as d = 5.55 $\pm$ 3.57~kpc using the parallax measurement in the recently published GAIA~EDR3 catalog \citep{2021A&A...649A...1G}. A total of 96 X-ray bursts have been reported from \source in the MINBAR catalog \citep{2020ApJS..249...32G}, implying a burst rate of 0.1 bursts per hour, which is calculated using the available exposure time while the source is active.


\begin{figure*}
	\includegraphics[scale=0.45]{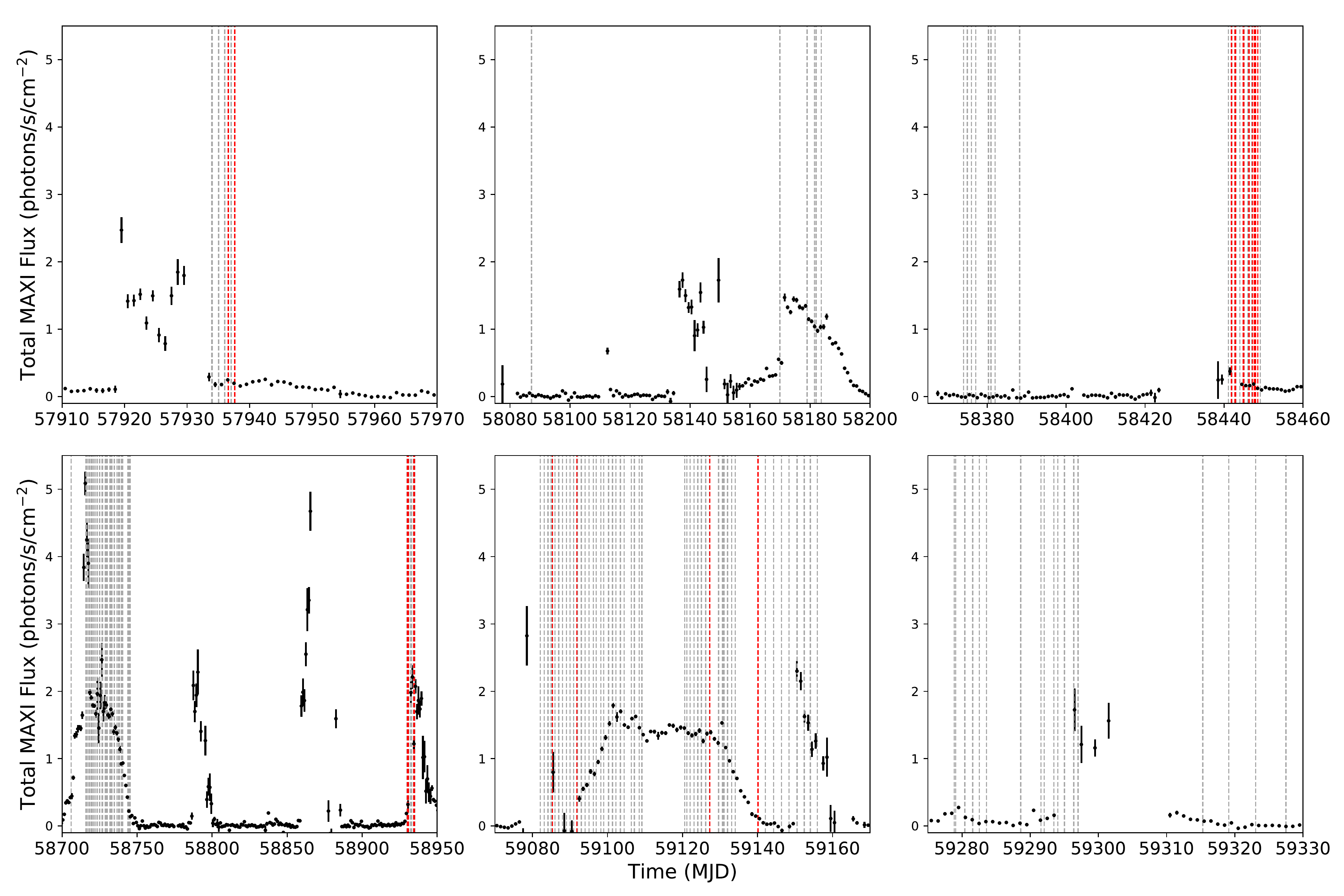}
    \caption{Long term lightcurve of \source as observed by MAXI in the 2--20.0~keV band. The panels show  different time intervals when \nicer observed the source. The MAXI data is shown with black dots, the grey vertical dashed lines indicate the \nicer observation times. The red dashed lines show the times when a thermonuclear X-ray burst is detected. Note that some bursts which occurred with short recurrence are not discernible in the figure.}
    \label{fig:long_term_lc}
\end{figure*}


\section{Observations and Data Analysis}
\label{sec:Observation}

Starting from June 20th, 2017 \nicer observed \source for a total of 620~ks with 134 observations. We used all of the observational data of \source from \nicer, publicly available through the High-Energy Astrophysics Science Archive Research Centre (HEASARC)\footnote{\url{https://heasarc.gsfc.nasa.gov}}. The ObsIDs used in this study cover the following ranges 0050340104--109, 1050340101--125, 2050340101--129, 3050340101--159 and 4050340101-121. Note that ObsIDs starting with a zero were collected during instrument validation and are not publicly available through HEASARC. Applying the standard filtering criteria results in a total clean exposure time of 347~ks, using HEASOFT v.6.27.2, NICERDAS v7a, and the calibration files as of 2020/07/27. We applied barycentric correction to the event files using the source coordinates\footnote{\url{http://simbad.u-strasbg.fr/simbad/sim-ref?bibcode=2018yCat.1345....0G}}  (J2000) R.A. 19h11m16.05s DEC. +00$^{\circ}35^{\prime} 05.8^{\prime\prime}$ and JPLEPH.430 ephemerides \citep{2014IPNPR.196C...1F}.

\begin{figure}
	\includegraphics[width=\columnwidth]{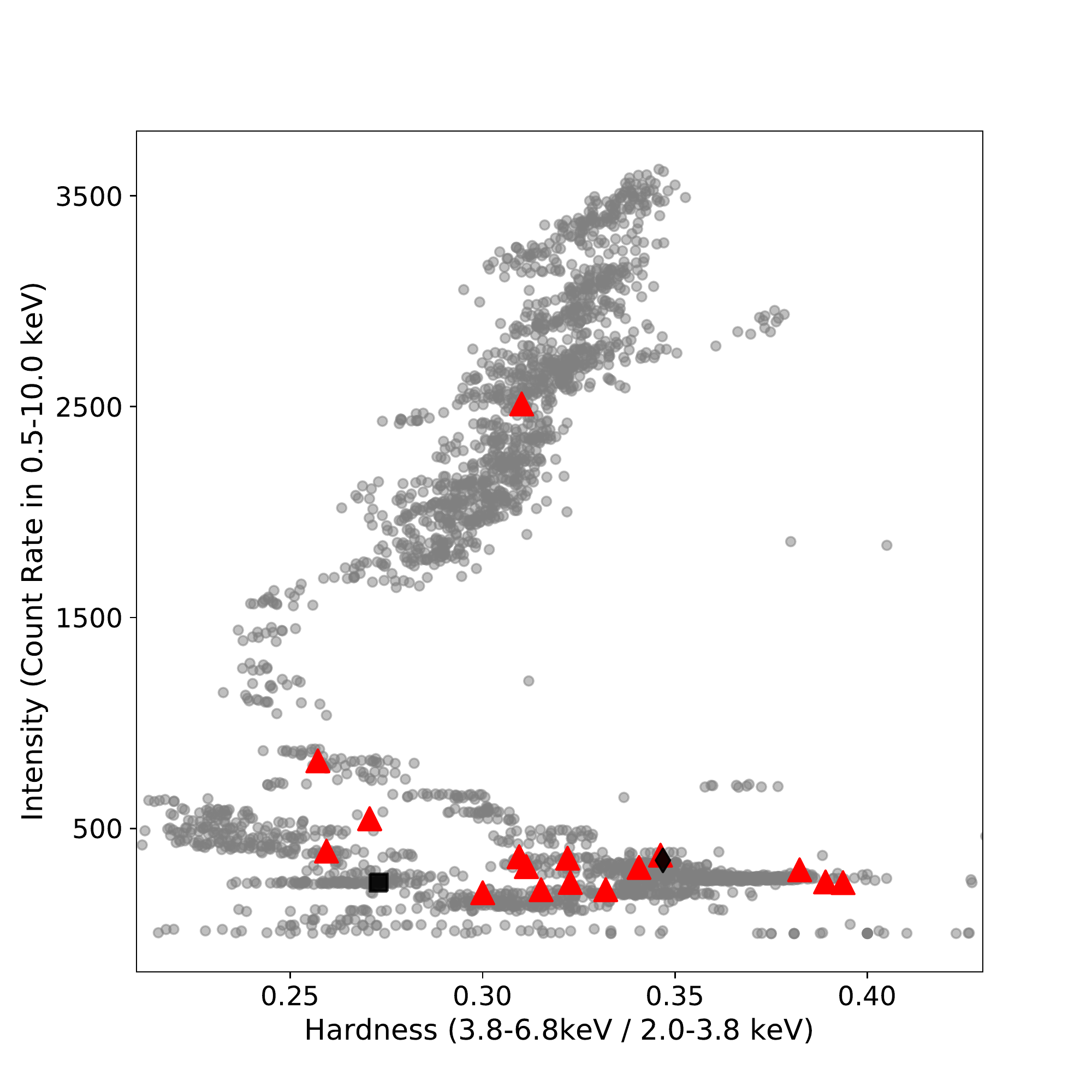}
    \caption{Hardness-Intensity diagram extracted from all \nicer observations of \source. Hardness ratio is defined as the ratio of the count rate in 3.8--6.8 keV and the 2.0--3.8 keV bands. The locations of the detected X-ray bursts are indicated by red triangles. Also in black diamond and square we show the locations of the short recurrence bursts 15-16, and 21-22, respectively.}
    \label{fig:hardness}
\end{figure}

\autoref{fig:long_term_lc} shows the longterm lightcurve of \source as observed by MAXI in the 2--20~keV band during \nicer observations. Individual panels show the time intervals when a \nicer observation was performed. It can be seen that \nicer observations are typically clustered around outbursts of the source. We also generated a hardness intensity diagram of the source based on all of the \nicer observations used here. For this purpose we generated 0.5--10~keV, 2.0--3.8~keV, and 3.8--6.8~keV lightcurves of all the clean event files with a time resolution of 128~s and computed the hardness ratio by dividing the count rates in the 3.8--6.8 keV by the count rates in the 2--3.8~keV band. The resulting hardness intensity diagram is shown in \autoref{fig:hardness}.

We searched for thermonuclear X-ray bursts within all of the unfiltered data and found 22 such events. Some of the basic properties of these bursts are given in \autoref{tab:bursts} and 0.5--10~keV lightcurves are given in Appendix \ref{app:lcbursts}. We also show the locations of the bursts in the hardness-intensity diagram using data obtained just prior to each burst (see \autoref{fig:hardness}). Note that the time burst 12 occurred is filtered out with the standard filtering criteria, we therefore used unfiltered events for this burst. 

We calculate the start, rise and decay e-folding times of the bursts using 0.5--10~keV lightcurves with a temporal resolution of 0.5~s. The start time is defined as the time-bin just before the first moment the burst rate increases to above 4~$\sigma$ of the average count rate, which is calculated from data obtained over the prior 25~s. The rise time of each burst, is defined as the time for a burst to reach within 5\% of the peak count-rate starting from the burst start. Finally, the decay e-folding time is defined as the time for the count rate to decrease by a factor $e$ after the peak. For each burst these values are given in \autoref{tab:bursts}. Here and throughout the paper, we quote one sigma statistical uncertainties of all the measurement unless otherwise stated.

\begin{table*}
\caption{Some characteristic properties of all the detected with \nicer thermonuclear X-ray bursts from \source. Parameters are derived from 0.5--10.0~keV lightcurves with a time resolution of 0.5~s, therefore the uncertainties in the rise and decay times are 0.5~s.}
\begin{tabular}{cccccccc}
\hline 
Burst No. & MJD (TDB)  & OBSID & Peak-Rate$^a$ & Pre-burst Rate$^b$ & Rise-Time & Decay e-folding time \\
 & & & counts/s & counts/s & s & s \\
\hline
\hline 

1 &57936.584843 & 0050340108 & 1750$\pm$60 & 226$\pm$7 & 7.0  & 11.5 \\
2 &57937.615500& 0050340109 & 2620$\pm$80 & 272$\pm$3 & 3.0 & 17.5 \\
3 &58441.896973 & 1050340117 & 3550$\pm$90& 845$\pm$3 & 4.5 & 13.5 \\
4 & 58442.801000 & 1050340118 & 3490$\pm$90 & 571$\pm$3 & 5.5 & 9.5 \\
5 & 58442.932044& 1050340118 & 3300$\pm$90& 581$\pm$3& 4.5 & 14.0 \\
6 & 58444.845108 & 1050340120 & 2780$\pm$80& 396$\pm$2 & 6.0 & 21.0 \\
7 & 58444.978103 & 1050340120 & 2700$\pm$80& 386$\pm$3& 4.5 & 22.0 \\
8 & 58446.319741 & 1050340122 & 3250$\pm$80&356$\pm$2& 3.5 & 15.5 \\
9 & 58447.182637& 1050340123 & 3030$\pm$80& 283$\pm$2 & 3.0 & --$^c$ \\
10 &58447.686326& 1050340123 & 2780$\pm$90&233$\pm$2& 7.8 & 14.5 \\
11 &58447.949824& 1050340123 & 2830$\pm$90& 231$\pm$2& 6.0 & 17.0 \\
12 & 58448.526803 & 1050340124 & 2780$\pm$90& 216$\pm$2& 6.5 & 18.0 \\
13 & 58930.162945& 3050340101 & 2720$\pm$80 & 325$\pm$3 & 5.5 & 21.5 \\
14 & 58930.735030 & 3050340101 & 2640$\pm$80& 349$\pm$3 & 9.5 & 23.0 \\
15 & 58934.487522 & 3050340105 & 2760$\pm$80& 367$\pm$3& 3.5 & 23.5 \\
16 & 58934.493274 & 3050340105 & 1450$\pm$ 60& 369$\pm$3& 9.0 & 10.5\\ 
17 &58934.807525& 3050340105 & 2730$\pm$80& 387$\pm$3& 8.0 & 17.0 \\
18 &59085.311252 & 3050340111 & 8330$\pm$130& 385$\pm$ 4& 3.0 & 6.5 \\
19 &59091.838177 & 3050340117 & 2900$\pm$80 & 443$\pm$3 & 3.0 & 19.0 \\
20 &59127.285573& 3050340142 & 5730$\pm$130 & 2125$\pm$60 & 2.0 & 9.0 \\
21 &59140.123223 & 3050340150 & 2110$\pm$70 & 240$\pm$2 & 3.5 & 10.0 \\
22 & 59140.128449& 3050340150 & 2820$\pm$80 & 235$\pm$2& 5.0 & 9.0 \\
\hline
\end{tabular}
\label{tab:bursts}\\
\footnotesize{$^a$ Pre-burst count rates are subtracted. }\\
\footnotesize{$^b$ Calculated as the average count rate 100~s prior to the burst start time. Uncertainties reflect the standard error of the average of all the count rates used.}\\
\footnotesize{$^c$ Observation stops before the burst ends.}\\
\end{table*}

\begin{figure*}
	\includegraphics[width=\columnwidth]{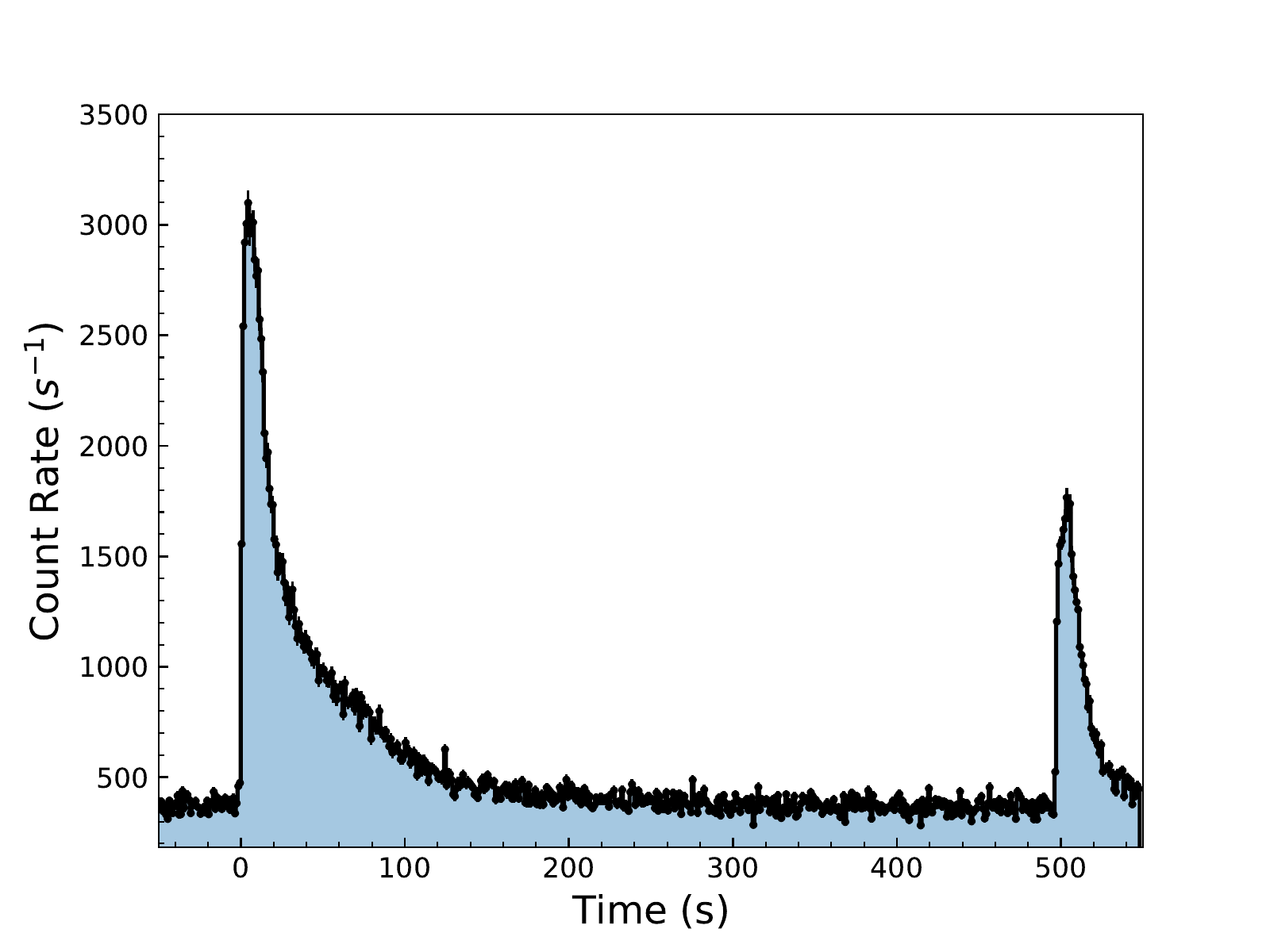}
	\includegraphics[width=\columnwidth]{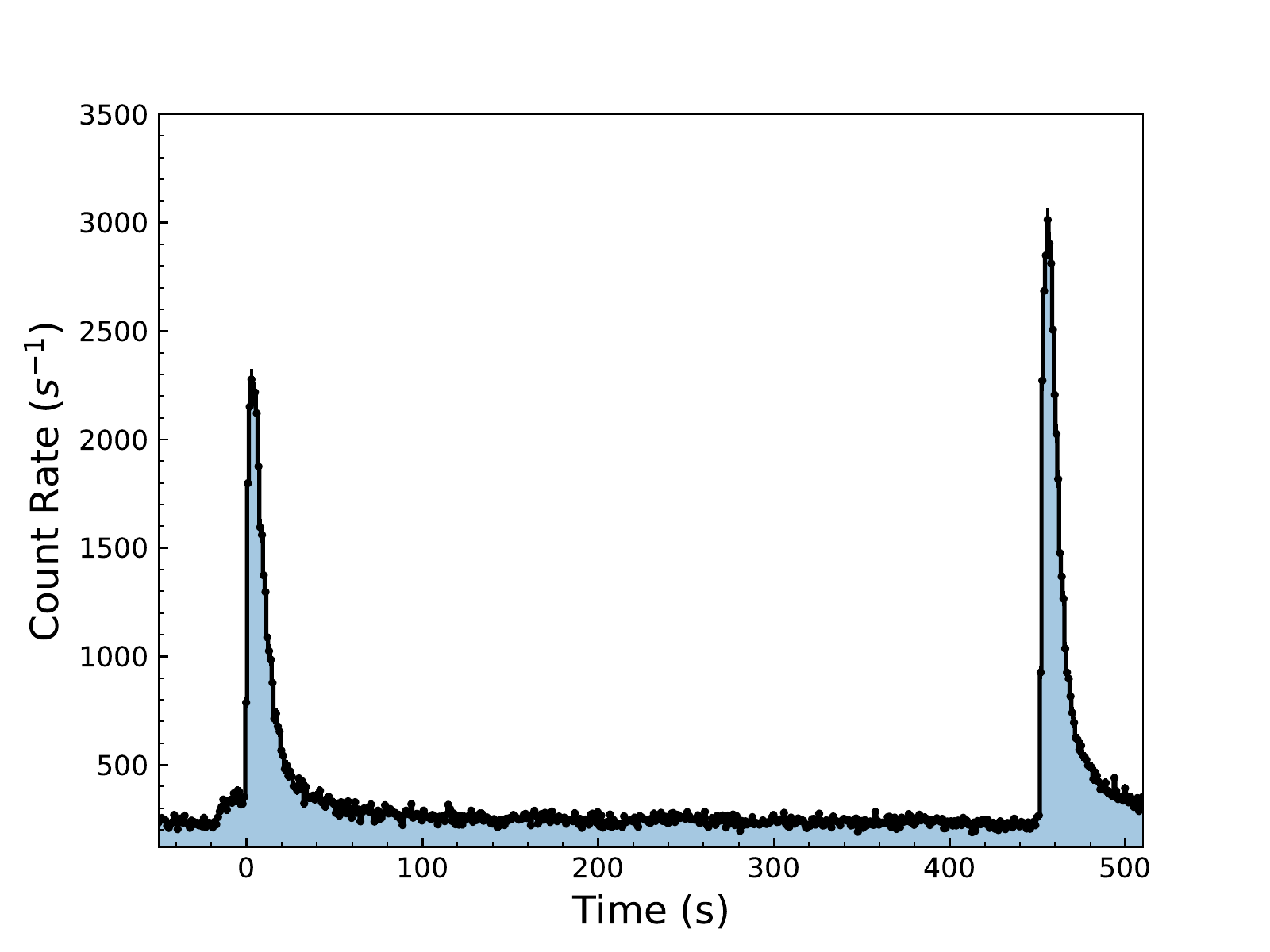}
    \caption{Lightcurves in the 0.5-10.0 keV range of bursts of 15, 16 (left panel) and 21 and 22 (right panel). The time resolution in these lightcurves is 1~s and pre-burst count rates are not subtracted.}
    \label{fig:double_bursts}
\end{figure*}

\subsection{Search for Burst Oscillations}
\label{sec:oscillations}

We searched for burst oscillations in the 0.5--3.0, 3.0--10.0, 0.5--10.0~keV bands starting 20~s before each burst till the end of the burst, which was assumed to happen when the count rate reaches 5\% of the peak, without subtracting the pre-burst rate. We performed a search for each 4~s long segment within 5 Hz of the known oscillation frequency (551$\pm$5 Hz) using the Z$^2_m$ statistic \citep{buccheri} with the number of harmonics, m to be one. We slid 4~s long search window with a one-second increment and repeated the search in the same frequency range for each time segment. We then constructed a dynamic power spectral density for the entire search interval. We do not find any episode in any burst with Z$^2_{m=1}$ power corresponding to the single-trial significance more than $4 \sigma$. Following \cite{2017ApJ...834...21O} we also calculated the fractional rms amplitudes of burst oscillations for each burst and energy interval. The fractional rms amplitudes we calculate using the average values obtained from two time segments just following the peak of each burst range from 0.039--0.064, 0.036--0.072, 0.028--0.048 for 0.5--3.0~keV, 3.0--10.0~keV, and 0.5--10.0~keV ranges, respectively. Only in the brightest burst, burst 18, the upper limits are even smaller 0.020, 0.028, 0.017, respectively for the same energy ranges.


\subsection{Spectroscopy of the Pre-burst Emission}

In order to characterize the persistent emission from the source, we extracted X-ray spectra with an exposure time of 100~s, 120~s prior to the start of each burst. We estimated the background using the version 7b of \texttt{nibackgen3C50} \footnote{\url{https://heasarc.gsfc.nasa.gov/docs/nicer/tools/nicer\_bkg\_est\_tools.html}} tool calculated for each observation \citep{2021arXiv210509901R}. We removed the Focal Plane Modules 14 and 34 from our analysis and used the response and ancillary response files within the NICER CALDB release \texttt{xti20200722}, adjusted to our selection of the modules. 

We fit each of these X-ray spectra using \emph{Sherpa} \citep{2001SPIE.4477...76F} distributed with the CIAO v4.13 with an absorbed disc blackbody plus a power-law model. Note that we also tried several other models commonly used for these sources. However, overall this model provided the best fits with the least number of parameters. For the disc blackbody model we assumed a face-on disc, so the $\cos{\theta}$ term in the normalization of this model is assumed to be 1. We used the {\it tbabs} model assuming interstellar abundances to take into account the interstellar absorption \citep{2000ApJ...542..914W}. We first fit all the data keeping the hydrogen column density free in each observation. We then, calculated an error weighted average of all the hydrogen column density measurements as N$\rm{_H}$ = 0.49$\times10^{22}~\rm{cm}^{-2}$ and used this value throughout the study. Note that there are different  N$\rm{_H}$  values for \source reported by \cite{1999A&A...347L..51C, 2014MNRAS.441.1984C, 2016MNRAS.461.3847G, 2018ApJ...855L...4K,2018ApJ...859L...1B, 2020ApJS..249...32G}. The inferred values cover the range from 0.34 to 0.6  $\times10^{22}~\rm{cm}^{-2}$ by \cite{1999A&A...347L..51C} and \cite{2018ApJ...855L...4K,2018ApJ...859L...1B}, respectively, in agreement with the value used here. We present our results in Table \ref{tab:pers} for each X-ray spectrum extracted just prior to each burst.

\subsection{Time Resolved Spectroscopy}
\label{sec:time_resolved}

In order to inspect time variation of spectral properties during the bursts  we extracted time resolved X-ray spectra. For this purpose we followed the methods outlined in \cite{2012ApJ...747...76G, 2021ApJ...910...37G}. From the start of each burst, up to the peak we extracted X-ray spectra with exposure times of 0.5~s. From the peak, depending on the count rate we extracted X-ray spectra with exposure times 0.5, 1.0, or 2.0~s, following a similar approach to \cite{Galloway2008} and \cite{2012ApJ...747...76G}. 
As before we performed the spectral analysis using \emph{Sherpa}  \citep{2001SPIE.4477...76F} distributed with the CIAO v4.13 together with custom python scripts (using Astropy  \citealt{2018AJ....156..123A}, NumPy \citealt{van2011numpy}, Matplotlib \citealt{Hunter:2007}, and Pandas \citealt{mckinney-proc-scipy-2010}). After subtracting only the instrumental and diffuse sky background as calculated by the \texttt{nibackgen3C50} tool, we grouped each spectrum to have at least 50 counts per channel in the 0.5$-$10.0~keV range. 

For the fitting of the burst spectra we followed three approaches. The first one is the classical approach, and involves the use of a fixed pre-burst emission for the emission from the system and a blackbody function for the burst emission.  For a second approach we used the \fa method which involves the use of a scaling factor to the pre-burst emission to provide statistically better fits. Finally, following the results of the \fa method we also employed a model taking into account the reprocessing of the burst emission by the accretion disk following the approaches used by \cite{2004MNRAS.351...57B,2004ApJ...602L.105B}. 

For the classical blackbody approach, we fit the resulting burst spectra with a blackbody function and the disc blackbody plus a power-law model, which is fixed to its best fit pre-burst values as given in \autoref{tab:pers}. 
Independent of the resulting $\chi^2$ values from the simple blackbody fits, we re-fit the data with the addition of the \fa parameter, a scaling factor to multiply the pre-burst emission. We then used the f-test to determine whether the introduction of the \fa factor was statistically required. In cases where the chance probability of the decrease in the $\chi^2$ values are higher than 5\%, we fixed the \fa parameter at unity, so that the pre-burst emission remained constant throughout the burst, and only used a blackbody function in the determination of the spectral parameters. If the chance probability is lower than 5\% then we kept the results based on the \fa approximation for the bursts. 

Finally, to understand whether reflection of burst emission by the accretion disc can account for the soft excess, we employed existing reflection models 
\footnote{\url{https://heasarc.gsfc.nasa.gov/xanadu/xspec/models/bbrefl.html}} \citep{2004ApJ...602L.105B, 2004MNRAS.351...57B}. These models basically take into account the reflection spectrum from an accretion disc illuminated by the burst emission, which is assumed to have a Planckian shape. In addition to the normalization factor, there are two free parameters: the log of the ionization parameter, $\xi$, and the $kT$ of the illuminating blackbody. The ionization parameter is defined as $\xi$ = 4$\pi\times F/n$, where $F$ is the flux (in \fluxcgs) of the blackbody and $n$ is the hydrogen number density of the reflector (which is assumed to be a constant density slab). Although we tried all of the available variations of the \emph{bbrefl} model within XSPEC library \citep{Arnaud1996} we present the results for a specific version of the model where the abundance in the disc is assumed to be solar and the disc has a hydrogen number density of $n$ = 10$^{18}\,\rm{cm}^{-3}$, which is more appropriate for disks around neutron stars. The tabulated values for the incident temperature in the model we used covers the range kT~=~0.25--3.5~keV, whereas the range for the ionization parameter is within 1.9--3.44 in steps of 0.05 for each parameter. We first fit the X-ray spectra obtained at the peak flux moment of each burst. We find that in each case the addition of the reflection model provides a better fit than a the fixed background approach. We then applied the reflection model to all of the X-ray spectra we extracted from all of the bursts where an \fa component is determined to be required.

\begin{table*}
\caption{Best fit model results for pre-burst X-ray spectra of \source using an absorbed disk blackbody plus a power-law model.}
\begin{tabular}{cccccccc}
\hline 
\hline 
Burst No.&kT & Norm$_{\rm{DBB}}$ & $\Gamma$ & Flux* & $\chi^2$ / dof \\
& (keV)&(R$^2_{\rm km}/\rm{D}_{10 {\rm kpc}}^{2}$)& &($\times10^{-9}$~\fluxcgs)& & \\
\hline    
1&0.87$\pm$0.10&12.3$\pm$7.0&1.14$\pm$0.07&1.4$\pm$0.2&374.88 / 280\\
2&1.26$\pm$0.17&5.0$\pm$2.3&1.31$\pm$0.04&1.6$\pm$0.2&316.26 / 299\\
3&0.62$\pm$0.01&473$\pm$38&1.63$\pm$0.03&3.5$\pm$0.2&410.70 / 391\\
4&0.42$\pm$0.01&859$\pm$118&1.74$\pm$0.03&2.6$\pm$0.1&353.22 / 351\\
5&0.45$\pm$0.01&698$\pm$105&1.74$\pm$0.03&2.6$\pm$0.2&368.72 / 354\\
6&0.32$\pm$0.02&969$\pm$307&1.82$\pm$0.03&1.8$\pm$0.1&384.25 / 306\\
7&0.28$\pm$0.02&1446$\pm$484&1.85$\pm$0.03&1.8$\pm$0.1&322.44 / 302\\
8&0.27$\pm$0.05&1086$\pm$741&1.82$\pm$0.03&1.7$\pm$0.1&352.52 / 296\\
9&2.24$\pm$0.12&1.3$\pm$0.4&2.18$\pm$0.10&1.3$\pm$0.2&359.42 / 280\\
10&2.01$\pm$0.09&2.0$\pm$0.5&2.18$\pm$0.19&1.1$\pm$0.2&267.50 / 245\\
11&2.13$\pm$0.12&1.4$\pm$0.4&2.06$\pm$0.14&1.1$\pm$0.2&263.39 / 236\\
12&1.86$\pm$0.08&2.6$\pm$0.6&2.20$\pm$0.21&1.0$\pm$0.2&266.16 / 226\\
13&2.49$\pm$0.13&1.2$\pm$0.3&2.13$\pm$0.09&1.6$\pm$0.3&282.28 / 302\\
14&0.16$\pm$0.03&9121$\pm$9154&1.74$\pm$0.02&1.8$\pm$0.1&282.22 / 303\\
15&2.51$\pm$0.12&1.4$\pm$0.3&2.2$\pm$0.09&1.8$\pm$0.4&347.39 / 325\\
16&2.54$\pm$0.12&1.3$\pm$0.3&2.1$\pm$0.09&1.9$\pm$0.3&361.09 / 327\\
17&2.40$\pm$0.24&0.7$\pm$0.3&1.9$\pm$0.07&1.9$\pm$0.3&352.93 / 325\\
18&0.46$\pm$0.01&722$\pm$72&1.73$\pm$0.06&1.5$\pm$0.1&317.81 / 270\\
19&0.37$\pm$0.02&659$\pm$168&1.79$\pm$0.03&2.0$\pm$0.1&388.55 / 321\\
20&1.26$\pm$0.02&108.5$\pm$6.4&1.33$\pm$0.03&10.0$\pm$0.4&728.50 / 567\\
21&1.44$\pm$0.33&1.2$\pm$0.9&2.10$\pm$0.06&1.0$\pm$0.1&234.11 / 246\\
22&1.12$\pm$0.47&1.7$\pm$4.6&2.06$\pm$0.05&1.0$\pm$0.1&253.96 / 244\\
\hline
\end{tabular}
\label{tab:pers}\\
\footnotesize{$*$ Unabsorbed 0.5$-$10~keV flux.}\\
\end{table*}

\begin{table*}
\caption{Spectral parameters obtained at the peak flux moment for each burst with or without the application of the \fa method are shown. Fluence of each burst is also presented. The fluences are calculated using the results of the \fa method.}
\begin{tabular}{ccccccccccc}
\hline
\hline

Burst No.& \fa ~at Peak& \multicolumn{2}{c}{Peak Flux $^*$} & \multicolumn{2}{c}{kT at peak (keV)}  & \multicolumn{2}{c} {BB Norm at peak } & Fluence$^+$ \\
  &  & with \fa & without \fa  & with \fa  & without \fa & with \fa & without \fa & with \fa \\
\hline
1 & 3.2$\pm$0.4 & 1.7$\pm$ 0.5 & 1.7$\pm$0.1 & 2.8$\pm$0.5 & 1.6$\pm$0.1 & 30$\pm$13 & 229$\pm$28&11.2$\pm$0.5 \\
2 & 3.0$\pm$0.4 & 4.2$\pm$0.5 & 3.6$\pm$0.3 & 2.3$\pm$0.2 & 1.9$\pm$0.1 & 137$\pm$26 & 257$\pm$27&24.2$\pm$0.7 \\
3 & 2.1$\pm$0.1 & 6.0$\pm$0.7 & 4.4$\pm$0.4 & 2.6$\pm$0.2 & 1.9$\pm$0.1 & 132$\pm$21& 307$\pm$33&49$\pm$0.9 \\
4 & 2.1$\pm$0.2 & 4.7$\pm$0.5 & 3.9$\pm$0.3 & 2.2$\pm$0.1 & 1.8$\pm$0.1 & 189$\pm$28 & 345$\pm$35&37$\pm$0.7 \\
5 & 2.0$\pm$0.2 & 4.4$\pm$0.4 & 3.8$\pm$0.3 & 2.1$\pm$0.1 & 1.8$\pm$0.1 & 195$\pm$28& 326$\pm$33&37$\pm$0.7 \\
6 & 2.3$\pm$0.2 & 4.4$\pm$0.5 & 3.7$\pm$0.3 & 2.3$\pm$0.2 & 2.0$\pm$0.1 & 151$\pm$24 & 264$\pm$29&41.7$\pm$0.9 \\
7 & 2.1$\pm$0.2 & 3.6$\pm$0.3 & 3.3$\pm$0.3 & 2.1$\pm$0.1 & 1.8$\pm$0.1 & 184$\pm$27& 288$\pm$30&42.0$\pm$0.8 \\
8 & 4.3$\pm$0.3 & 9.8$\pm$1.5 & 5.0$\pm$0.5 & 3.8$\pm$0.4 & 2.2$\pm$0.1 & 58$\pm$13 & 212$\pm$25&60.6$\pm$1.6 \\
9 & 3.5$\pm$0.3 & 6.8$\pm$0.9 & 4.7$\pm$0.5 & 3.0$\pm$0.3 & 2.1$\pm$0.1 & 91$\pm$17 & 212$\pm$25&15.9$\pm$0.9 \\
10 & 4.0$\pm$0.4 & 6.4$\pm$1.0 & 4.6$\pm$0.5 & 3.0$\pm$0.3 & 2.3$\pm$0.2 & 83$\pm$19 & 160$\pm$22&41.5$\pm$1.2 \\
11 & 4.7$\pm$0.5 & 6.2$\pm$1.1 & 4.1$\pm$0.4 & 3.1$\pm$0.4 & 2.0$\pm$0.1 & 72$\pm$19 & 255$\pm$33&37.3$\pm$1.1 \\
12 & 4.1$\pm$0.3 & 7.1$\pm$0.9 & 4.7$\pm$0.4 & 3.3$\pm$0.3 & 2.3$\pm$0.1 & 69$\pm$12 & 178$\pm$16&32.0$\pm$1.3 \\
13 & 3.0$\pm$0.3 & 4.6$\pm$0.6 & 3.6$\pm$0.3 & 2.6$\pm$0.3 & 1.9$\pm$0.1 & 97$\pm$21 & 248$\pm$28&48.9$\pm$1.1 \\
14 & 2.4$\pm$0.2 & 4.4$\pm$0.5 & 3.7$\pm$0.3 & 2.4$\pm$0.1 & 2.0$\pm$0.1 & 119$\pm$20 & 214$\pm$24&41.1$\pm$0.9 \\
15 & 3.1$\pm$0.3 & 5.9$\pm$0.8 & 4.2$\pm$0.3 & 3.0$\pm$0.3 & 2.1$\pm$0.1 & 80$\pm$16 & 191$\pm$16&41.6$\pm$1.3 \\
16 & 1.8$\pm$0.2 & 1.4$\pm$0.2 & 1.3$\pm$0.1 & 1.8$\pm$0.2 & 1.7$\pm$0.1& 115$\pm$29 & 165$\pm$26&10.9$\pm$0.3 \\
17 & 3.1$\pm$0.3 & 4.3$\pm$0.5 & 3.4$\pm$0.3 & 2.4$\pm$0.2 & 2.0$\pm$0.1& 121$\pm$22& 215$\pm$23&40.7$\pm$1.0 \\
18 & 5.8$\pm$0.4 & 10.0$\pm$1.6 & 4.9$\pm$0.4 & 3.1$\pm$0.3 & 1.9$\pm$0.1 & 121$\pm$27 & 323$\pm$37&47.7$\pm$1.4 \\
19 & 2.3$\pm$0.2 & 5.5$\pm$0.7 & 4.2$\pm$0.4 & 2.7$\pm$0.2 & 2.1$\pm$0.1 & 107$\pm$18 & 208$\pm$24&45.7$\pm$1.0 \\
20 & 2.5$\pm$0.1 & 7.7$\pm$1.0 & 5.1$\pm$0.4 & 2.9$\pm$0.3 & 1.8$\pm$0.1 & 115$\pm$25 & 466$\pm$ 47&53.3$\pm$1.4 \\
21 & 2.4$\pm$0.3 & 2.6$\pm$0.3 & 2.3$\pm$0.2 & 1.9$\pm$0.2 & 1.7$\pm$0.1 & 170$\pm$30 & 261$\pm$26&10.7$\pm$0.3 \\
22 & 3.5$\pm$0.3 & 3.7$\pm$0.4 & 3.0$\pm$0.2 & 2.2$\pm$0.2 & 1.7$\pm$0.1 & 138$\pm$23 & 381$\pm$36&18.6$\pm$0.5 \\
\hline
\end{tabular}
\label{tab:peaks}\\
\footnotesize{$*$ Unabsorbed bolometric flux in units of $\times10^{-8}$~\fluxcgs.}\\
\footnotesize{$+$ Fluence in units of $\times10^{-8}$~\flencecgs.}\\
\end{table*}


\section{Results}
\label{Sec:Results}

We here report the detection of 22 X-ray bursts observed from \source by \nicer using all of the observations obtained since July 2017 till April 2021. We note that, as evident in the upper right corner panel of \autoref{fig:long_term_lc}, 10 of these X-ray bursts occurred in a seemingly failed outburst of the source observed in November 2018, where the source stayed in the low-flux hard state for the entire duration of the outburst.

Burst 20 has been observed when the source flux is at the highest level compared to the rest of the bursts. In fact, all of the other bursts we detected happened during the beginning or the end of the outbursts. Only bursts 15, 16 and 17 happened at around the peak of the March 2020 outburst, but according to MAXI lightcurve these three bursts happened when the source intensity decreased almost by half for a short time interval. 

Bursts 15 and 16 are also interesting because of the short recurrence they show. These bursts are separated by only 496~s. Burst 17, which is also observed the same day, is separated by 7.5 hours from the preceding burst. Similarly on 18 October 2020, we observe another short recurrence time burst event. In this case the bursts are separated by 451~s. The lightcurves of these bursts are shown in \autoref{fig:double_bursts}. We see that for March 2020 event the initial burst was much brighter than the second, while the October 2020 event has the opposite order. 

\subsection{Time Resolved Spectroscopy}
\subsubsection{Application of Different Background Approaches}

\begin{figure}
	\includegraphics[width=\columnwidth]{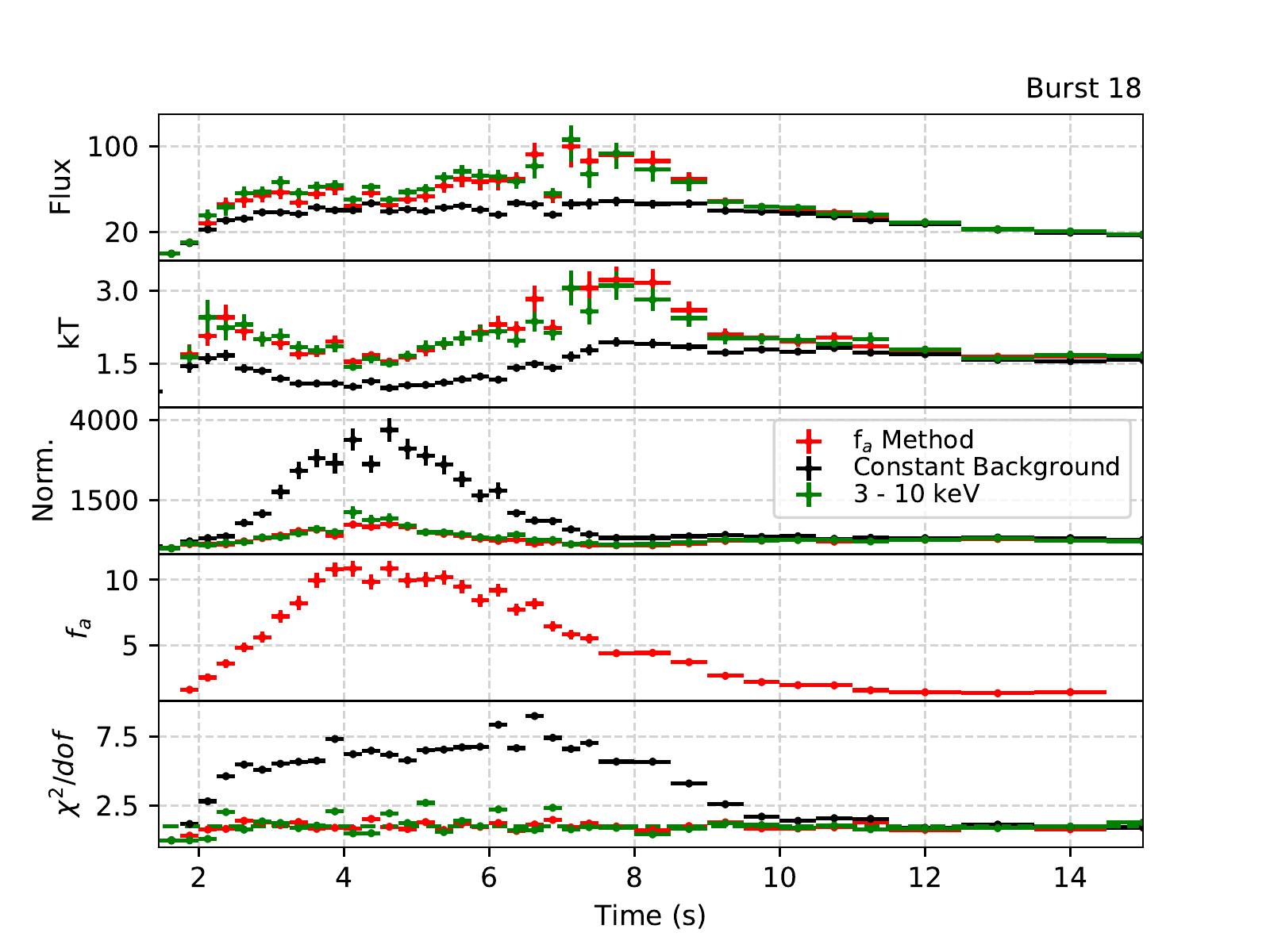}
	
	\includegraphics[width=\columnwidth]{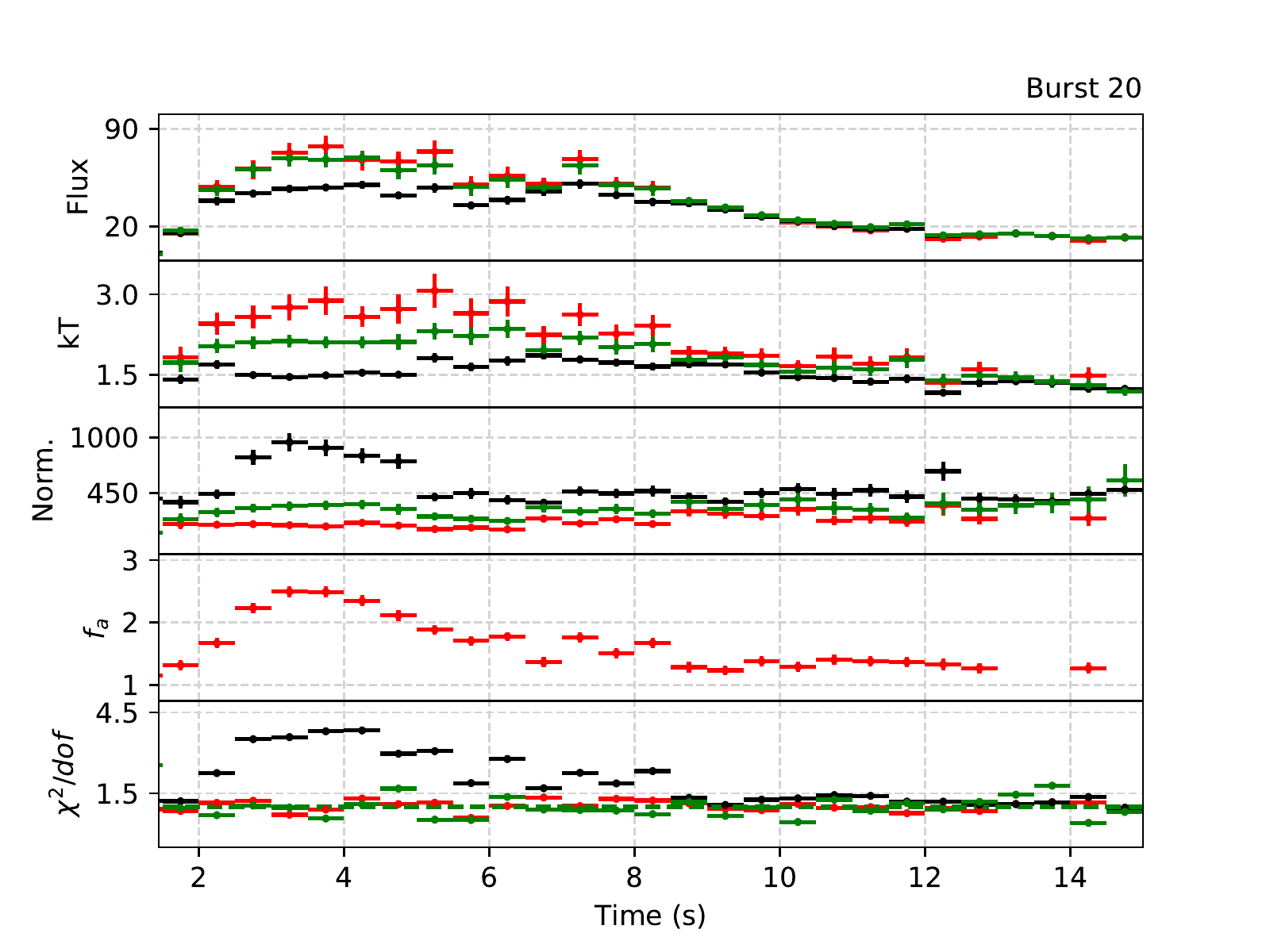}
    \caption{Time evolution of spectral parameters are shown for Burst 18 and 20. We show, from top to bottom: the bolometric X-ray flux, the blackbody temperature, the blackbody normalization (in units of R$^2_{\rm km}/\rm{D}_{10 {\rm kpc}}^{2}$), \fa and finally, the fit statistic. The red symbols show the results of the \fa method, black and green symbols show the results for constant background approach but in the 0.5--10.0~keV or 3--10.0~keV energy ranges, respectively.}
    \label{fig:burst_plots_b18}
\end{figure}

Combining the unique soft X-ray sensitivity with the relatively large number of bursts observed from the source gives us a chance to test different approaches used to fit time resolved X-ray spectra of bursts. The resulting spectral evolution for each burst are summarized in Figures \ref{fig:burst_plots_1}, \ref{fig:burst_plots_2}, \ref{fig:burst_plots_3}, \ref{fig:burst_plots_4} in Appendix \ref{app:sp_time_ev}. These figures show time evolution of flux, blackbody temperature and the blackbody normalization as inferred from the fits as well as the $\chi^2$ values. For a better comparison, in each figure we show the results for both fixed background approach and the \fa method, with best fit \fa values shown in a separate panel. As is evident from figures in Appendix \ref{app:sp_time_ev}, especially around the peaks, the fixed background approximation does not provide a statistically acceptable fit. To test the energy dependency of the fits, we also applied the fixed background model in the 3.0--10.0~keV range. Such an approach resulted in better fits indicating that the poorer fits in the 0.5--10.0~keV range results mostly due to excess emission in the soft X-rays. The resulting spectral evolution using only the data in the 3.0--10.0~keV range is shown for bursts 18 and 20 in \autoref{fig:burst_plots_b18} as an example. 

\begin{figure}
	\includegraphics[width=\columnwidth]{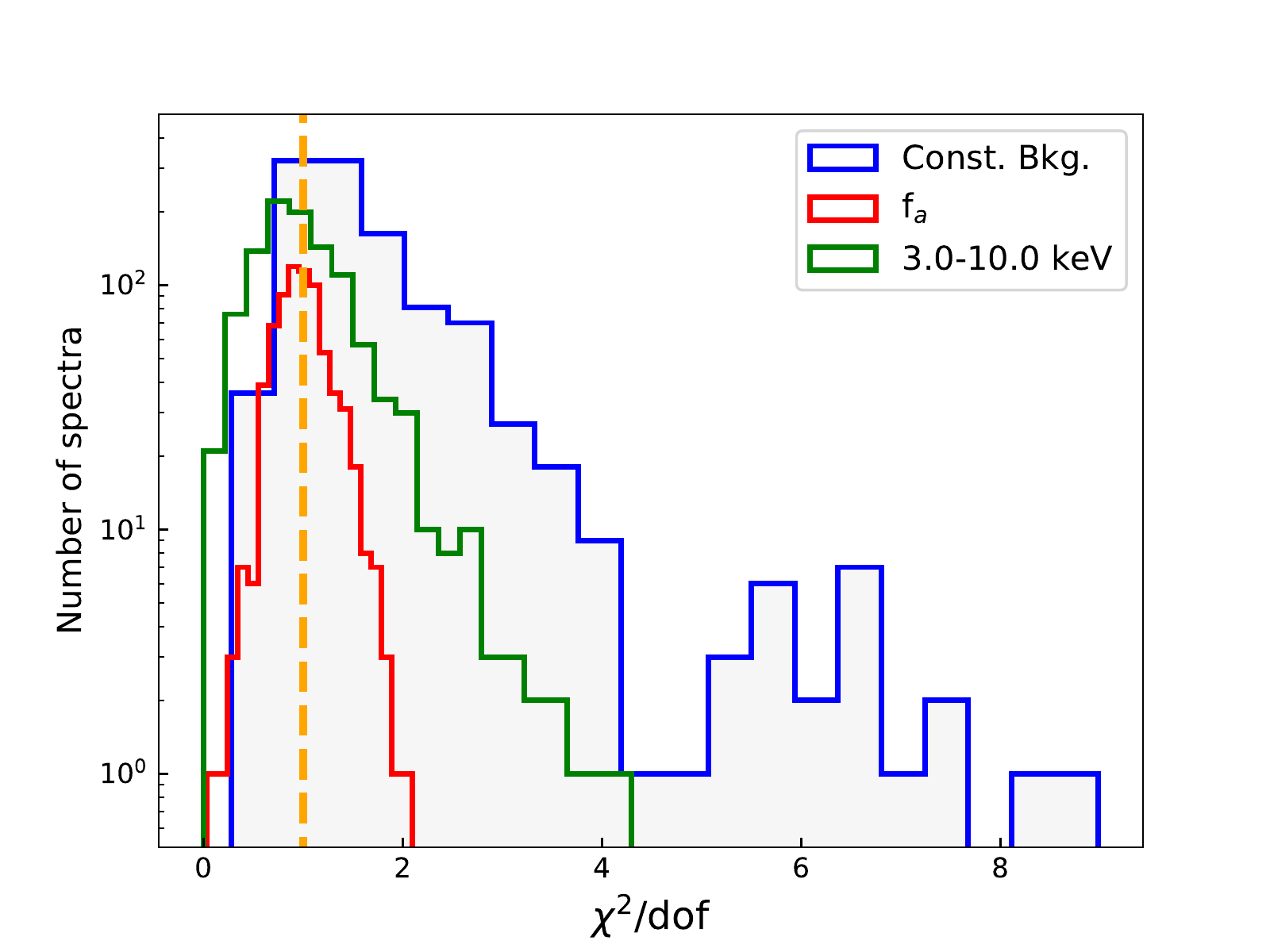}
    \caption{Histograms of all the $\chi^2$/dof values using constant background or the \fa method as well as the resulting statistics values when we only fit to the data in the 3--10~keV range. Only the results where the burst flux is above 10$^{-9}$ \fluxcgs~are included here. The statistical improvement, in cases where an adding an \fa parameter is the preferred model can be seen. Orange vertical line shows the case where $\chi^2$/dof=1.}
    \label{fig:chi_hist}
\end{figure}

\begin{figure}
	\includegraphics[width=\columnwidth]{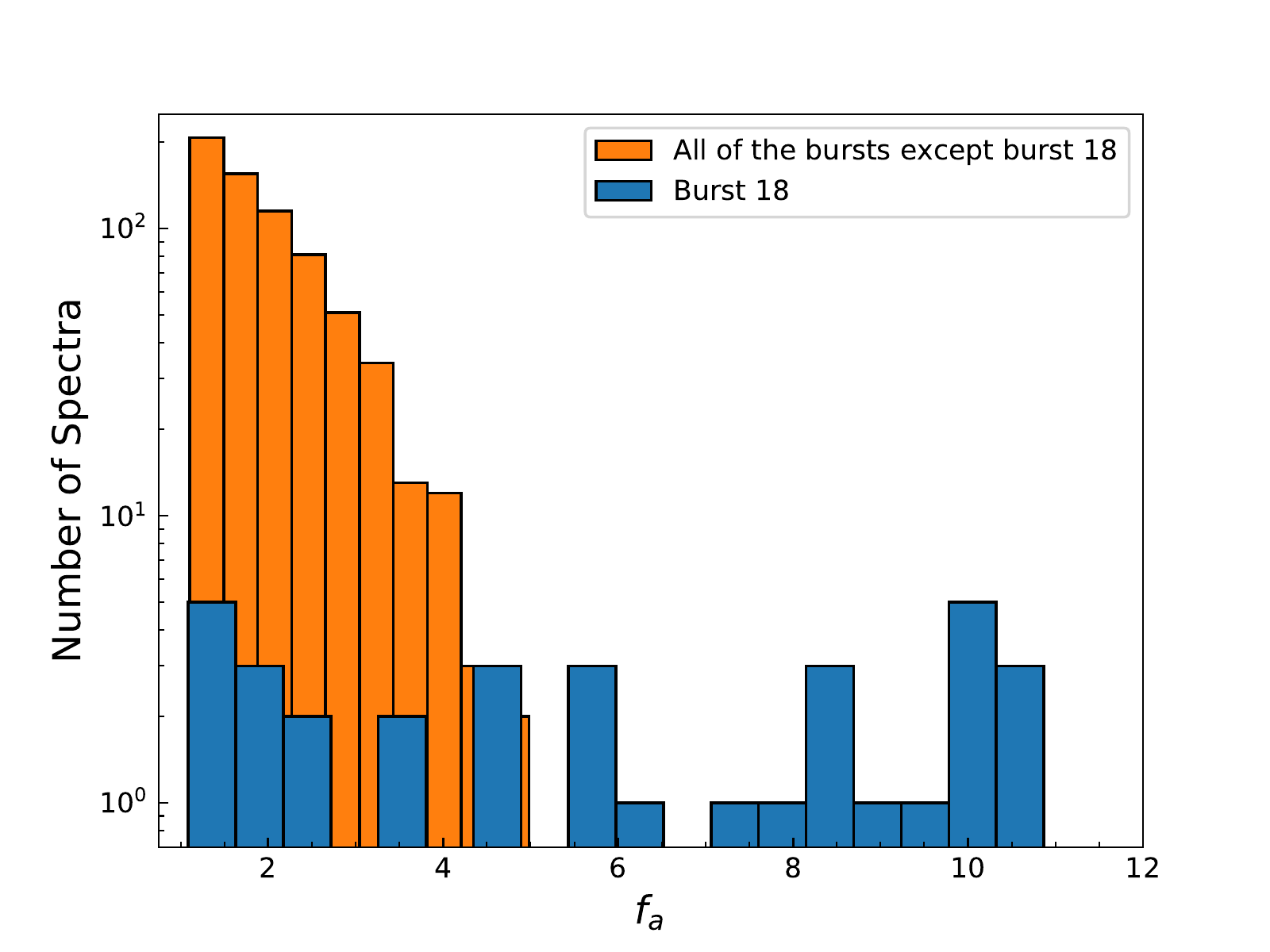}
    \caption{Distribution of \fa values including all of the X-ray spectra where the flux is above 10$^{-9}$~\fluxcgs\,except the \fa values obtained from burst 18, is shown in orange. We also show the \fa values inferred from only burst 18 with blue, where a PRE is observed.}
    \label{fig:fa_hist}
\end{figure}

In \autoref{fig:chi_hist} we show the resulting distributions of $\chi^2$ values for X-ray spectra where the bolometric burst flux is above 10$^{-9}$ \fluxcgs. Such a flux limit is selected because at this level the burst flux becomes comparable to the pre-burst flux of the source (see \autoref{tab:pers}). The $\chi^2$ distributions show that fixed background approach to the time resolved X-ray spectra in the 0.5--10.0~keV range do not provide statistically acceptable fits for a large number of X-ray spectra. In total, we have 1041~X-ray spectra down to the specified flux limit and in 68\% of this sample (707 spectra) adding an \fa component caused a statistically significant improvement in the fits. Figure \ref{fig:fa_hist} shows the histogram of the best fit \fa values, where the bolometric flux and \fa are greater than 10$^{-9}$ \fluxcgs~ and 1.0, respectively. Note that in \autoref{fig:fa_hist}, \fa values for burst 18 are shown in a different color to emphasize the fact that the largest \fa values are obtained when the burst shows evidence for PRE. 

\begin{figure}
	\includegraphics[width=\columnwidth]{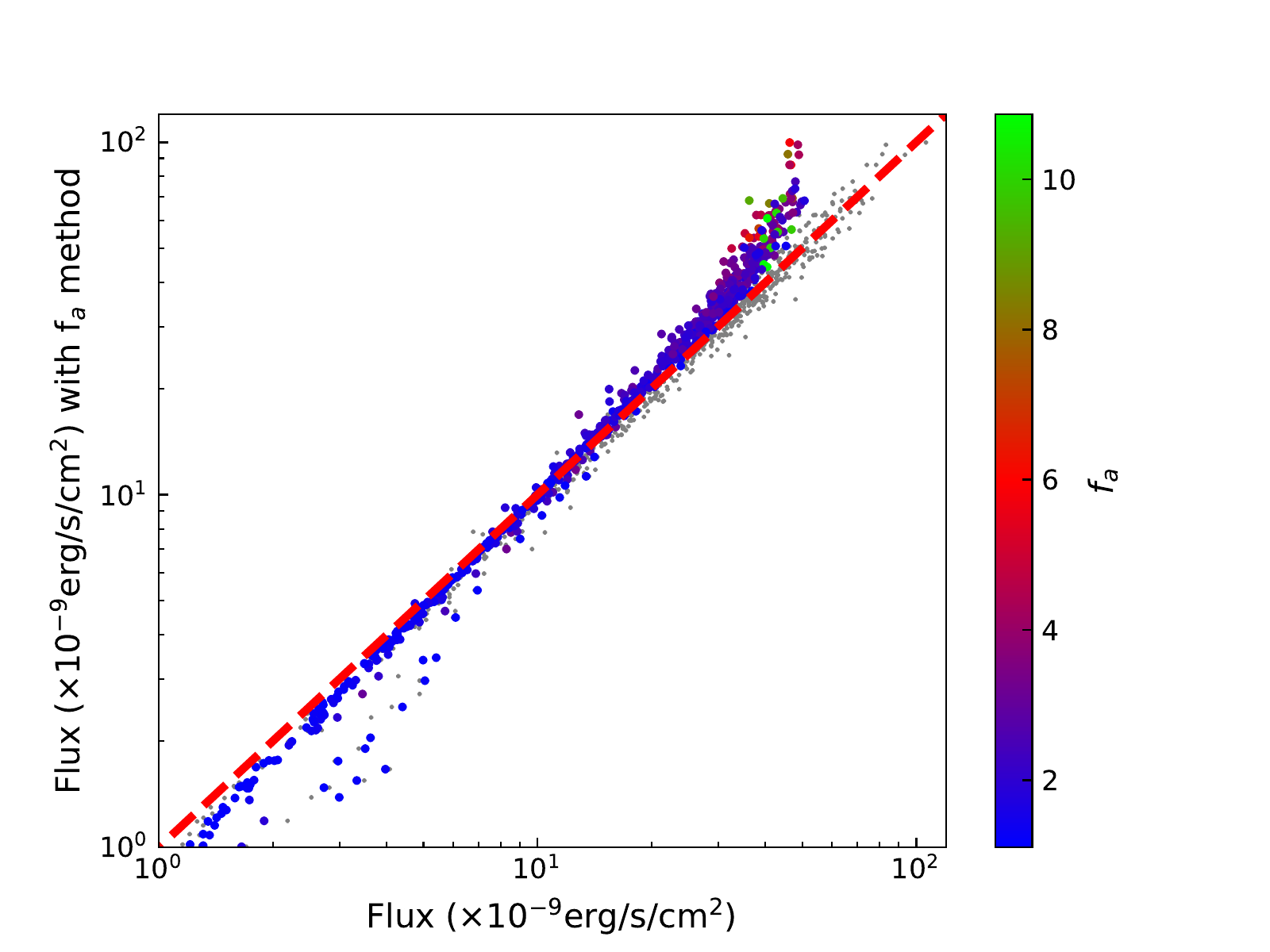}
	\includegraphics[width=\columnwidth]{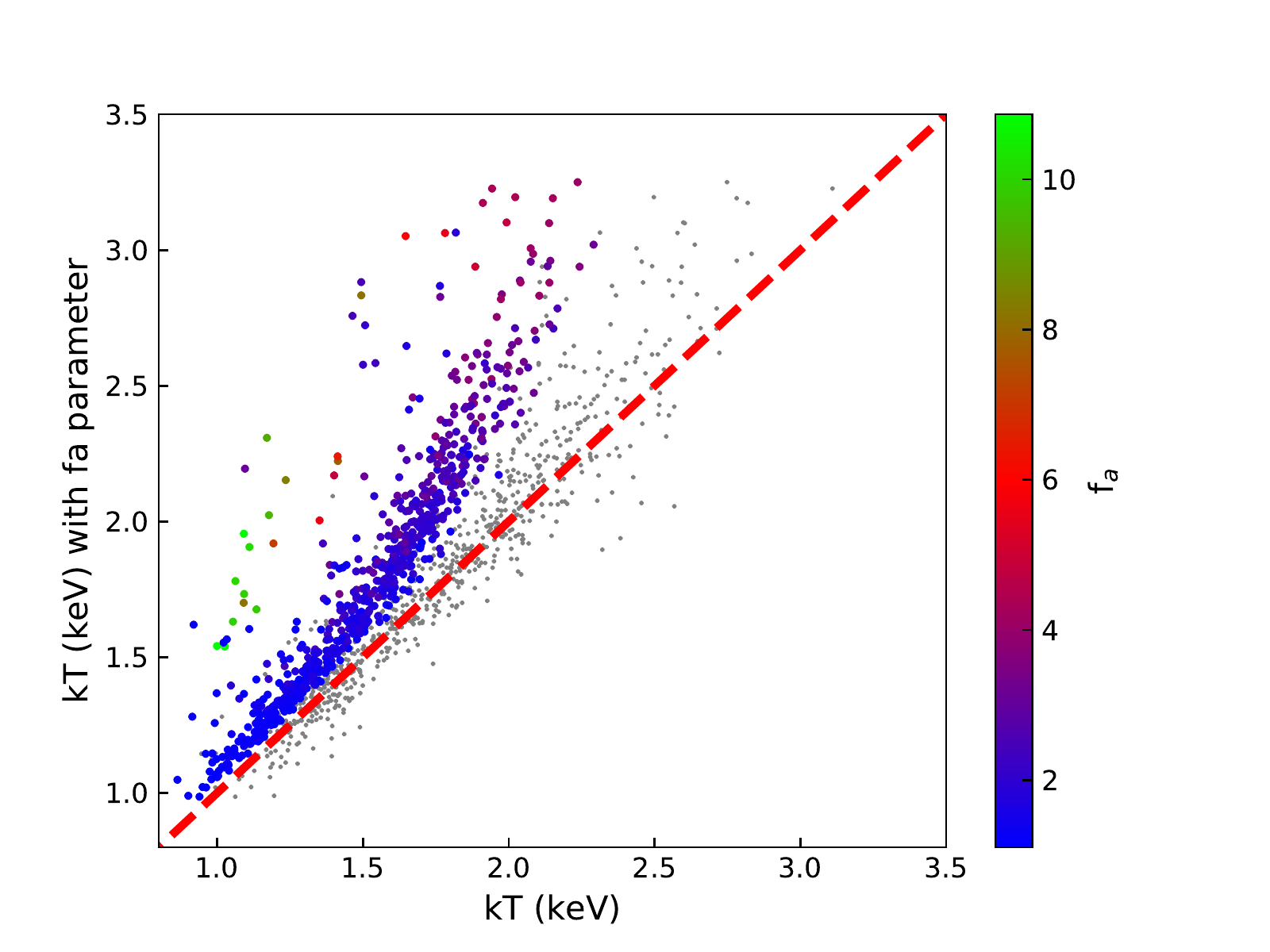}		
	\includegraphics[width=\columnwidth]{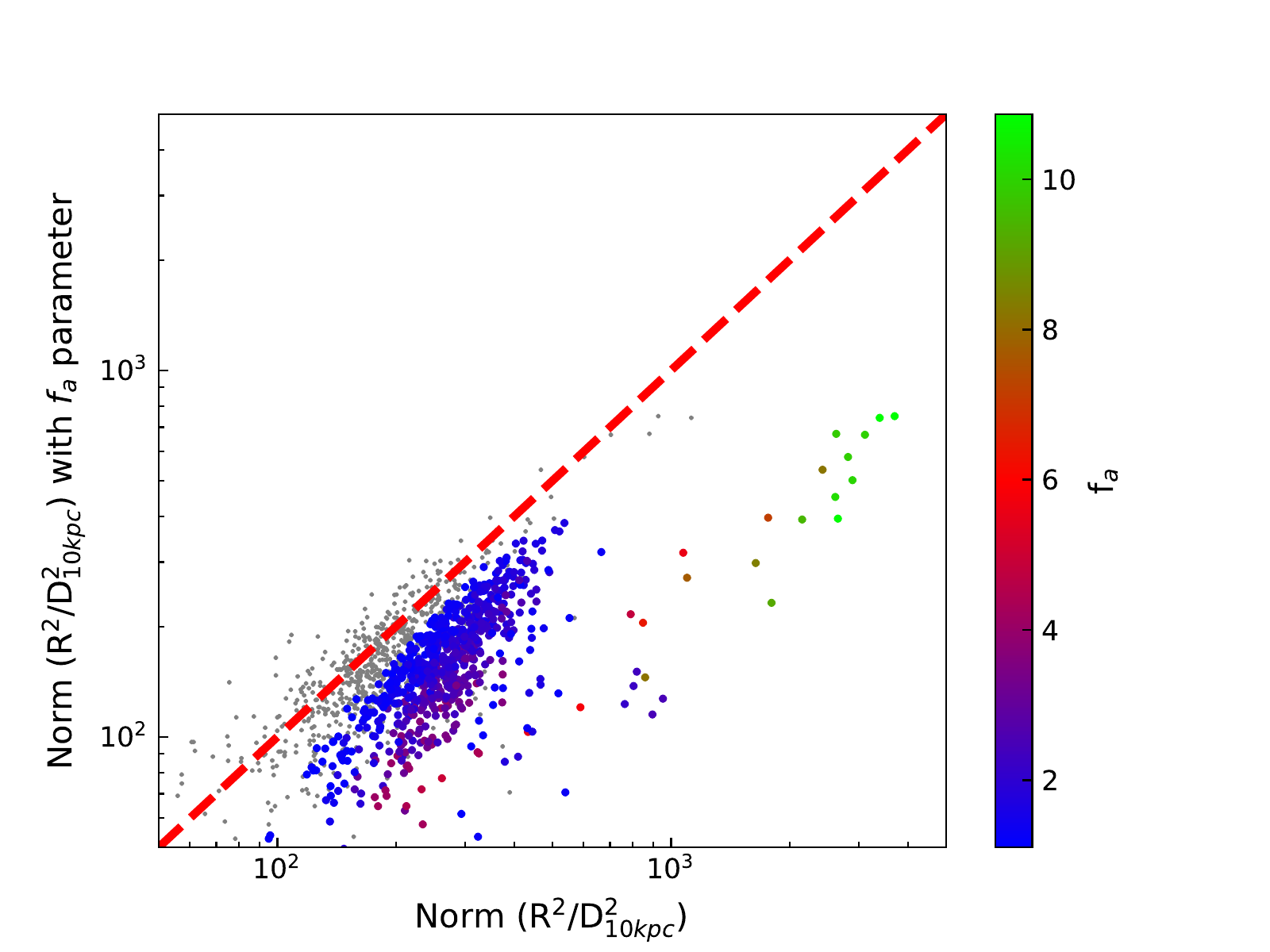}
    \caption{From top to bottom comparison of flux, blackbody temperature and blackbody normalization values obtained with or without applying the \fa method~(from top to bottom) as a function of \fa value. In each panel we also include the results from fits to only the 3--10~keV range with grey dots, which show a much better agreement with the results obtained from \fa method.}
    \label{fig:sp_comp}
\end{figure}

\begin{figure*}
	\includegraphics[width=\columnwidth]{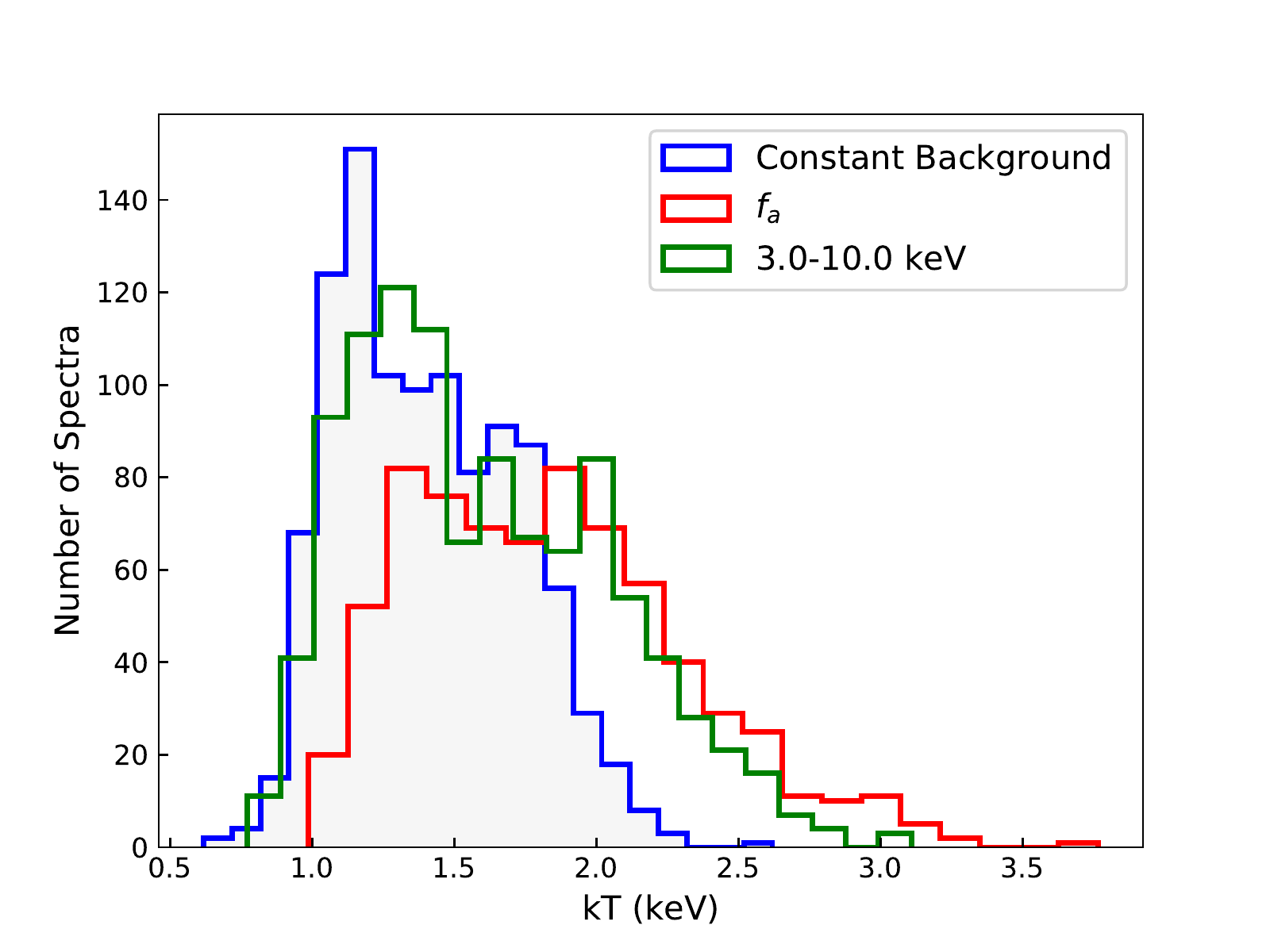}
	\includegraphics[width=\columnwidth]{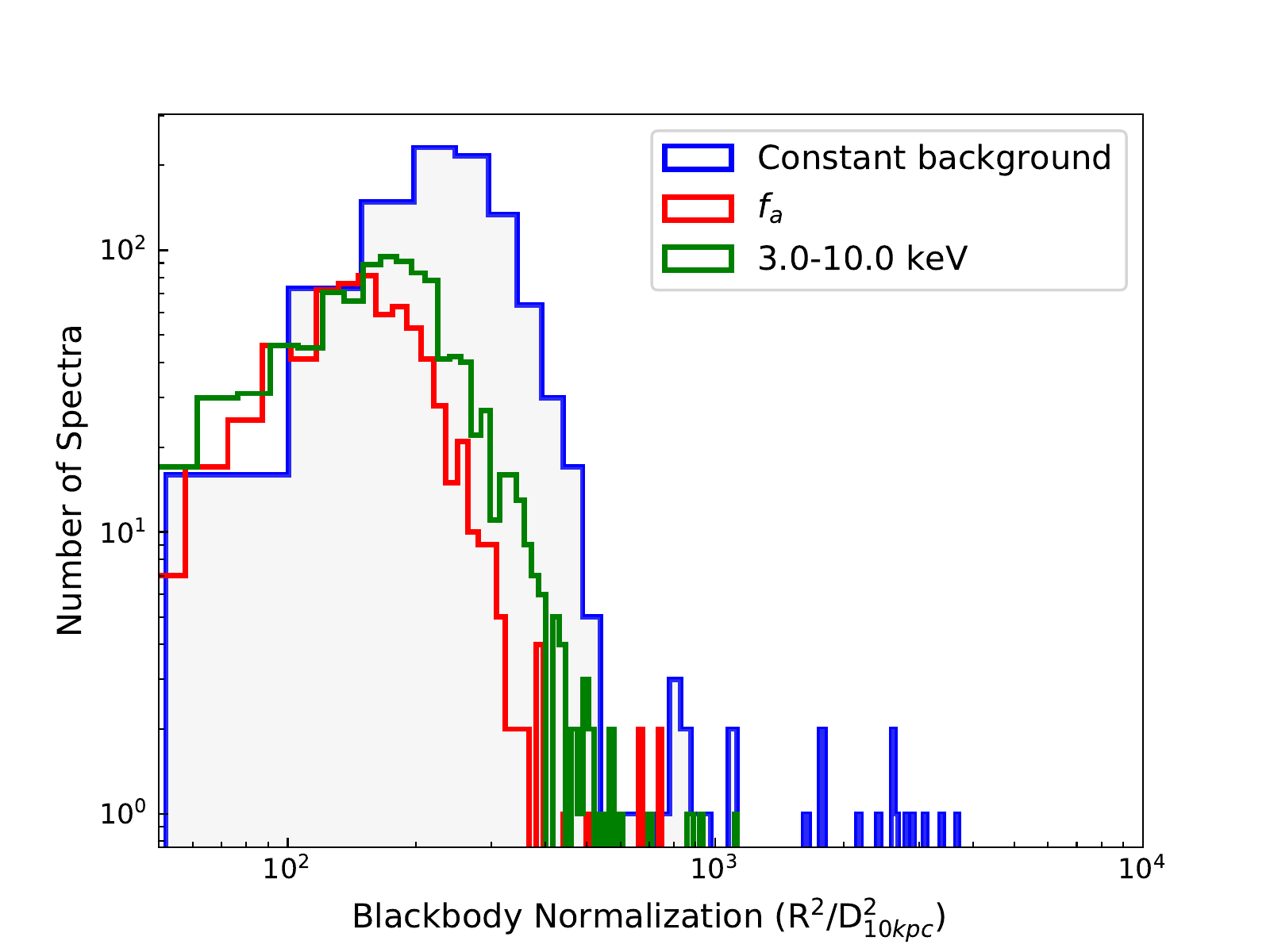}
    \caption{Histogram of blackbody temperature~(left panel) and normalization~(right panel) values with or without the \fa method. Each curve represents the constant background, the \fa, and energy range of 3.0--10.0~keV with the colors blue, red, and green, respectively. We include all of the X-ray spectra where the bolometric flux and \fa are greater than $10^{-9}$ \fluxcgs ~and 1.0, respectively.}
    \label{fig:ktnorm_hist}
\end{figure*}

In \autoref{fig:sp_comp} we show the best fit spectral parameters inferred using either the classical fixed background method or the \fa method. It can be seen that because the 0.5--10~keV X-ray spectra extracted from bursts are broader than a pure blackbody, using the fixed background approach results in blackbody temperatures that are significantly lower and normalization values that are significantly larger than the parameters as inferred using the \fa method. The color coding in \autoref{fig:sp_comp} shows that the difference in temperature and normalization grows with the magnitude of \fa. In the same figure, we also show with grey color the results inferred from fitting only the 3--10~keV band. The spectral parameters obtained in this higher energy band do not show the variations in temperature and normalization found in the full band, implying that these approaches agree with each other much better. These results suggest that the best fit spectral parameters obtained in the RXTE/PCA era were mostly free of the soft X-ray excess \cite[e.g.,][]{Galloway2008,2012ApJ...747...76G,2012ApJ...747...77G,2016ApJ...820...28O},  however a detailed analysis is beyond the scope of this paper and may not be possible given the decreasing effective area of \nicer towards higher energies. The fact that \nicer has a larger effective area in the soft X-ray band likely further alters the best fit spectral parameters. The same trend can also be seen in \autoref{fig:ktnorm_hist} where we show the histograms of blackbody temperature and normalization values for the three different fitting approaches.


Such significant variations in the spectral parameters can even affect the identification of a burst as a PRE event. Using the fixed background approach, we find two bursts showing evidence for a PRE. In burst 18 the normalization of the blackbody reaches to~36 km, which then decreases down to about 9.7~km, assuming a source distance of 6.0~kpc. In burst 20, the blackbody normalization reaches up to
18~km before normalizing to about 12~km after touchdown. By all definitions \citep{Galloway2008,2012ApJ...747...77G}
these values show evidence for a PRE. However in both cases during these episodes the fixed background approach results in significantly worse fits to the data if the whole \nicer band (0.5--10.0~keV) is used. If the \fa method is employed, the models provide much better fits to the data but this also strongly affects the inferred values in the blackbody normalization and temperature. The evidence for a PRE in burst 20 disappears completely, while for burst 18 a spectral evolution as expected from a PRE event can still be seen. The variation in the spectral parameters can be seen in \autoref{fig:burst_plots_b18}. We note that when only the 3--10~keV band is used the spectral evolution closely matches the evolution inferred from the \fa method. 

We also present in \autoref{tab:peaks} the best fit parameters obtained for each burst at the peak flux moment using the two methods as well as the fluences of each burst. We calculated the fluences by integrating all the flux values found via the \fa method, starting from the beginning of a burst till the flux is less than 10\% of the peak following \cite{2021ApJ...910...37G}.

\begin{figure}
	\includegraphics[width=\columnwidth]{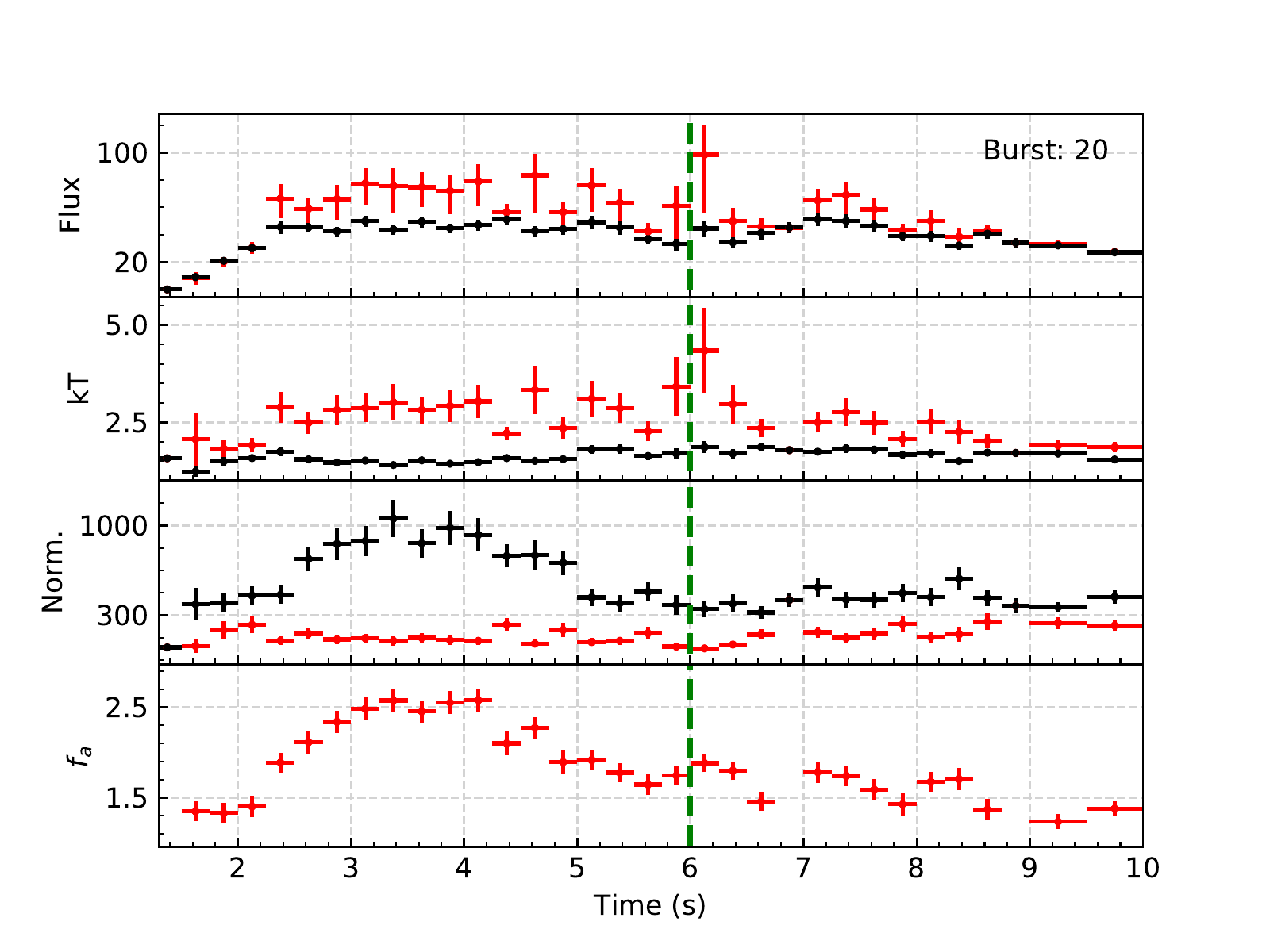}
    \caption{Spectral evolution observed in Burst 20. The location of the secondary peak is marked with a green dashed line. Time resolution used here is 0.25~s which shows more detail around the secondary peak observed in the lightcurve.}
    \label{fig:burst_plots_b20s_peak}
\end{figure}

From the \autoref{fig:burst_lc_3} it can be seen that burst 20 shows evidence for a secondary peak $\approx$ 6~s after the start of the burst. From the lightcurve alone the secondary peak resembles the secondary peaks previously observed from 4U~1608$-$52 by \nicer \citep{Jaisawal2019,2021ApJ...910...37G}. Time resolved spectroscopy of the burst does not show a significant spectral evolution when the routine time resolution is used, which in this case is 0.5~s (see \autoref{fig:burst_plots_4}). To test whether somewhat more detail can be detected we also generated X-ray spectra with 0.25~s time resolution. The results of the \fa~method fits to these spectra are shown in \autoref{fig:burst_plots_b20s_peak}. While a jump in the blackbody temperature near the secondary peak can be seen, the  large statistical uncertainties prevent any conclusion.


A histogram of all the \fa values obtained are shown in \autoref{fig:fa_hist}. It can be seen that in a great majority  of the cases the best fit \fa values are within 1--3. Only in burst 18, \fa values as high as 11 are observed. These findings are mostly in line with the findings of \cite{2015ApJ...801...60W}, where it is shown that \fa values are smaller in bursts showing no evidence for a PRE. \cite{2018ApJ...867L..28F,2020NatAs...4..541F} performed simulations of accretion discs subjected to burst emission surrounding a neutron star for thick and thin accretion disc assumptions, respectively. In both cases they predict a significant increase in the mass accretion rate onto the neutron star, which is mainly driven by the Poynting-Robertson drag \citep{1937MNRAS..97..423R,1974ApJ...188..121B, 1989ApJ...346..844W,1992ApJ...385..642W}. The predicted increase in the accretion rate for an Eddington limited burst is as large as an order of magnitude in the case of a thin accretion disc. As a multiplicative factor to the pre-burst spectral model \fa values  greater than one, are typically attributed to increased accretion rates. Our results show that bursts from \source also show similar episodes of increased mass accretion rates. The maximum \fa values reached in bursts indicate increased episodes of mass accretion rate by a factor of 2 to 11. 

\begin{figure}
	\includegraphics[width=\columnwidth]{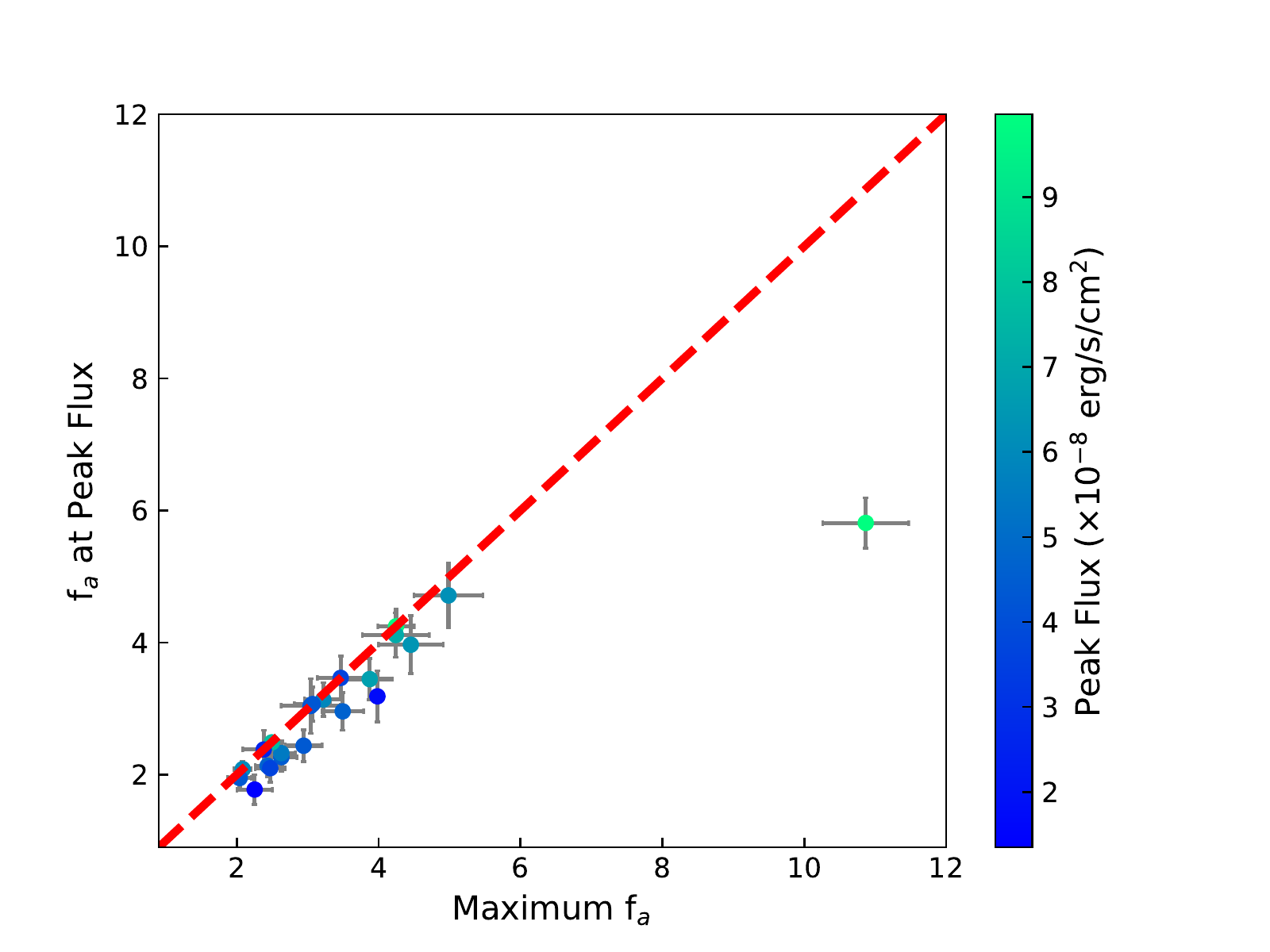}
    \caption{Relation between the maximum \fa value reached during a burst and the \fa at the peak flux moment. The color coding shows the peak flux reached in each burst. }
    \label{fig:peak_fa_max_fa}
\end{figure}

\begin{figure}
	\includegraphics[width=\columnwidth]{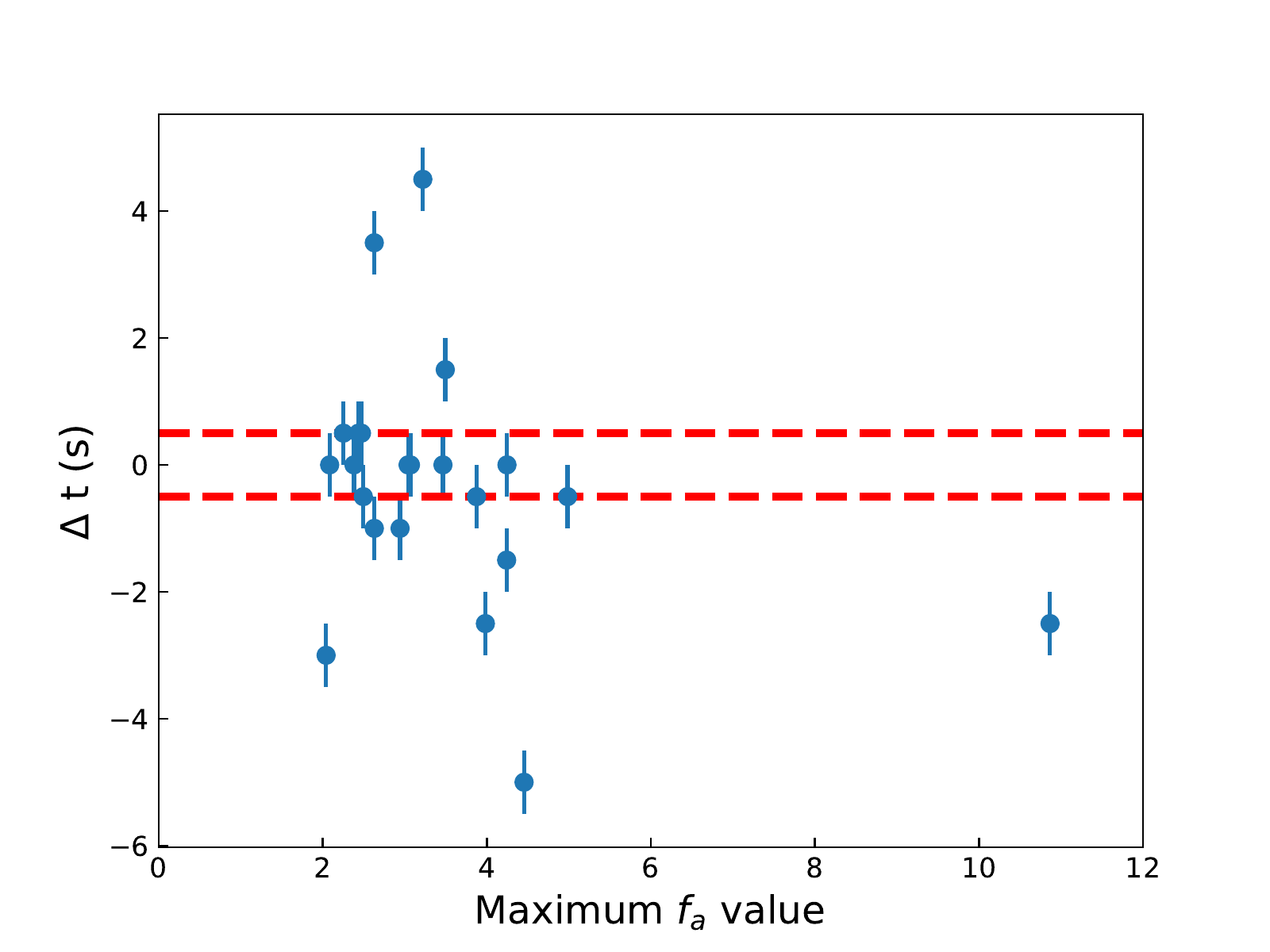}
    \caption{The time difference between the moment the accretion rate reached maximum (\fa value reached its maximum) and the peak flux moment of a burst as a function of the maximum \fa value. Horizontal red dashed lines show the $\pm$0.5~s time difference region. Since the integration time we used for the spectral analysis is 0.5~s it would be impossible to infer smaller time differences.}
    \label{fig:max_fa_time_lag}
\end{figure}

\begin{figure}
	\includegraphics[width=\columnwidth]{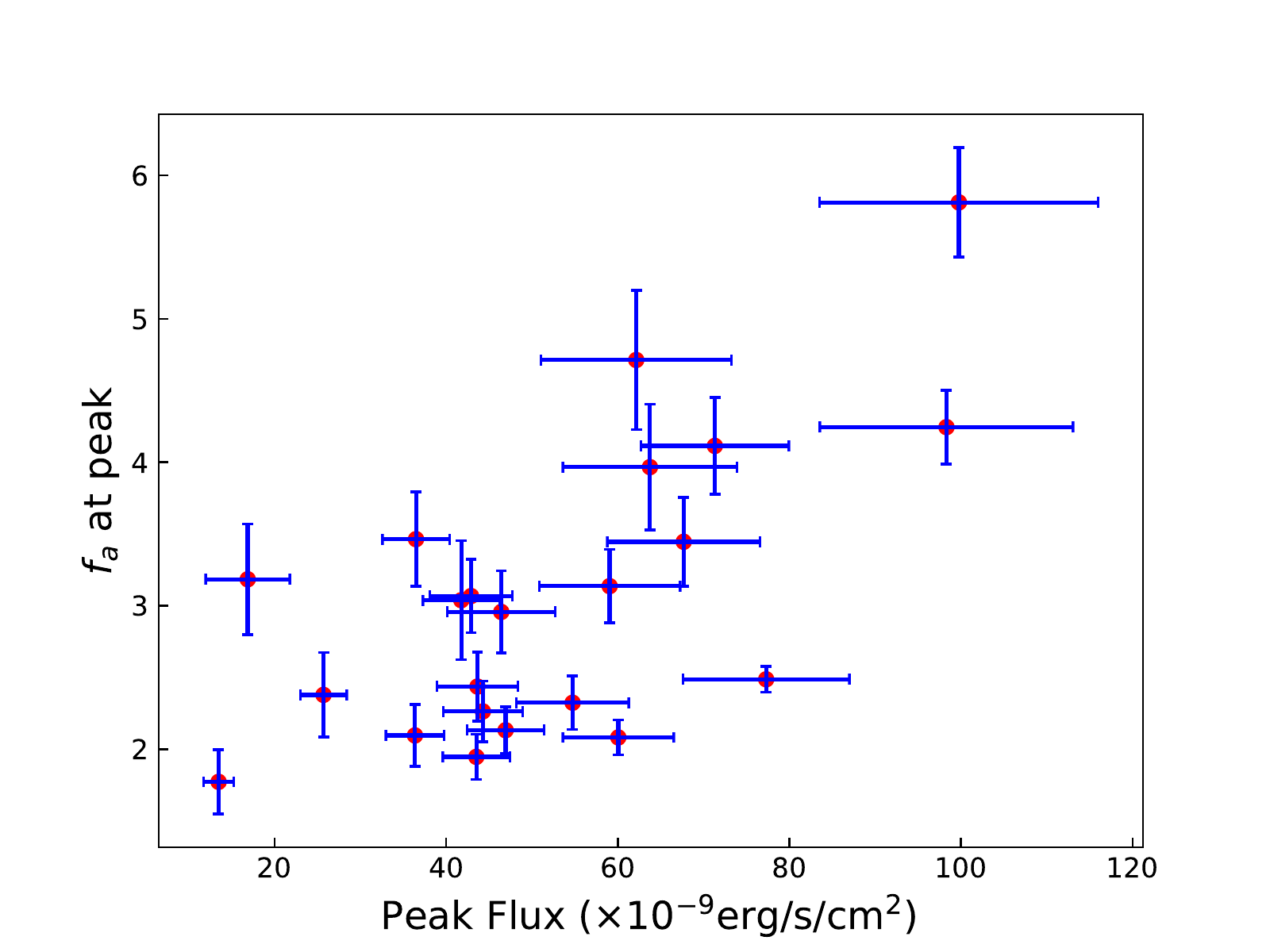}
    \caption{Correlation between the maximum \fa value and the peak flux in a burst.}
    \label{fig:peak_fa_peak_flux}
\end{figure}

\autoref{fig:peak_fa_max_fa} shows the relation between the maximum \fa value reached during the burst and the \fa value at the peak flux moment. We observe that the maximum \fa value reached during a burst is often at the peak flux moment. The Spearman's rank correlation coefficient between the \fa value at the peak flux moment of a burst and the maximum \fa reached is calculated as 0.96 with a p-value of $4\times10^{-12}$, excluding burst 18. In the case of burst 18 there is a remarkable difference in between the \fa value at the peak flux moment and the maximum \fa value reached, which roughly corresponds to the maximum of the observed blackbody normalization i.e, the maximum photospheric radius expansion moment. Although the \fa value at the peak flux and the maximum \fa reached during a burst are often correlated there are also differences. The time difference between the peak flux moment and the moment \fa reached its maximum may provide information on the accretion flow's response to the burst, if \fa probes the mass accretion rate on to the neutron star as suggested. Simulations performed by \cite{2020NatAs...4..541F} suggest that the increase in the mass accretion rate onto the neutron star precedes the peak flux moment. In \autoref{fig:max_fa_time_lag}, we show the time difference between the moment \fa value reached its maximum and the moment a burst reached its peak flux. Within the 22 bursts investigated here, the time difference is within the $\pm$0.5~s in 12 bursts, showing no significant time difference. In 7 bursts the \fa reaches its maximum value before the burst reaches its peak flux with an average time difference of 2.5~s. These differences are in line with predictions from simulations \citep{2020NatAs...4..541F} where the mass accretion rate onto the neutron star shows increase before the burst reaches the peak. Simulations also provide insight into the time difference between when Compton cooling of the plasma in the disc is taken into account and not. In the absence of Compton cooling the time difference between the peak flux of the burst and the maximum of the increase in the mass accretion rate is expected to be smaller. Since the efficiency of Compton cooling depends on the location of the corona \citep{2018SSRv..214...15D}, variations in the efficiency of Compton cooling may explain the two groups we observe here. Although not very strongly, we also found that the peak flux of a burst is correlated with the \fa value obtained at that moment. The correlation coefficient  between the peak flux of a burst and the \fa at peak is 0.66. It is obvious that the brighter the burst is, the larger the observed \fa, meaning the deviation from a pure blackbody model is stronger. This relation is shown in \autoref{fig:peak_fa_peak_flux}.   

\subsubsection{Application of Reflection Model}
\begin{figure}
	\includegraphics[width=\columnwidth]{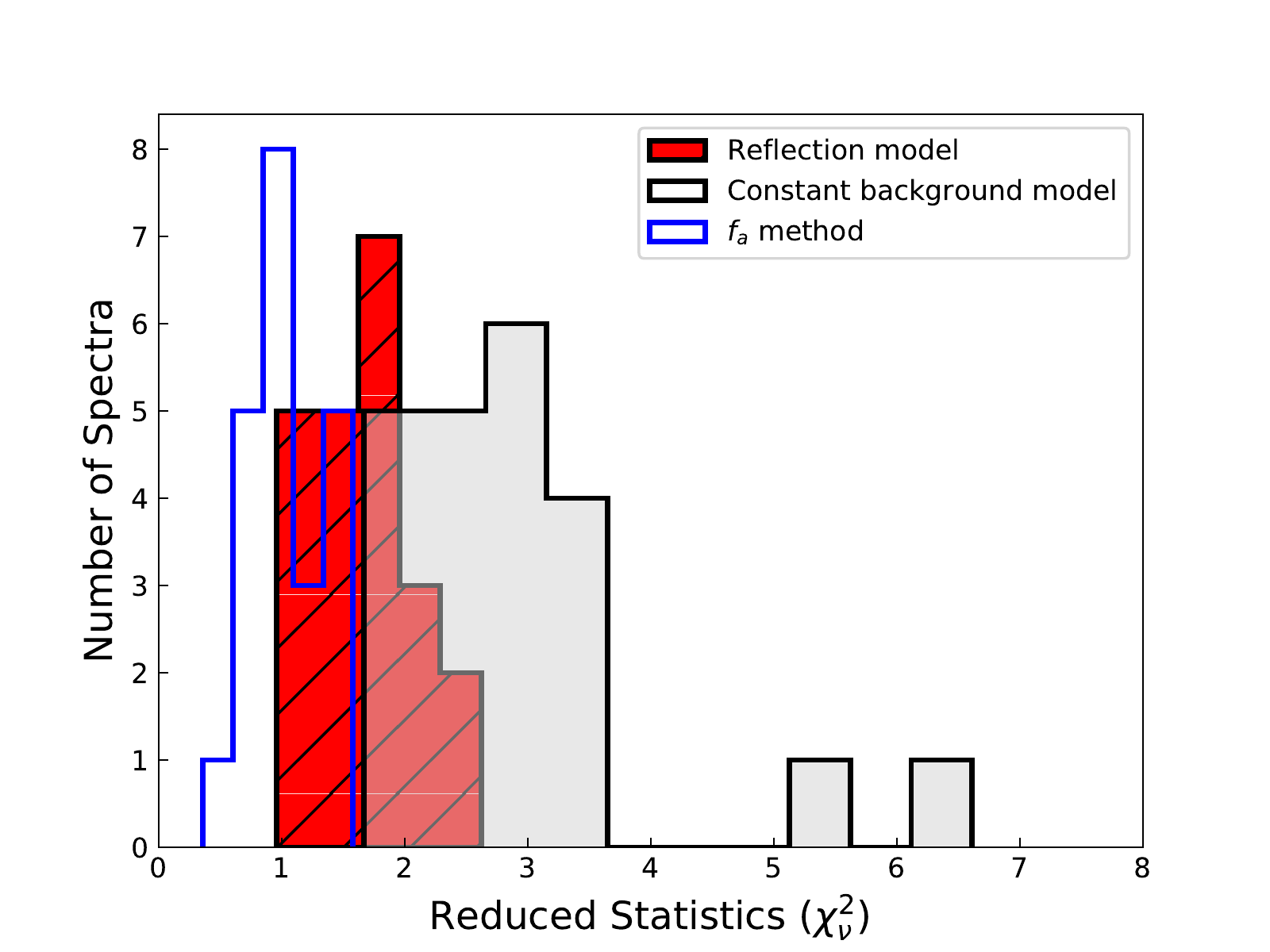}
    \caption{Histogram comparing of the reduced $\chi^2$ values obtained at the peak flux moment of each burst using simple blackbody, \fa, and reflection model methods.}
    \label{fig:refl_chi2}
\end{figure}

\begin{figure*}
	\includegraphics[width=\columnwidth]{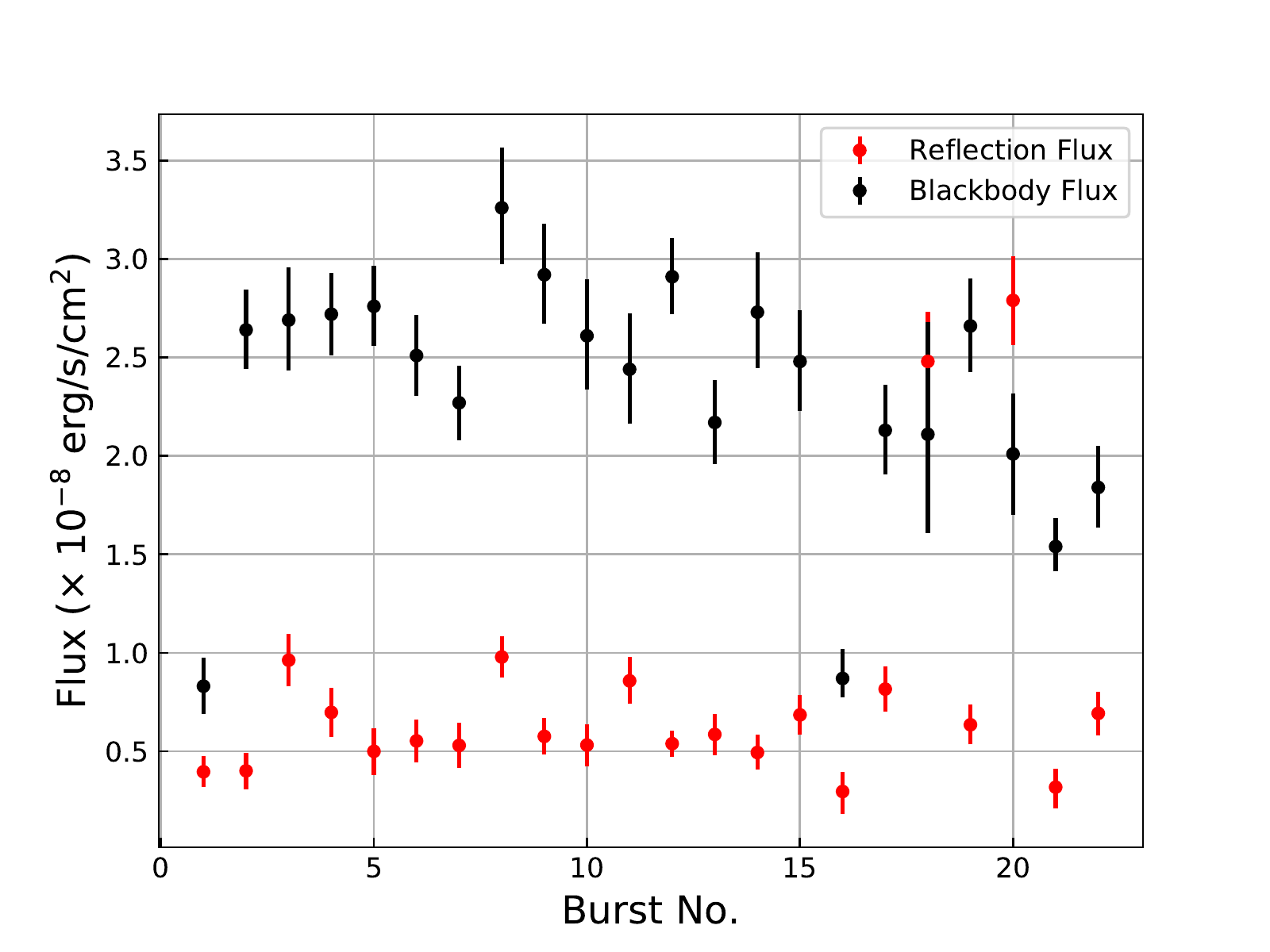}
		\includegraphics[width=\columnwidth]{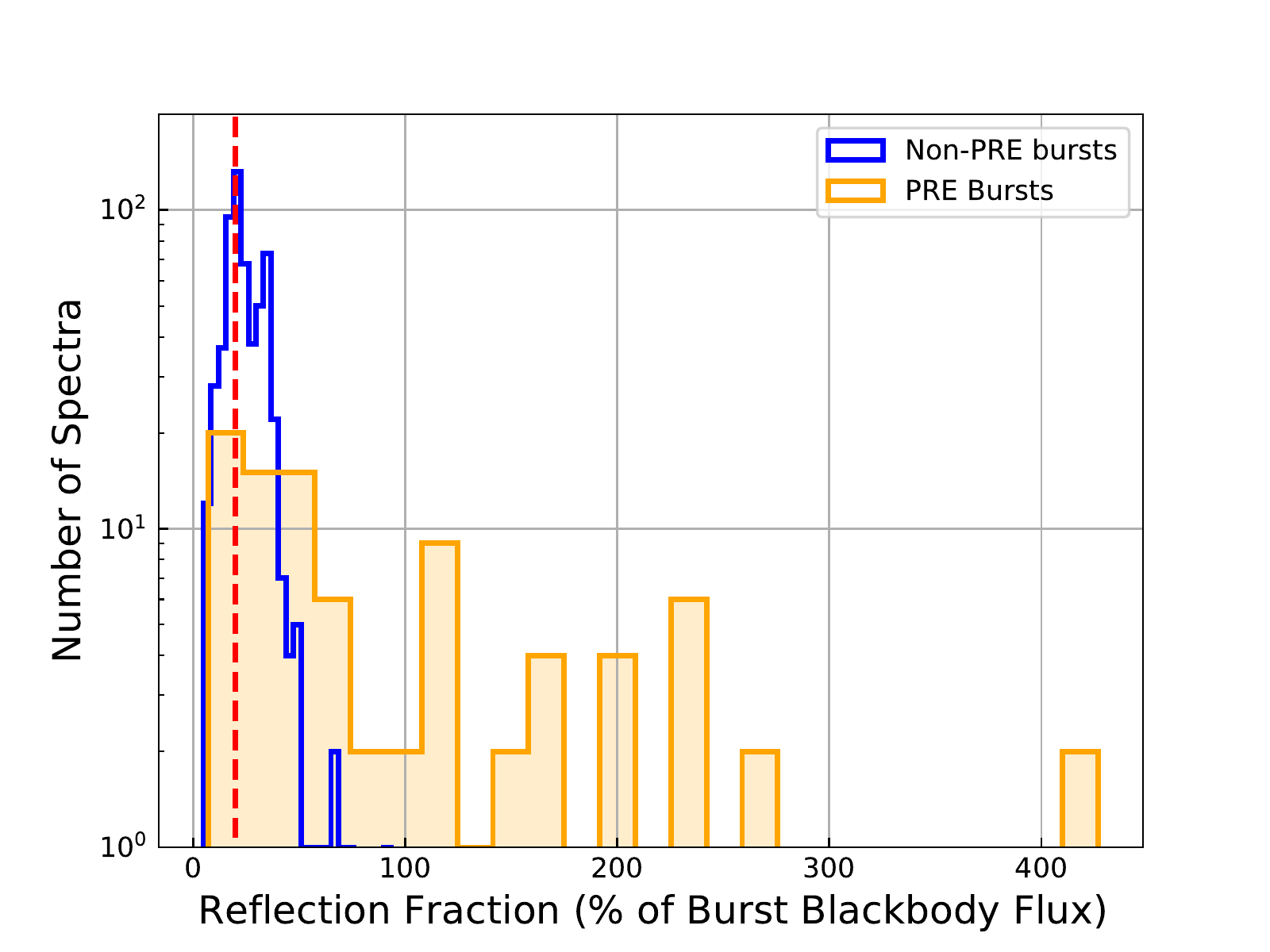}

    \caption{{\it Left Panel}: 0.5--10.0~keV fluxes of the burst blackbody emission and the reflection component at the peak flux moment of each burst. {\it Right Panel}: Flux of the reflection component in units of burst blackbody emission fraction for all the burst. Blue histogram shows the reflection fraction for non-PRE bursts, while the orange histogram shows the fraction for the bursts which show evidence for a PRE. Red vertical dashed line shows the peak of the distribution for the non-PRE bursts which is 20\%.}
    \label{fig:bb_refl_fraction}
\end{figure*}

Results of the reflection model fits show once again a significant improvement compared to blackbody fits with fixed background. We find that in each case the ionization parameter of the reflection model reaches the upper limit of the tabulated values, indicating that the accretion disc is strongly ionized by the burst emission. The resulting distribution of the $\chi^2$ values of the three different methods for the peaks of the bursts are shown in Figure \ref{fig:refl_chi2}. It can be seen that while the reflection model improves the goodness of the fit, \fa method still provides a statistically better result with one less free parameter. The fact that the ionization parameter of the reflection model is pegged at the largest value of the tabulated model means that we are basically fitting the bremsstrahlung continuum to model the soft excess, with only free parameter being the normalization or the flux of the reflection component, since the density in the disk is fixed. However, it is expected that the reflection fraction should be 20-30\% of the burst emission itself \citep{2018ApJ...855L...4K}, which depends on the inclination angle of the system as well as the inner disc radius. A comparison of the inferred burst and reflection model fluxes at the peak moments of each burst is shown in the left panel of \autoref{fig:bb_refl_fraction}. From our fits, we can infer that while this generally holds true for all of the bursts, in two bursts where simple blackbody approximation results show evidence for a PRE (burst 18 and 20), the inferred flux of the reflection model exceeds the burst emission itself. This result shows that the application of the reflection model at the peak of the bursts with PRE is actually not enough just by itself and a further soft component is required. 

As a second approach we also fitted all of the X-ray spectra where \fa was statistically required in all of the bursts. The ionization parameter values throughout the bursts still hit the upper limit of the tabulated models. However the flux ratio of the reflection models compared to the burst emission shows a very similar distribution to what we obtained from fitting only the peaks of the bursts. A histogram of the flux ratios of the reflection models and the burst blackbody emission is shown in the right panel of \autoref{fig:bb_refl_fraction}. We see that while in a great majority of the spectra in non-PRE bursts the flux of the reflection model is around 20\% of the burst emission itself, in bursts where there is some evidence for a PRE and especially at around the peak fluxes the fraction of the flux of the reflection models exceeds the incident flux from the burst blackbody indicating that the results are not physical. We can therefore conclude that especially at the peaks of these bursts the reflection model only helps to somehow improve the fits but is not enough by itself to result in a physically reasonable fit.  We note that for burst 18 we also tried to fit the spectra with both the reflection model as well as the \fa parameter. However, the existing spectral data does not allow us to constrain the parameters of the reflection model and the \fa at the same time.

\section{Discussion and Conclusions}
\label{sec:Conclusions}

Observations of low mass X-ray binaries showing X-ray bursts, like \source, 4U~1820\textminus30, SAX~J1808.4\textminus3658, Swift~J1858.6\textminus0814, MAXI~J1807$+$132, XTE~J1739\textminus285 and 4U~1608\textminus52 \citep{2018ApJ...855L...4K,2018ApJ...856L..37K,2019ApJ...885L...1B,2020MNRAS.499..793B,2021MNRAS.501..261A,2021ApJ...907...79B,2021ApJ...910...37G}  already demonstrated the power of NICER in probing the effects of X-ray bursts on their accretion environments. Here, we follow-up on these studies by investigating an ensemble of 22 X-ray bursts observed from \source across accretion states by \nicer to better understand the spectral evolution especially in the soft X-ray band. First of all, the existing \nicer data set from \source already shows some interesting bursts. As noted, there are two bursts (18 and 20) showing evidence for a PRE. However the use of the \fa method affects the inferred spectral parameters in a way that minimizes the evidence for a PRE. 

In burst 20 the lightcurve shows a significant secondary peak, roughly about 6~s after the burst start. However, the spectral evolution does not show a similarly significant change in the spectral parameters at the expected time. Within our sample this is the only burst which happens at a relatively high mass accretion rate as inferred from the flux of the pre-burst emission. Assuming the Eddington limit of the source as F$_{\rm Edd}=$ 10.44$\times10^{-8}$ \fluxcgs~\citep{2012ApJ...747...77G}, we can infer that the source was emitting roughly at 10\% Eddington limit level just before this burst. This level is in agreement with the bursts observed from 4U~1608$-$52 showing secondary peaks when the system was emitting at about 15\% of the Eddington limit \citep{2021ApJ...910...37G}. 

Bursts 15, 16 and 21 and 22 are also worth noting given that they can be classified as short recurrence bursts with separations of only 496~s and 451~s, respectively. These recurrence times are some of the shortest in the MINBAR catalog both for \source and for all the bursters in general \citep{2020ApJS..249...32G}. The minimum reported value is 233~s for 4U~1705--44 in the MINBAR catalog \citep{2020ApJS..249...32G, 2010ApJ...718..292K}.  Most recently, bursts 21 and 22 have been reported to happen 9.44 days after a superburst \citep{2021arXiv210807941L}. We note that burst 20 reported here happened only three days before the reported superburst. It is also interesting to further emphasize the fact that in the case of bursts 21 and 22, where the separation of the two events is only 451~s, the second burst is brighter than the first one. To the best of our knowledge, such a short recurrence event has not been reported before in the existing catalogs \citep{2020ApJS..249...32G, 2010ApJ...718..292K}. The peak count rate of burst 21 is only 75\% of burst 22, while the peak flux of the blackbody component when the \fa method is used is only 70\% of burst 22. The difference is even more significant when the fluences of each burst is compared, the fluence of burst 21 is only 54\% of burst 22. \cite{2007A&A...465..559B} reports a triple burst from EXO~0748-676, where the third burst is brighter than the second one but still somewhat dimmer than the first burst. However, prior to burst 21 there is data for only 600~s, which is not enough to test whether this was another triple burst event or not.

We also searched for burst oscillations in three different energy bands in all of observed X-ray bursts from \source. We found no significant burst oscillation within the \nicer sample. Our upper limits on the fractional rms amplitudes are typically around 0.05, which is similar to the limits presented in other studies \citep{2020ApJS..249...32G, 2017ApJ...834...21O} and smaller than the amplitudes of the previously reported oscillations. Within the MINBAR catalog roughly only 10\% (8 bursts) of all the bursts from \source showed significant burst oscillations and 6 of those were bursts showing PRE. In our case we only have two bursts showing evidence for a PRE. In burst, 18 our limits are as small as 0.02. Based on the amplitudes of previously reported oscillations we can rule out the existence of burst oscillations in that burst.

We performed time resolved spectroscopy of the bursts within the \nicer band to better understand the soft X-ray emission observed during the bursts from these systems. Shortly after \nicer started observations of X-ray bursters, evidence for a strong soft excess in the soft X-ray band of the time resolved spectra of bursts  have been reported \citep{2018ApJ...855L...4K,2018ApJ...856L..37K, 2021arXiv210713543B}. Hints about such an excess has already been known and could be studied in some cases \citep{2004ApJ...602L.105B, 2013ApJ...772...94W, 2015ApJ...801...60W}. We here provide a more systematic study of these deviations using all of the X-ray bursts observed so far from \source. We see that the application of the \fa model statistically improves the fits, indicating that the burst strongly affects the surrounding accretion disc. However, the \fa model by itself does not provide a detailed physical insight of the observed increase in the pre-burst emission. It is generally thought that the observed increase in the pre-burst emission is indicative of increased mass accretion rate \citep{2013ApJ...772...94W, 2015ApJ...801...60W} to the neutron star. In accordance with such expectations, recent simulations by \cite{2020NatAs...4..541F} predict a detectable yet temporary increase in the accretion rate onto the neutron star, mostly due to Poynting-Robertson drag during a burst. The maximum \fa values inferred here are in agreement with theoretical predictions. Simulations also predict the mass accretion rate to increase a few seconds before the burst reaches its peak. We find that in about 7 bursts we can observe a similar time difference. It is expected that the disc-burst interaction should be a function of burst luminosity and therefore the amount of soft excess should depend on, for example, the peak flux of a burst. Our results indicate that, although with some scatter, indeed there is such a relation (see \autoref{fig:peak_fa_peak_flux}). The deviation from a pure blackbody becomes strongest at the peak flux moment of each burst and the actual amount of deviation or soft excess depends on the peak flux of a burst. The \fa model nicely illustrates that point. We note that, although X-ray bursts show short time scale variations and therefore integration times used to extract spectra may have an affect of averaging different temperatures, the deviation observed here was also observed from 4U~1820$-$30 \citep{2018ApJ...856L..37K}, where the exposure times used for individual spectra were as short as 0.03~s.

There maybe several reasons for the burst emission to show deviations from a pure blackbody. First of all, atmospheric effects are expected to play a significant role \citep[for a review see e.g.,][]{2013RPPh...76a6901O}. All of the bursting neutron star atmosphere models \citep[see e.g.,][]{2011A&A...527A.139S,2004ApJ...602..904M, 2005A&A...430..643M} predict significantly broadened X-ray spectra compared to a pure blackbody. Furthermore, given the fact that these systems often contain rapidly rotating neutron stars, relativistic affects also have the potential to broaden the observed X-ray spectra \citep{2015ApJ...799...22B}. However, in both cases the predicted deviations from a blackbody emission are practically time independent, whereas the deviations we report here show significant variations during an X-ray burst.

One likely interpretation of the excess emission observed during X-ray bursts may be related to the reflection of the burst emission by the accretion disc. We used a tabulated reflection model assuming solar abundances to see if the reflection models can improve the fits to the X-ray spectra and provide an understanding of the reflection processes. Our results indicate that indeed reflection models do improve the fits and can account for the soft excess but not as well as the \fa model. The best fit ionization parameters of the used models were very high, allowing us to only put lower limits on that parameter. This indicates that the accretion disc is highly ionized. Still, the inferred 0.5--10.0~keV flux ratios of the best fit blackbody models and the reflection models are generally in line with what is expected (at around 20\% level), which shows that reflection may be a common feature in the X-ray spectra of bursts, and can be detected. In two bursts where we see some evidence of a PRE, the reflection model and blackbody model fractions reverses with reflection fractions much larger than the intrinsic burst flux. This indicates that during these bursts some other processes may also have a significant affect on the soft X-ray excess and therefore reflection model just by itself is not enough. 

\begin{figure*}
	\includegraphics[width=\columnwidth]{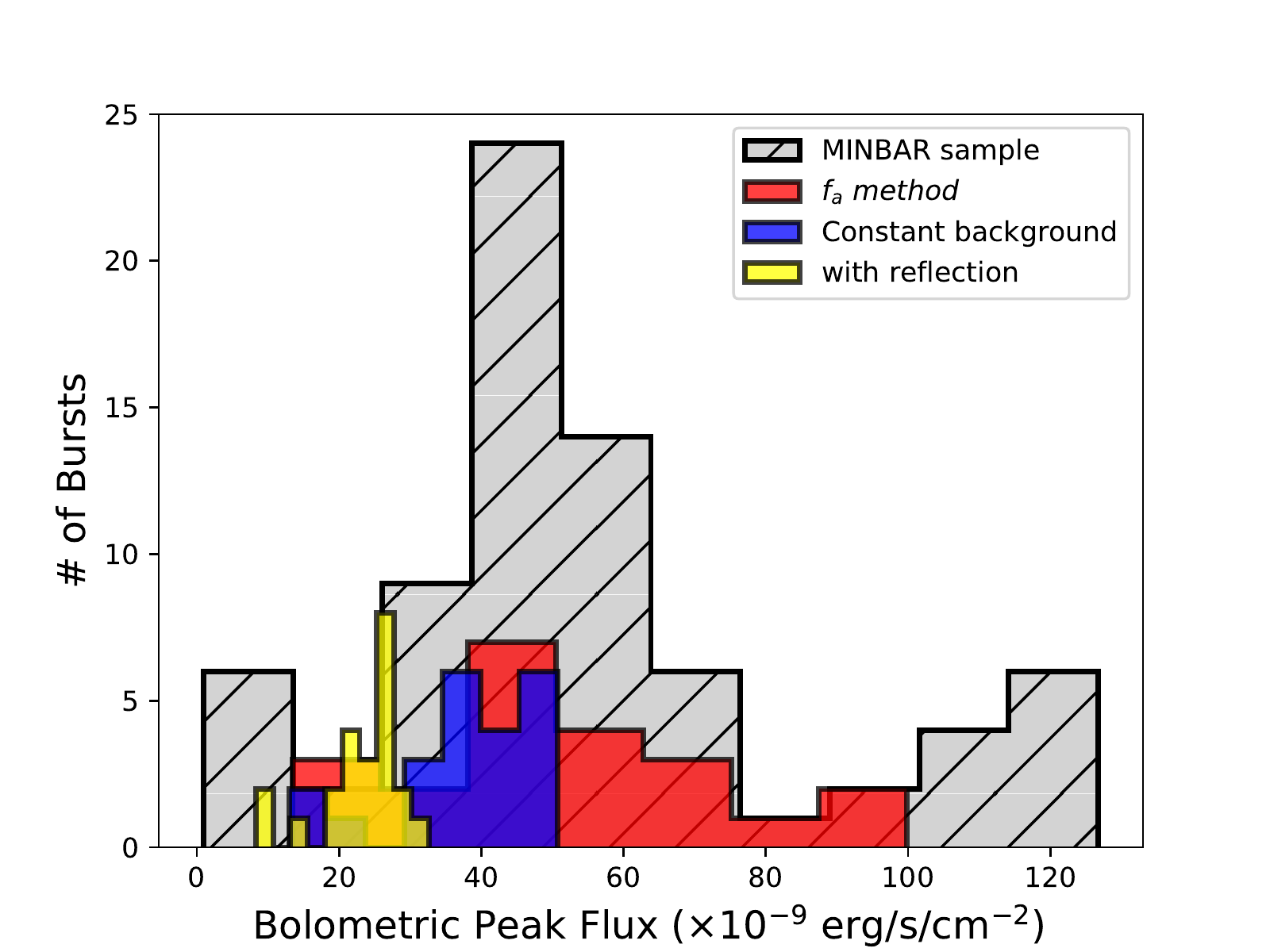}
	\includegraphics[width=\columnwidth]{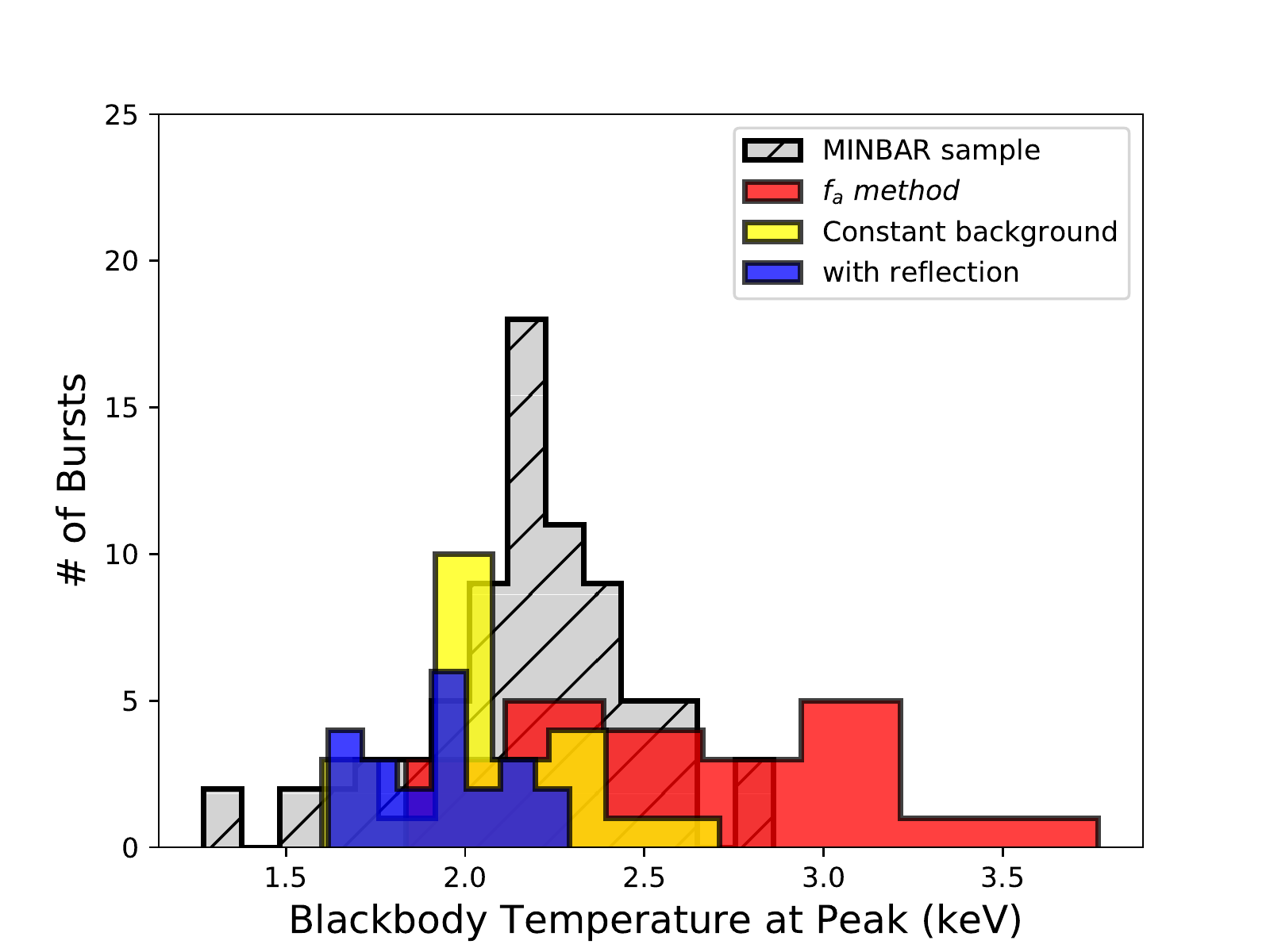}
    \caption{Comparison of spectral parameters obtained at the peak moments of each burst with archival bursts from MINBAR catalog \citep{2020ApJS..249...32G}. The histogram of spectral parameters for the archival bursts are shown with grey, while with blue the distribution of the spectral parameters as found from classical blackbody fits, with red the spectral parameters found from the \fa method, and with yellow the spectral parameters as inferred taking into account the reflection is shown. Note that from the MINBAR catalog we only used results of the bursts observed by RXTE/PCA.}
    \label{fig:compar_minbar}
\end{figure*}

Finally, we compared the best fit spectral parameters for the peaks of the bursts with the parameters obtained from the bursts in the MINBAR catalog \citep{2020ApJS..249...32G}. We show in \autoref{fig:compar_minbar} the bolometric flux and blackbody temperature values inferred at the peaks of the bursts in the MINBAR catalog observed mostly with RXTE/PCA together with the best fit values we find via different methods here. When combined with the results shown in Figures \ref{fig:sp_comp} and \ref{fig:ktnorm_hist}, it obvious that the use of different methods result in a significant variation in the inferred blackbody temperature and flux as compared to the archival measurements. It is likely that these results advocate the need for broad band studies that can readily be performed by simultaneous observations of \nicer and ASTROSAT \citep[][]{2017CSci..113..591Y} or with future large effective area detectors like STROBE-X \citep{2019arXiv190303035R} or eXTP \citep{2019SCPMA..6229502Z} in order to be able to both determine the spectral parameters of the bursts as well as their affects on the surrounding environments precisely. Even now, \nicer observations of X-ray bursts from sources with low hydrogen column density will provide a unique view on what impact do thermonuclear X-ray bursts have on their surroundings, thanks to its soft X-ray sensitivity and expanding archive of burst observations.

\section*{Acknowledgements}

We gratefully thank the anonymous referee for constructive comments and recommendations. We thank R\"umeysa Asl\i han Ert\"urk for her contribution on writing some of the scripts used in this paper. T.G. has been supported in part by the Scientific and Technological Research Council (T\"UB\.ITAK) 119F082 and the Turkish Republic, Presidency of Strategy and Budget project, 2016K121370. D.A. thanks the Royal Society for their support. This work was supported by NASA through the NICER mission and the Astrophysics Explorers Program.
This research has made use of the data and software provided by the High Energy Astrophysics Science Archive Research Center (HEASARC), which is a service of the Astrophysics Science Division at NASA/GSFC and the High Energy Astrophysics Division of the Smithsonian Astrophysical Observatory.

\section*{Data Availability}

All the data used in this publication is publicly available through NASA/HEASARC archives. 



\bibliographystyle{mnras}
\bibliography{example} 




\appendix 
\section{Light curves of bursts } 

Lightcurves of each burst as observed in the 0.5--10.0~keV range are given together with the burst start, peak and decaying e-folding times.

\label{app:lcbursts}
\begin{figure*}
	\includegraphics[scale=0.5]{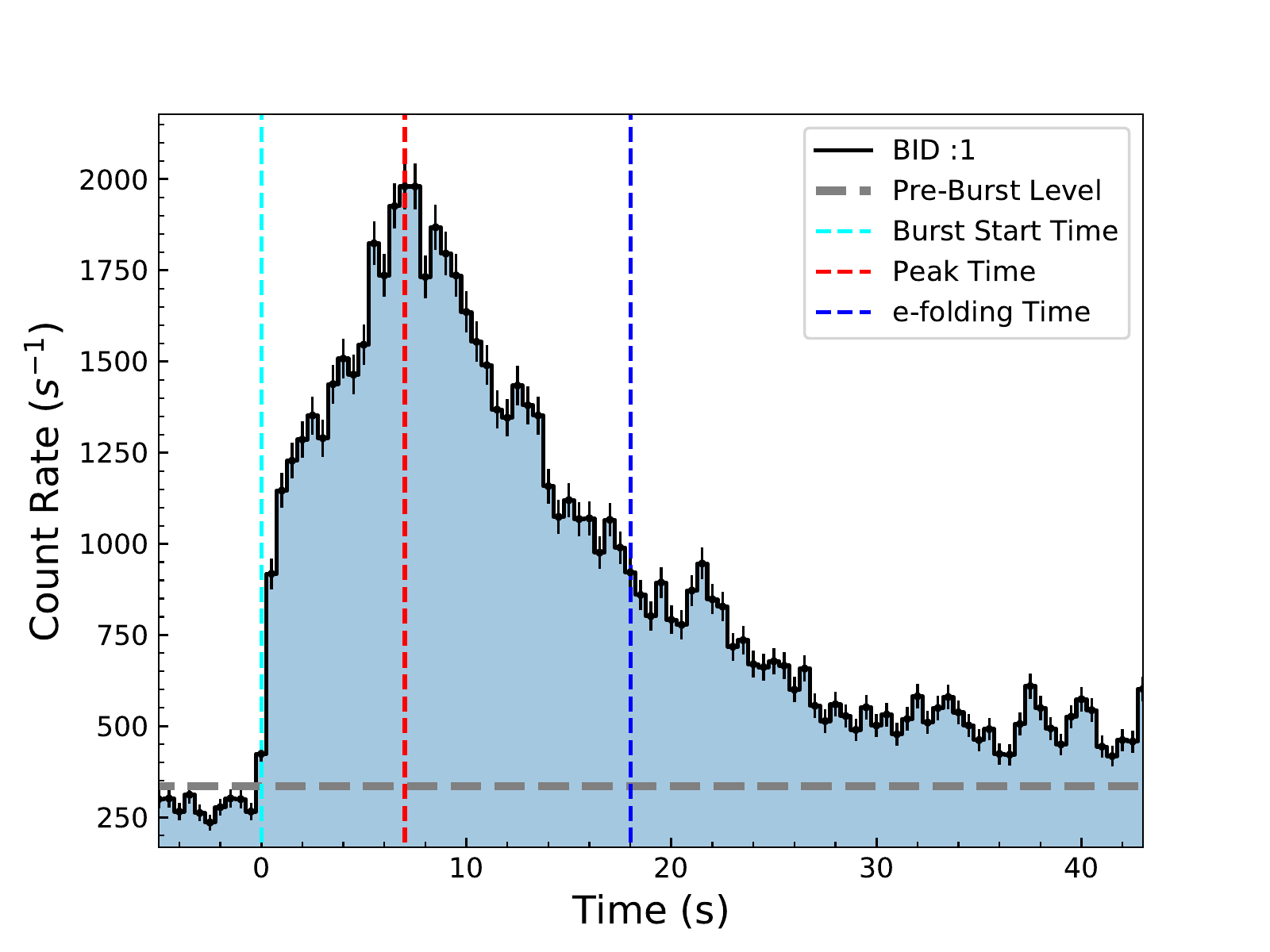}
	\includegraphics[scale=0.5]{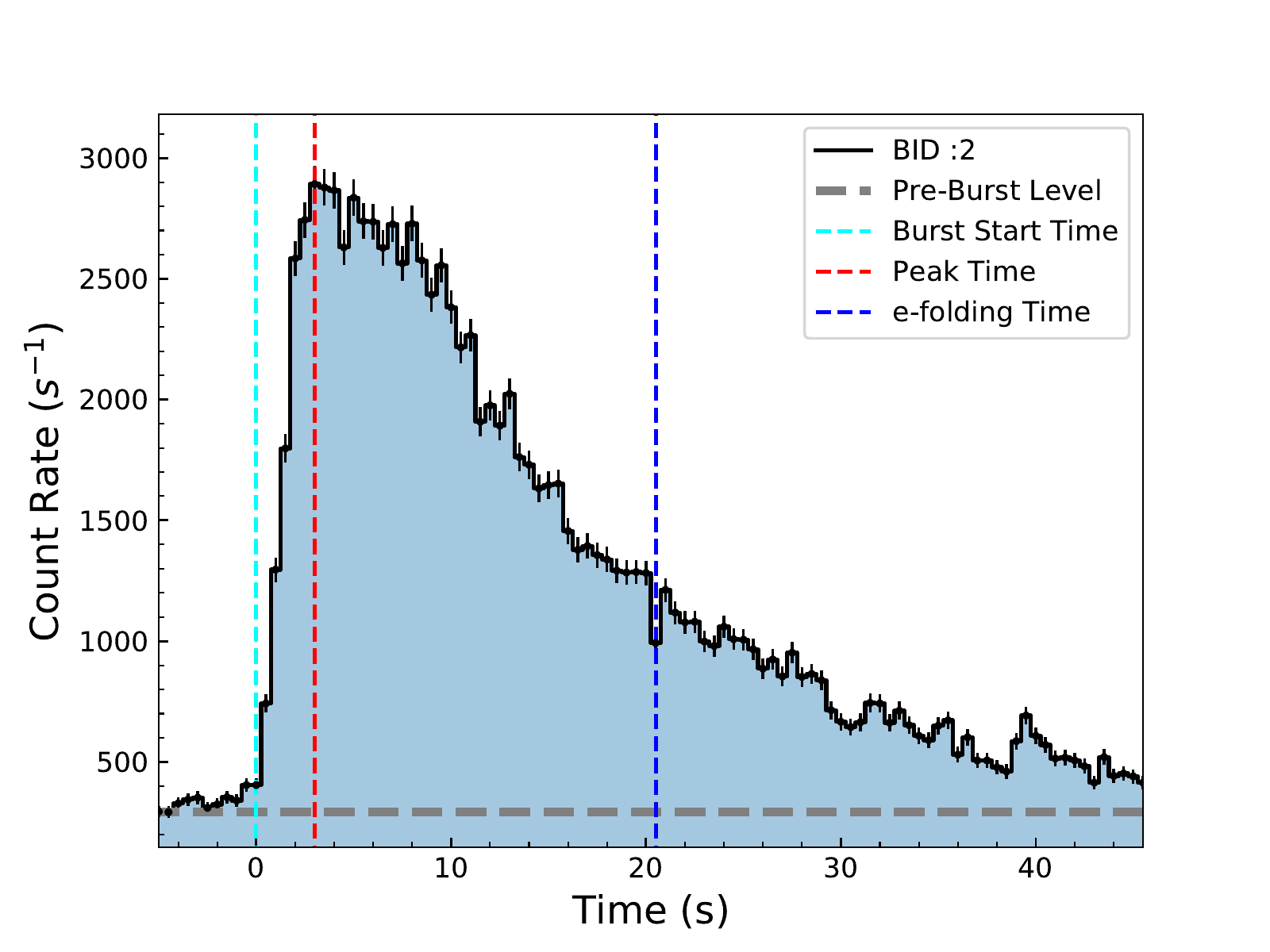}
	\includegraphics[scale=0.5]{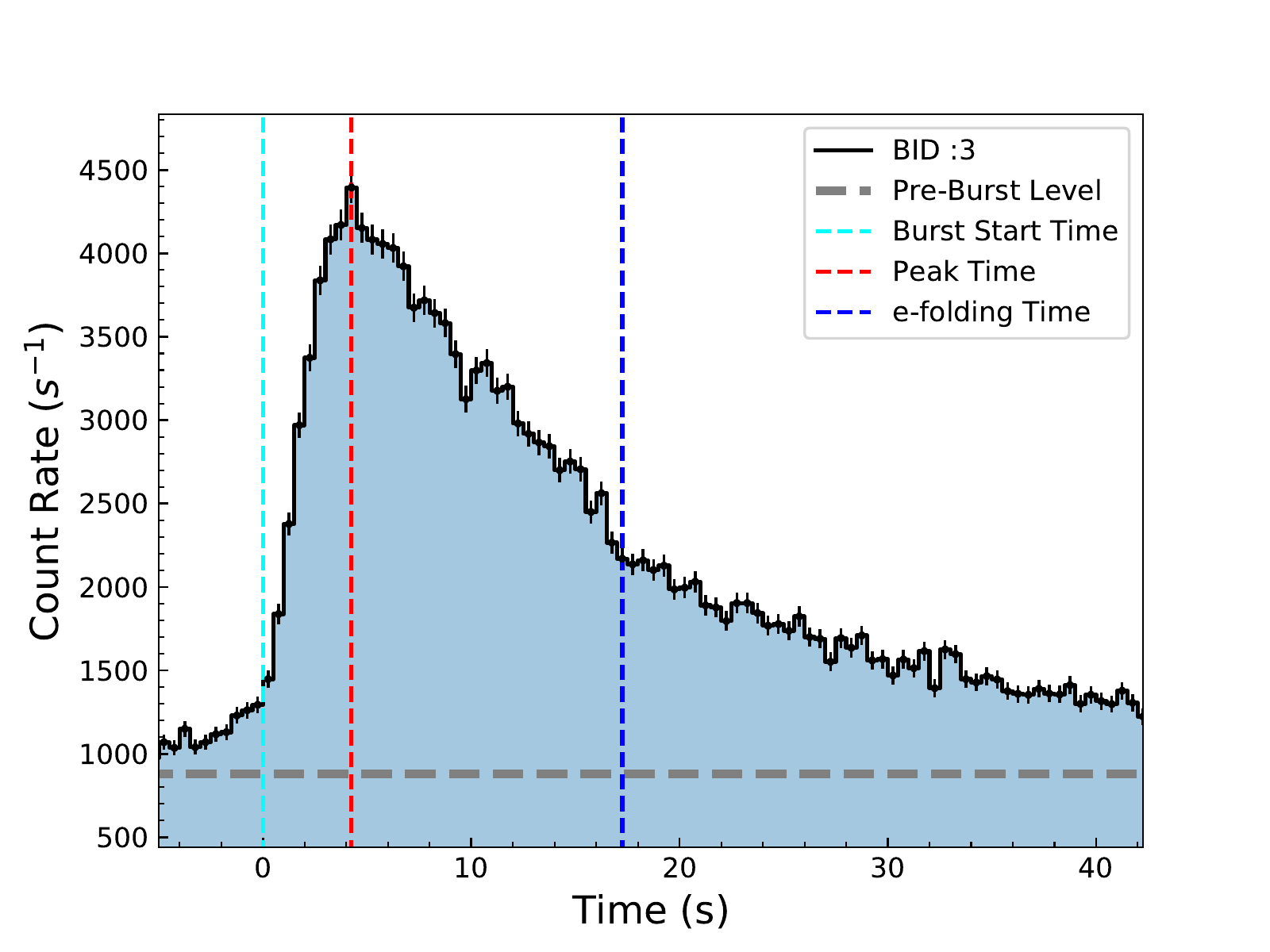}
	\includegraphics[scale=0.5]{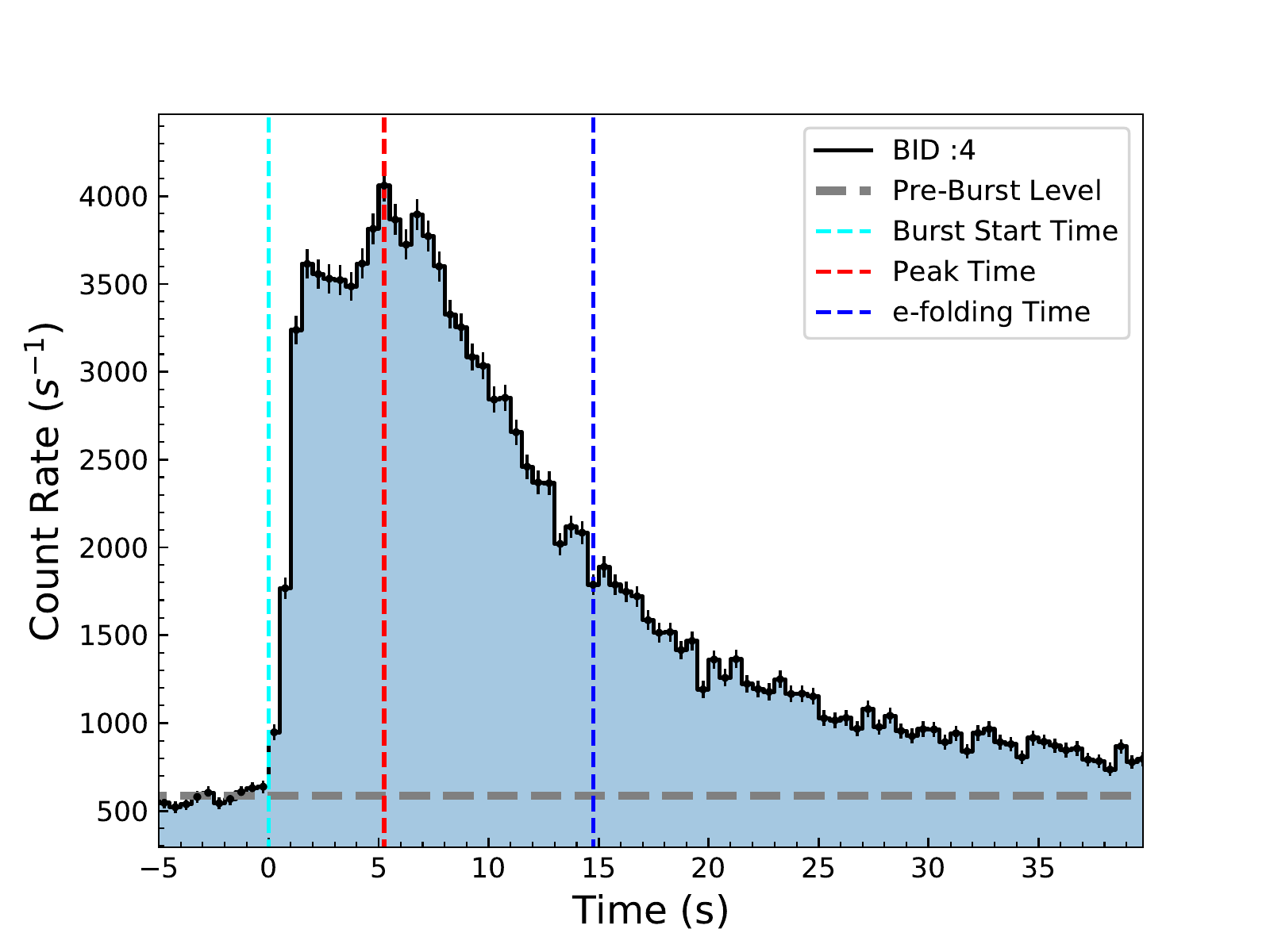}
	\includegraphics[scale=0.5]{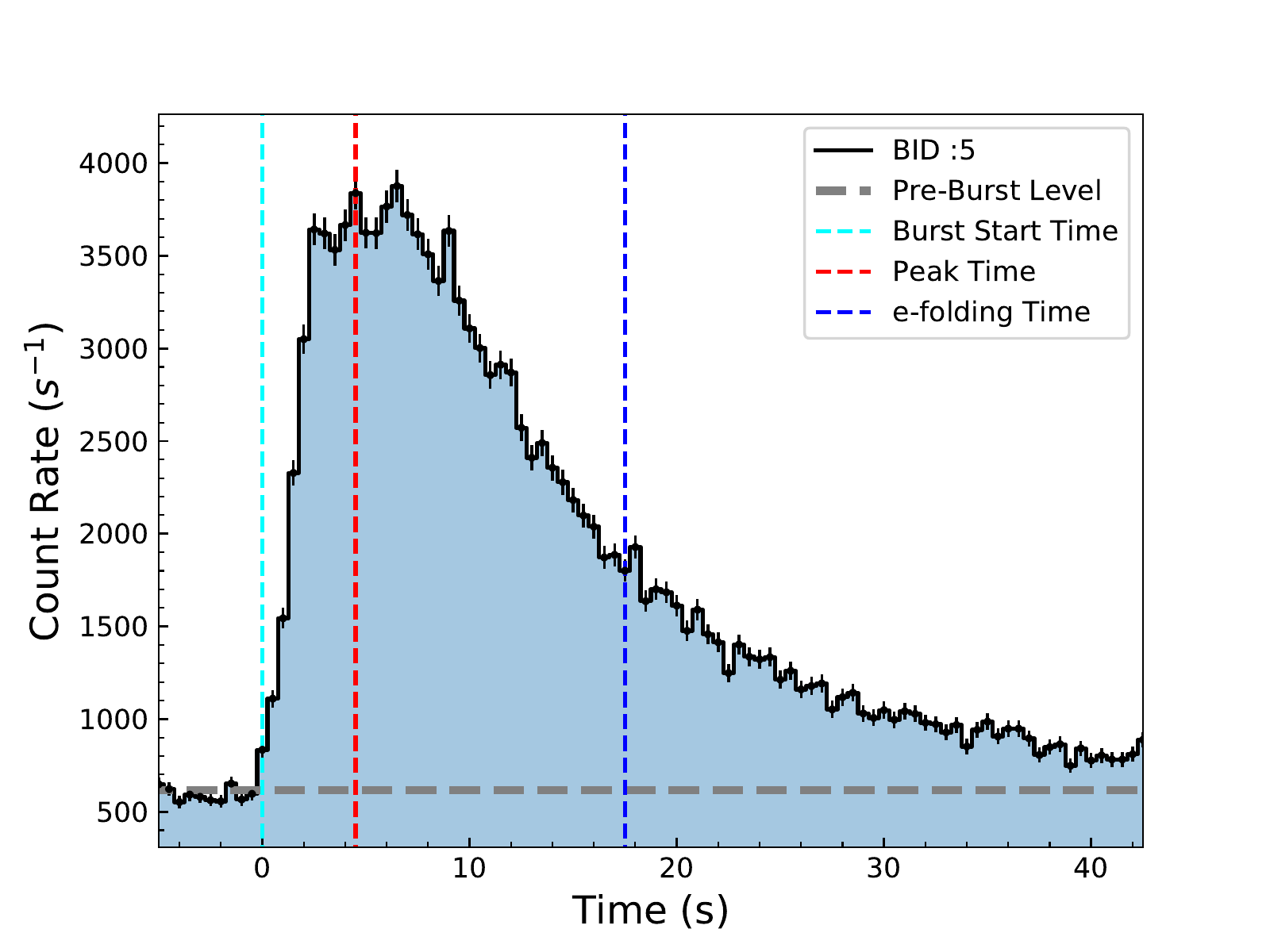}
	\includegraphics[scale=0.5]{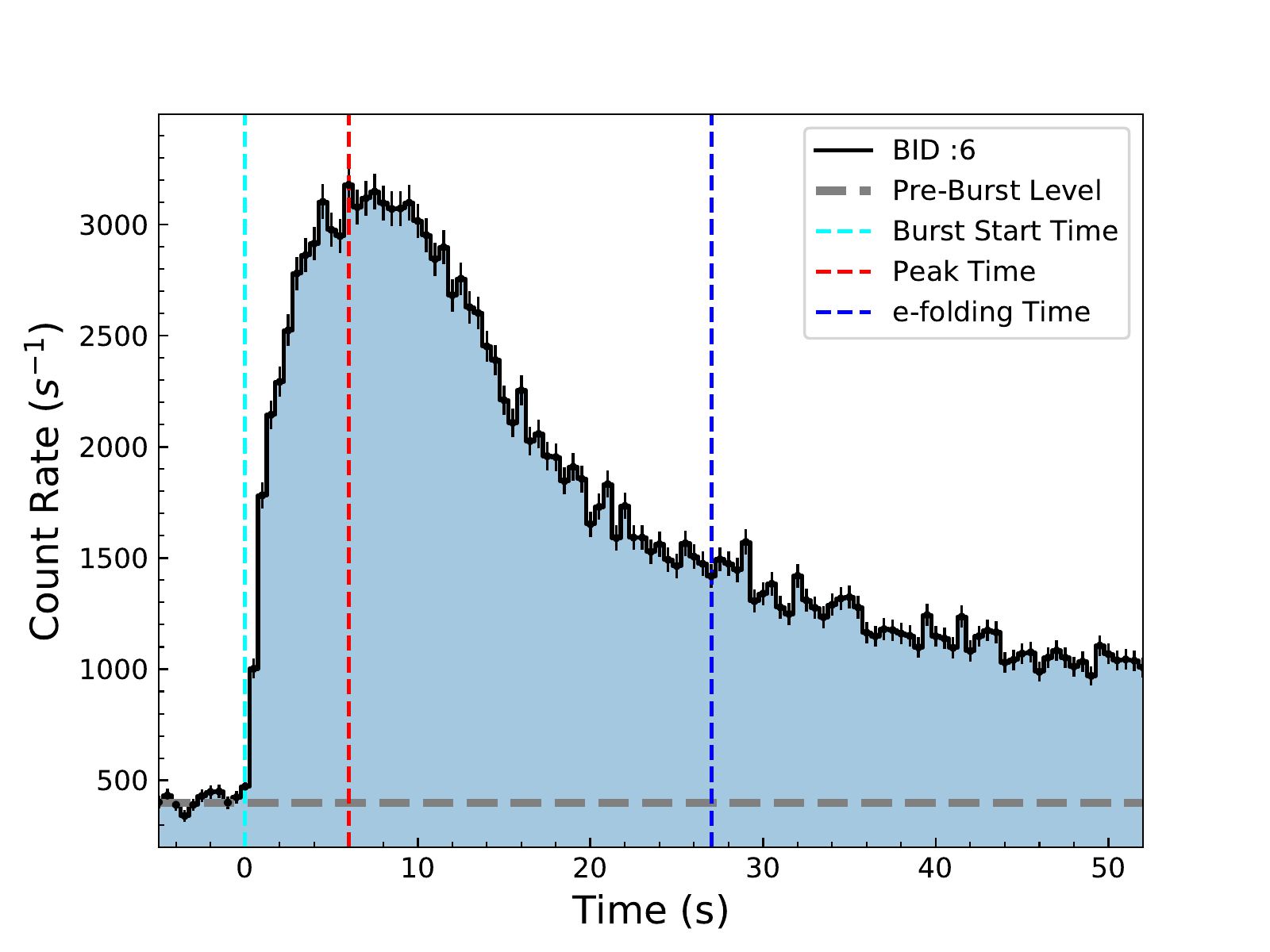}
	\includegraphics[scale=0.5]{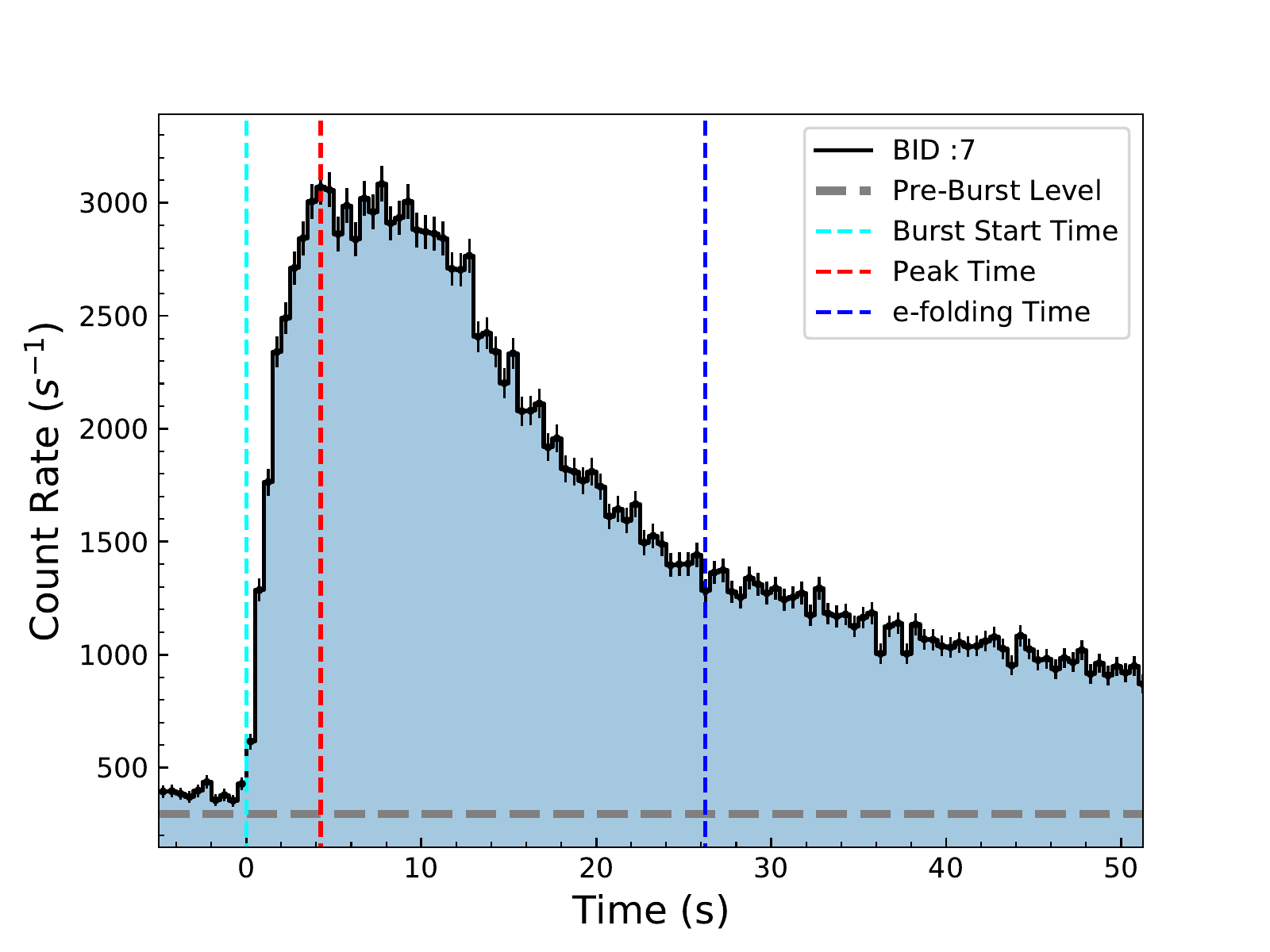}
	\includegraphics[scale=0.5]{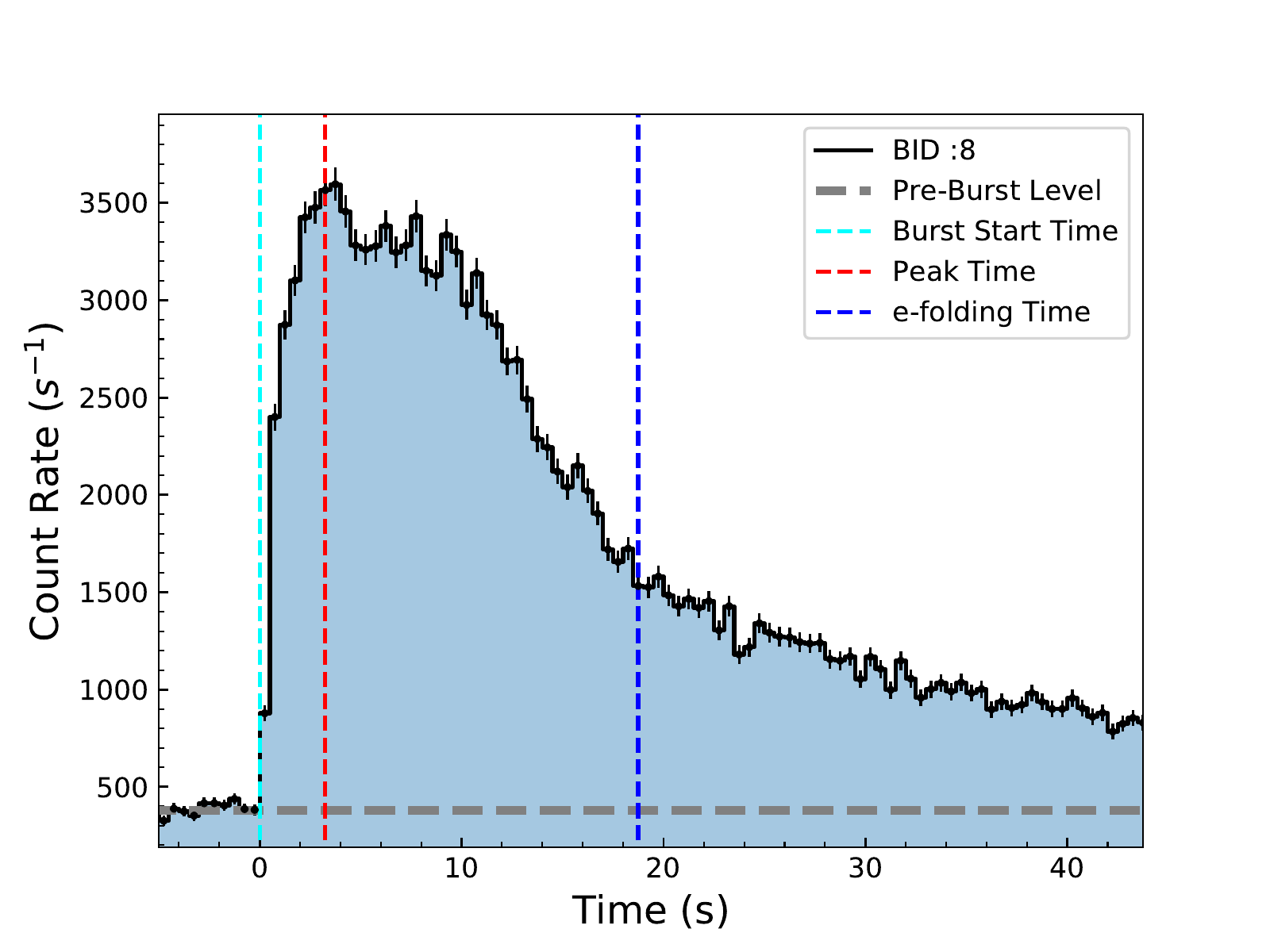}
    \caption{Background subtracted of 0.5$-$10.0~keV lightcurves of detected thermonuclear X-ray bursts with NICER.}
    \label{fig:burst_lc_1}
\end{figure*}

\begin{figure*}
	\includegraphics[scale=0.5]{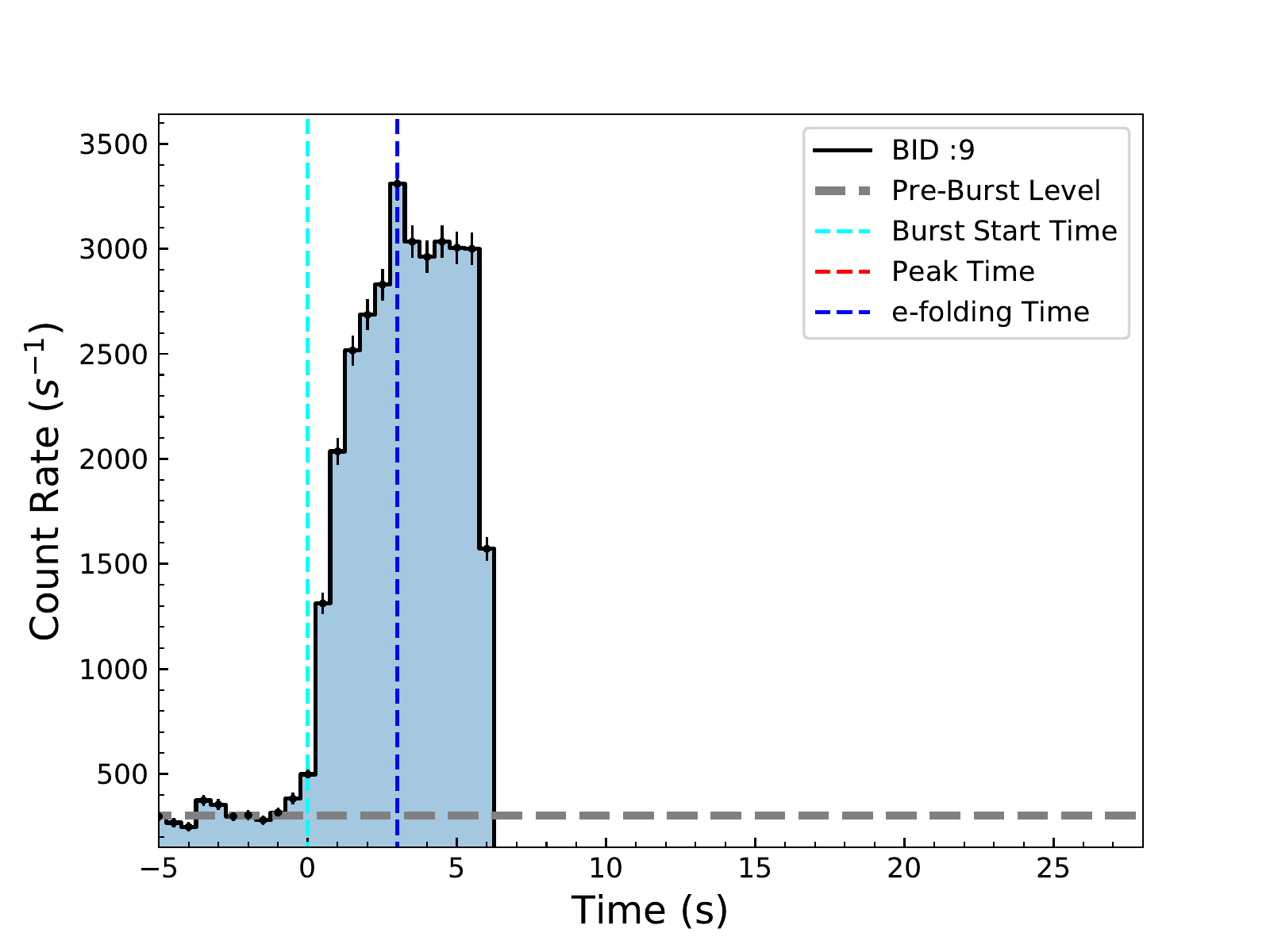}
	\includegraphics[scale=0.5]{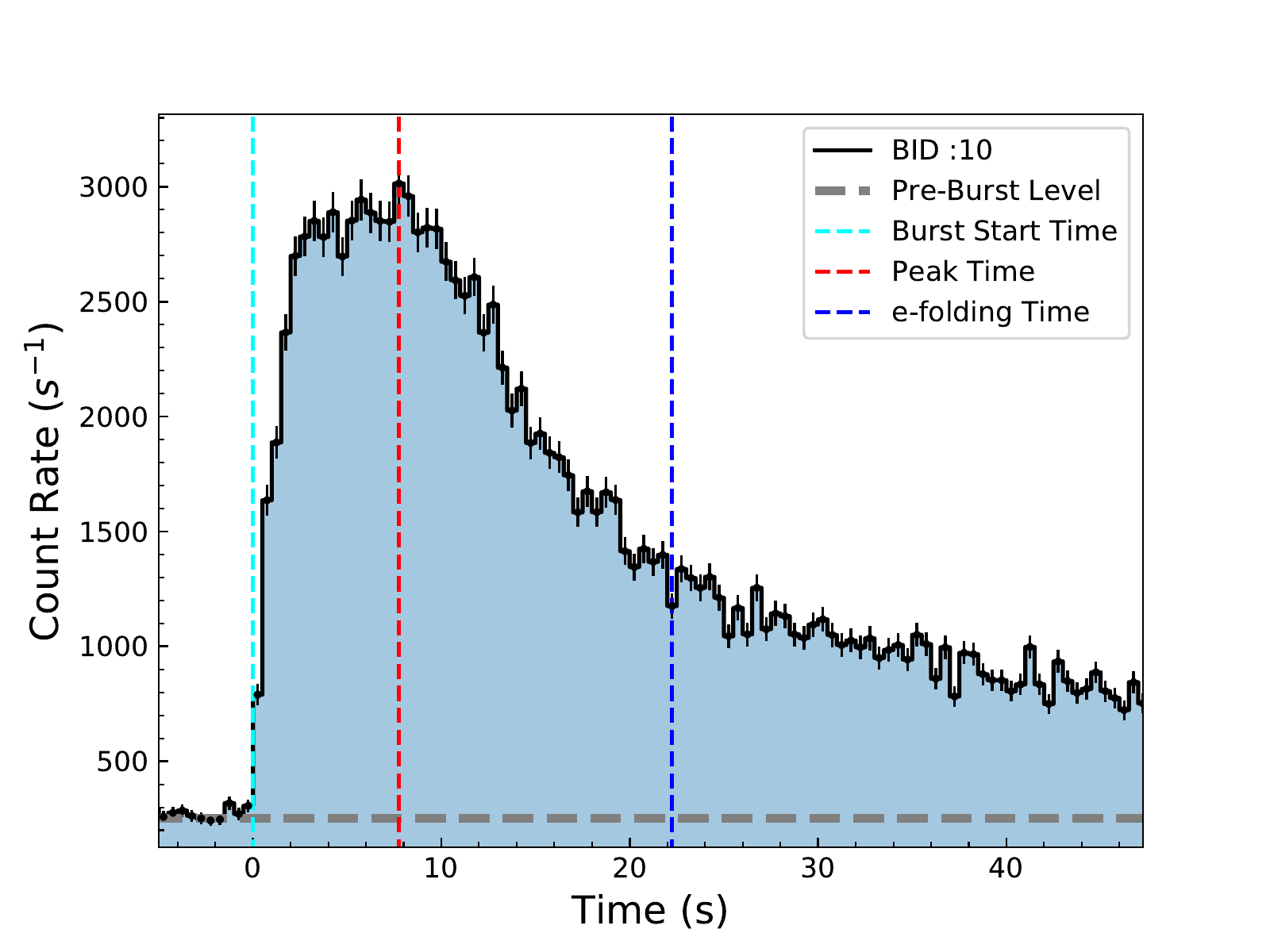}
	\includegraphics[scale=0.5]{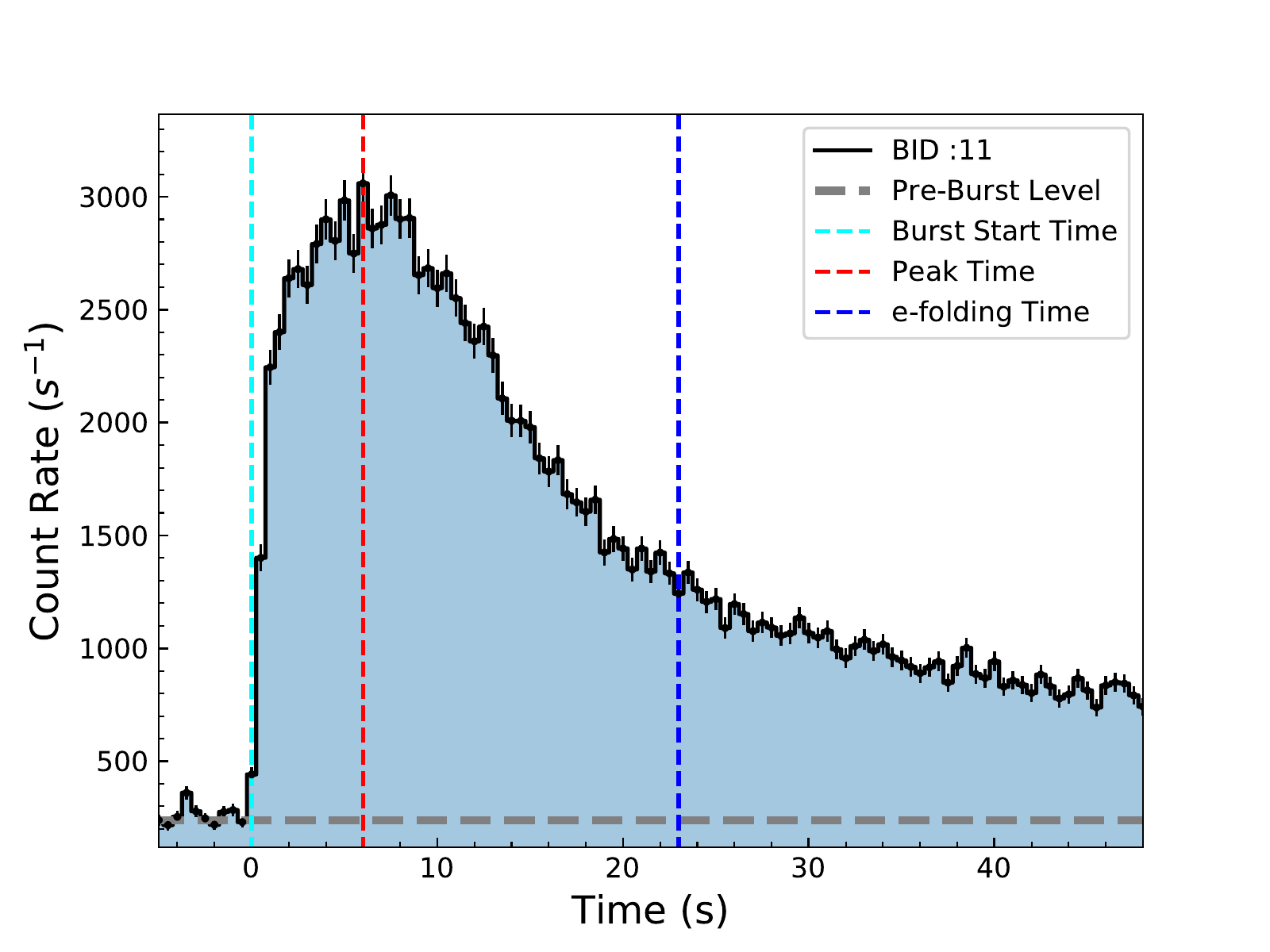}	\includegraphics[scale=0.5]{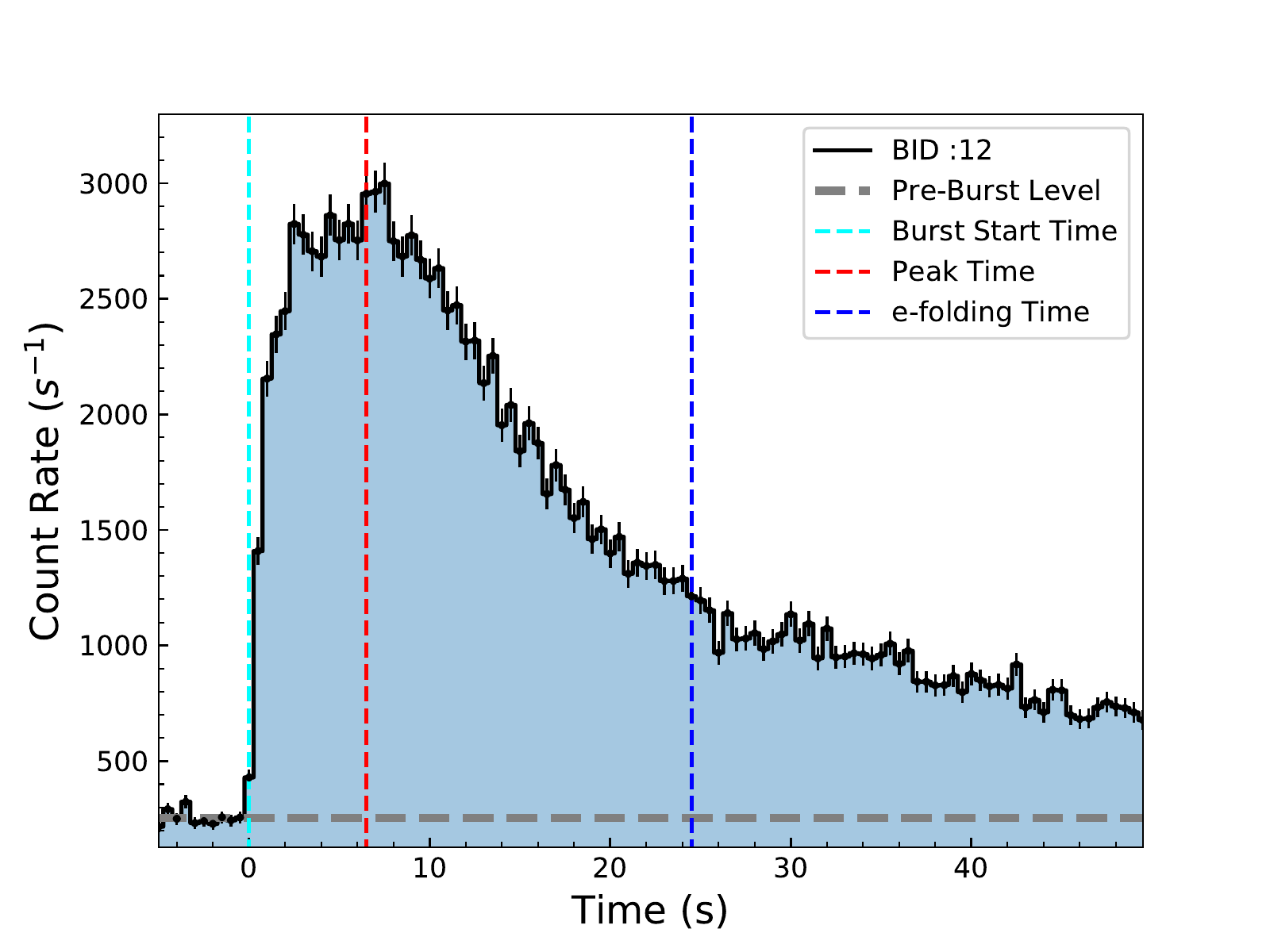}
	\includegraphics[scale=0.5]{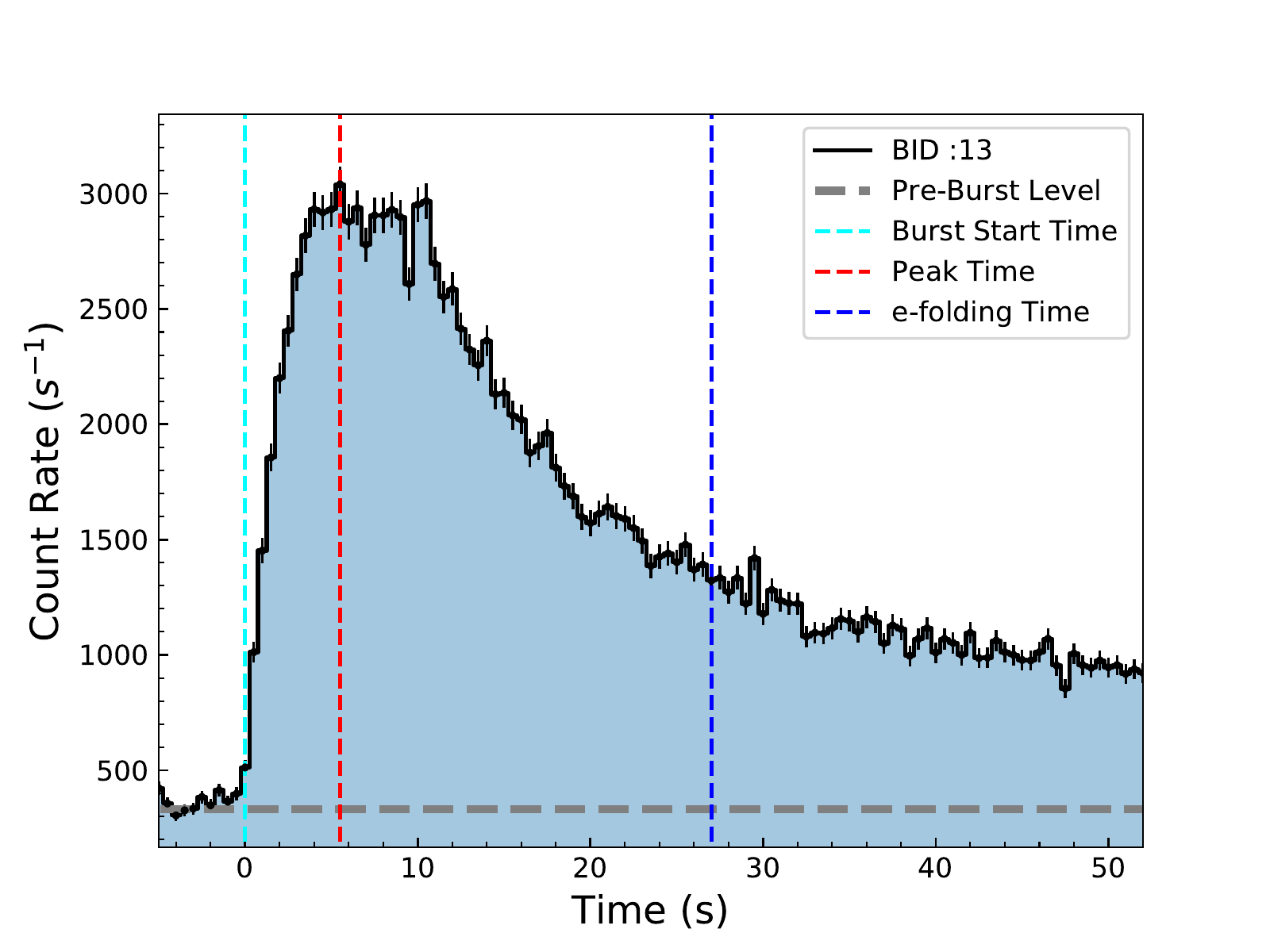}
	\includegraphics[scale=0.5]{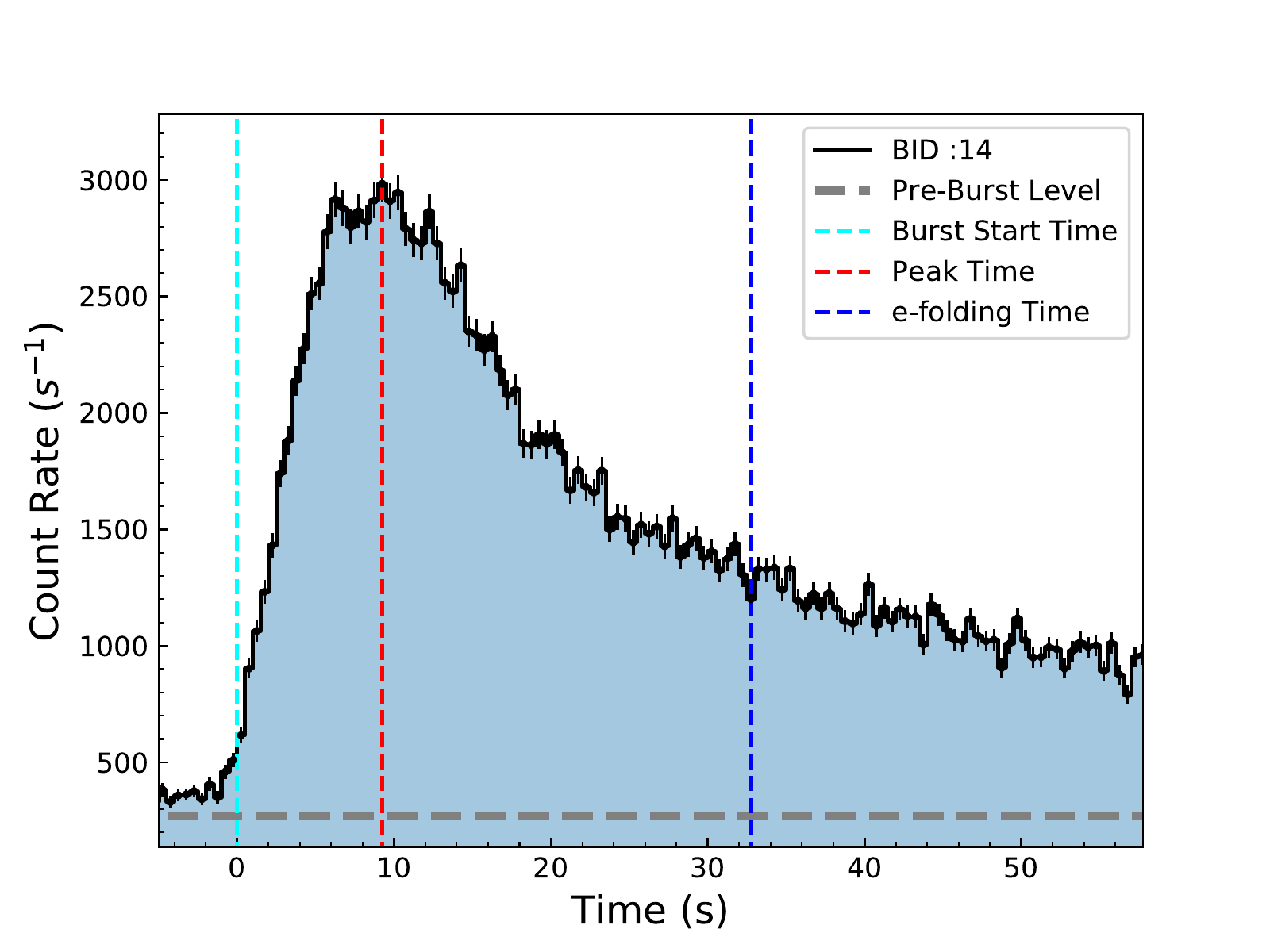}
	\includegraphics[scale=0.5]{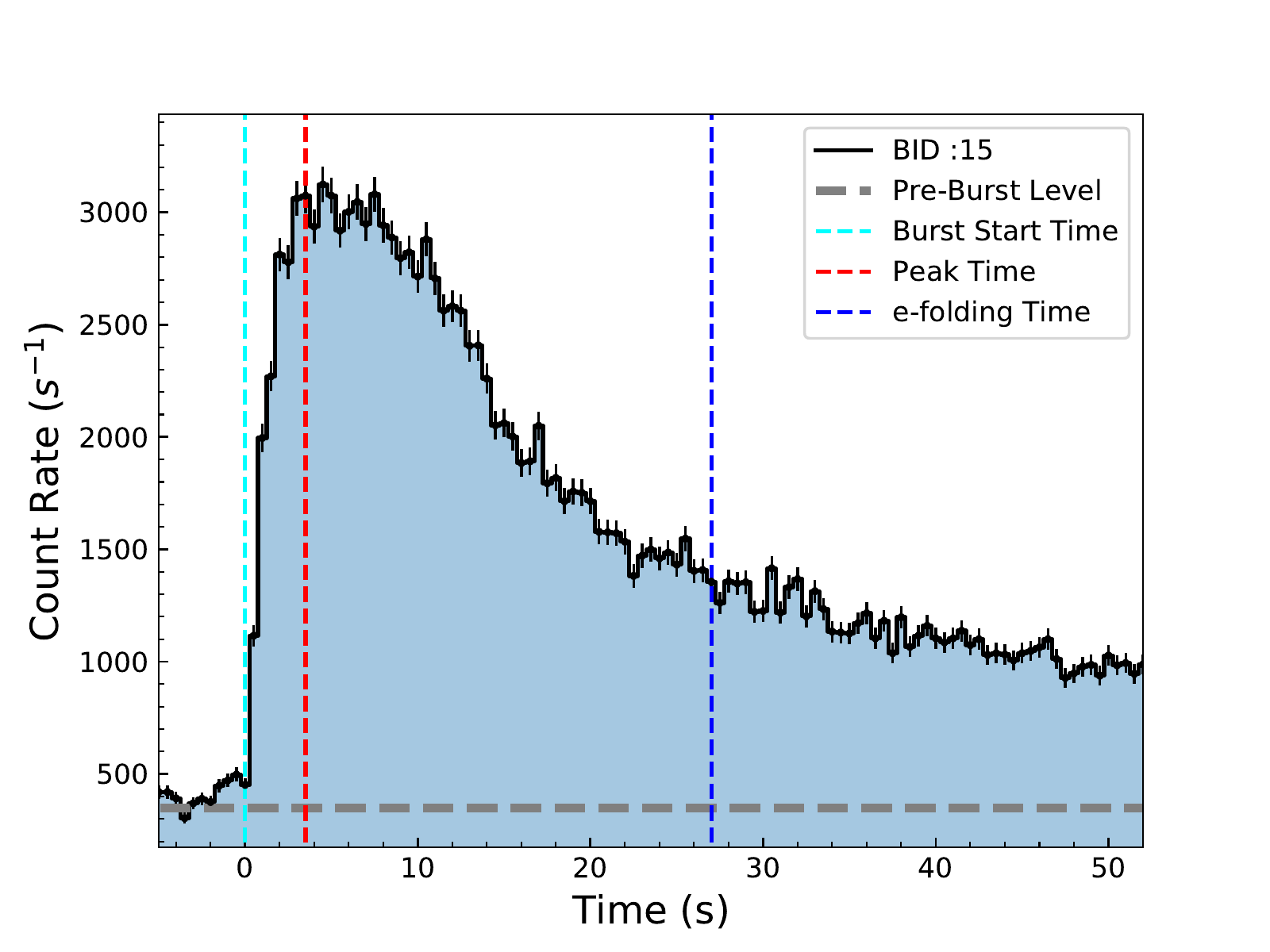}
	\includegraphics[scale=0.5]{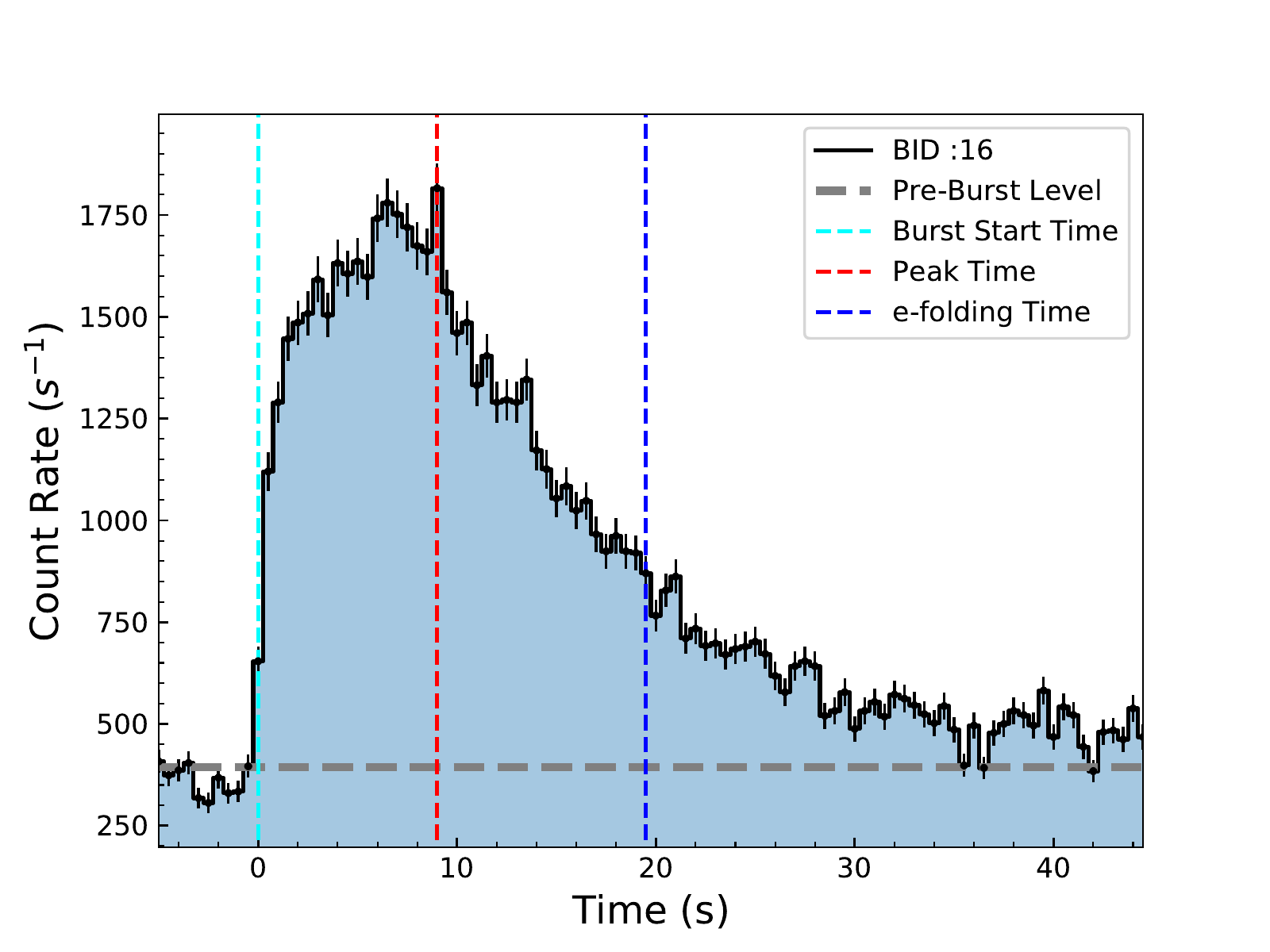}
    \caption{Same as \autoref{fig:burst_lc_1}. Note that the observation stopped just after the peak of burst 9.}
    \label{fig:burst_lc_2}
\end{figure*}

\begin{figure*}
	\includegraphics[scale=0.5]{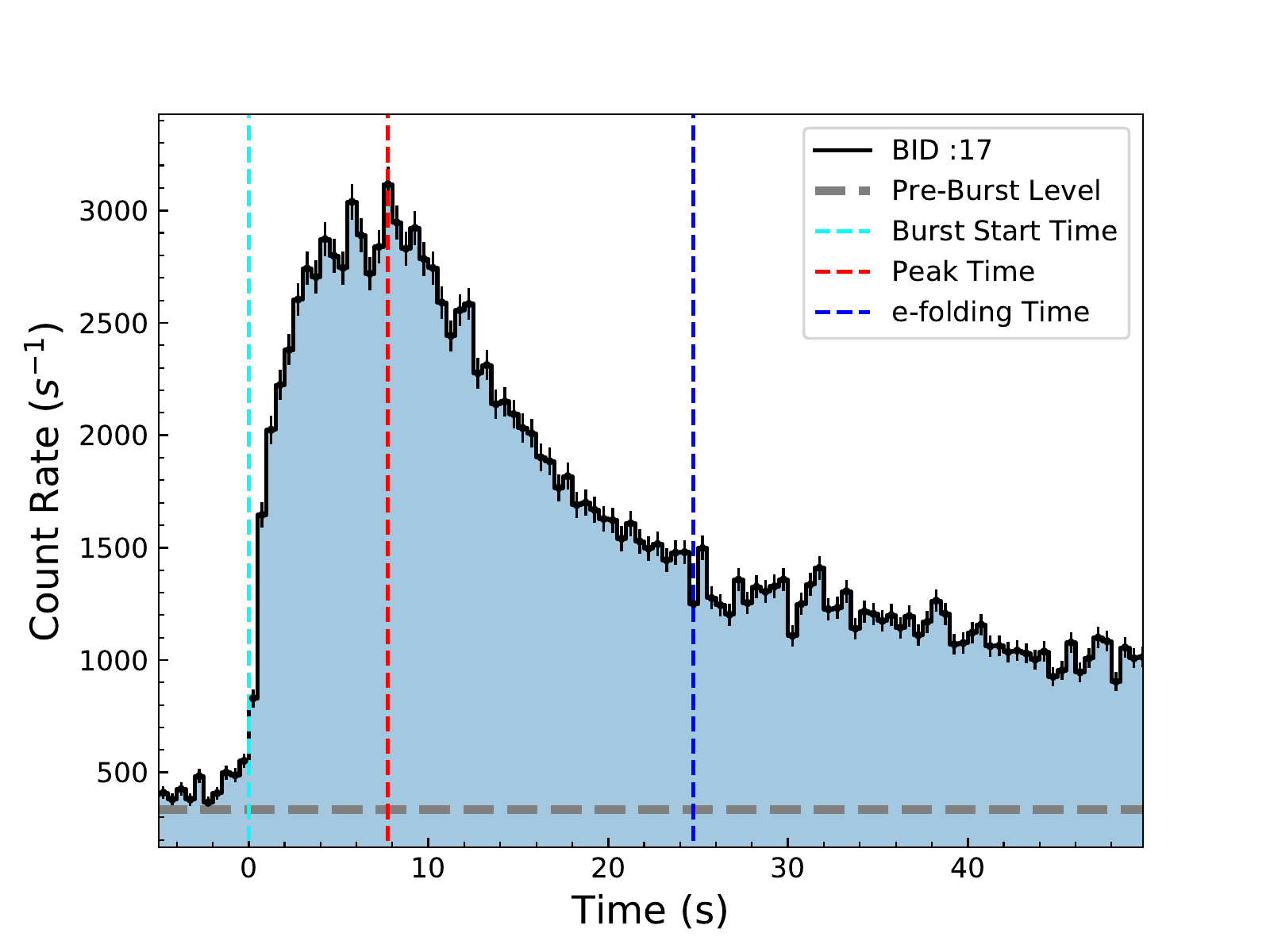}
	\includegraphics[scale=0.5]{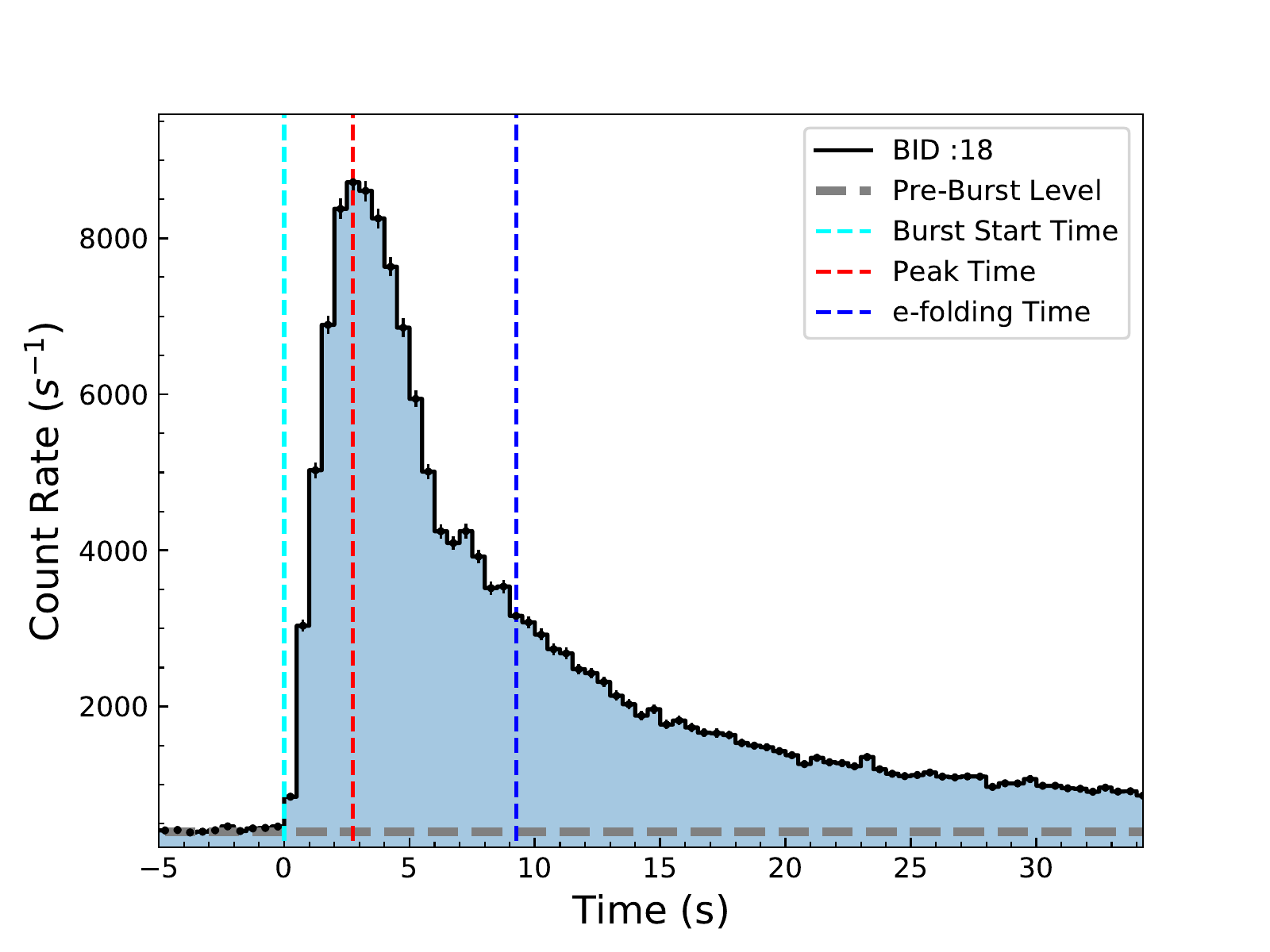}
	\includegraphics[scale=0.5]{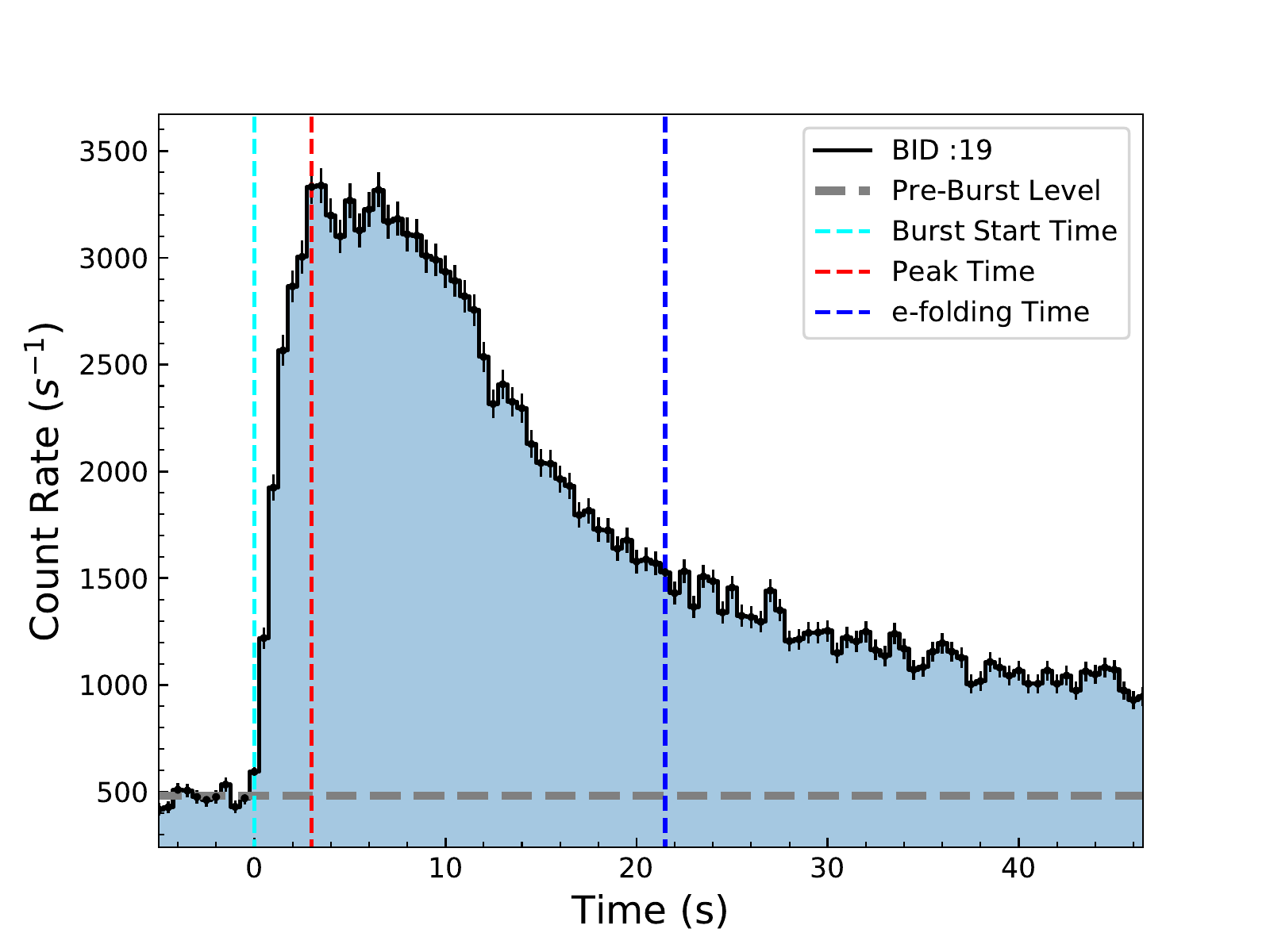}
	\includegraphics[scale=0.5]{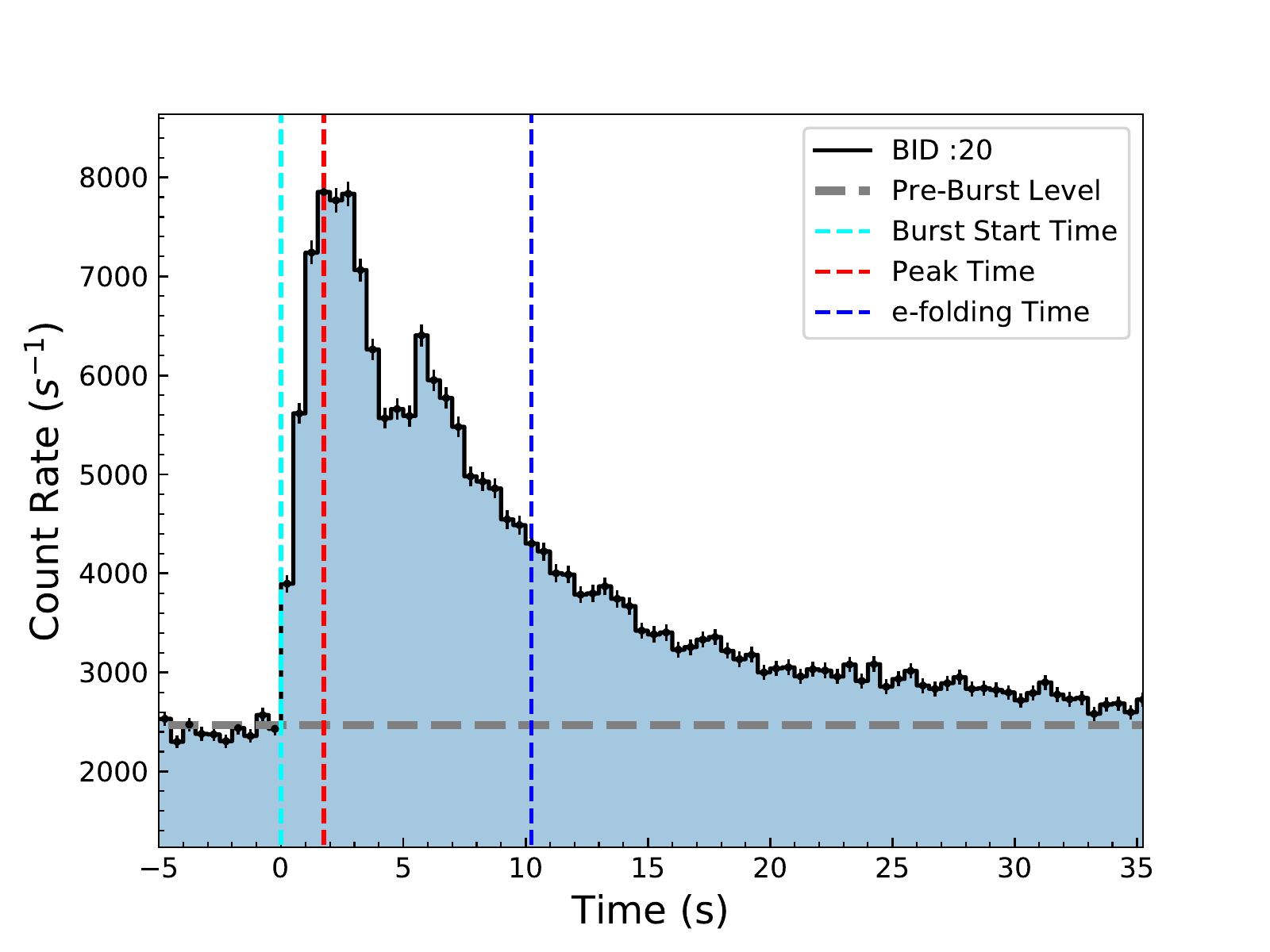}
	\includegraphics[scale=0.5]{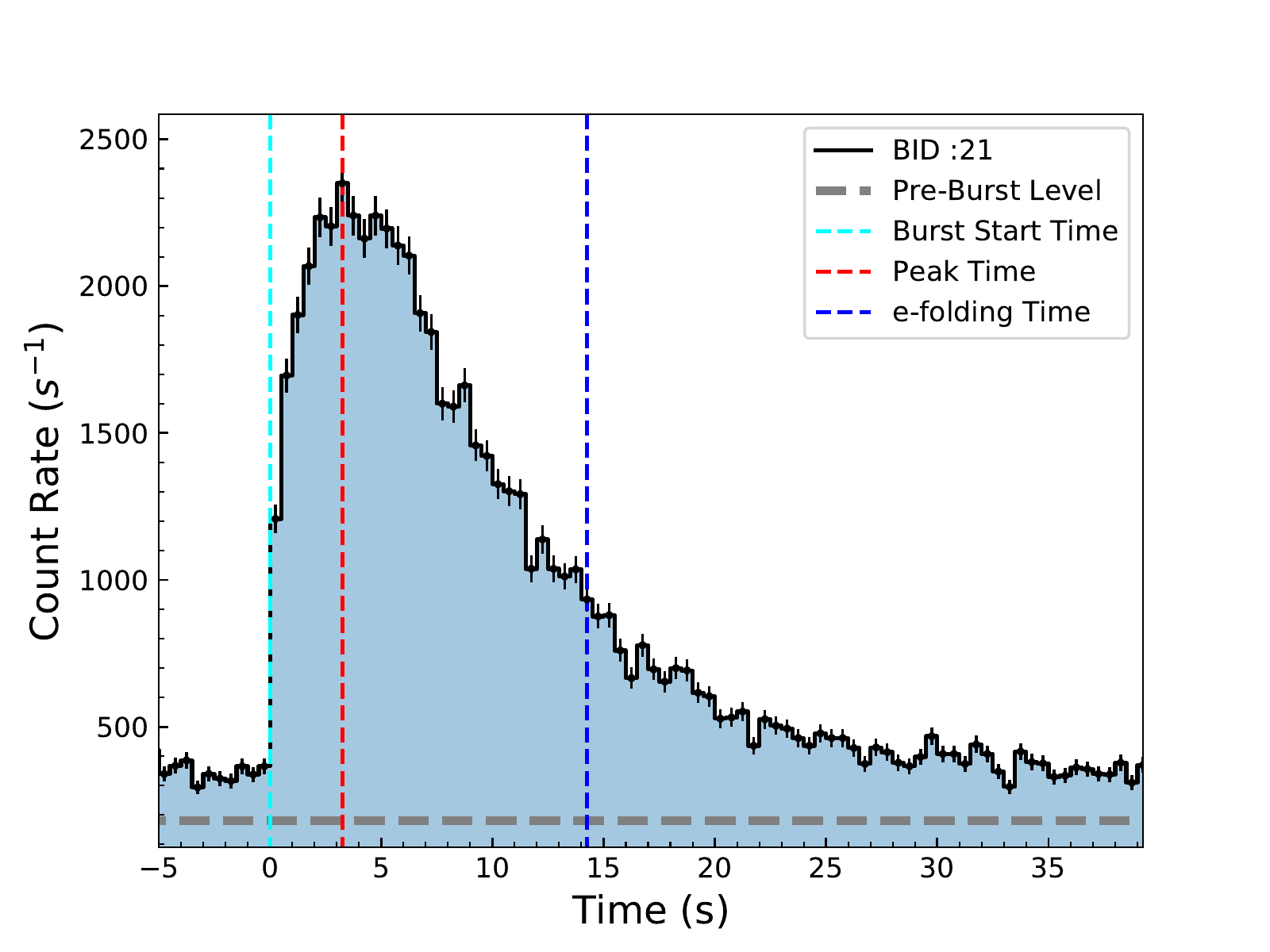}
	\includegraphics[scale=0.5]{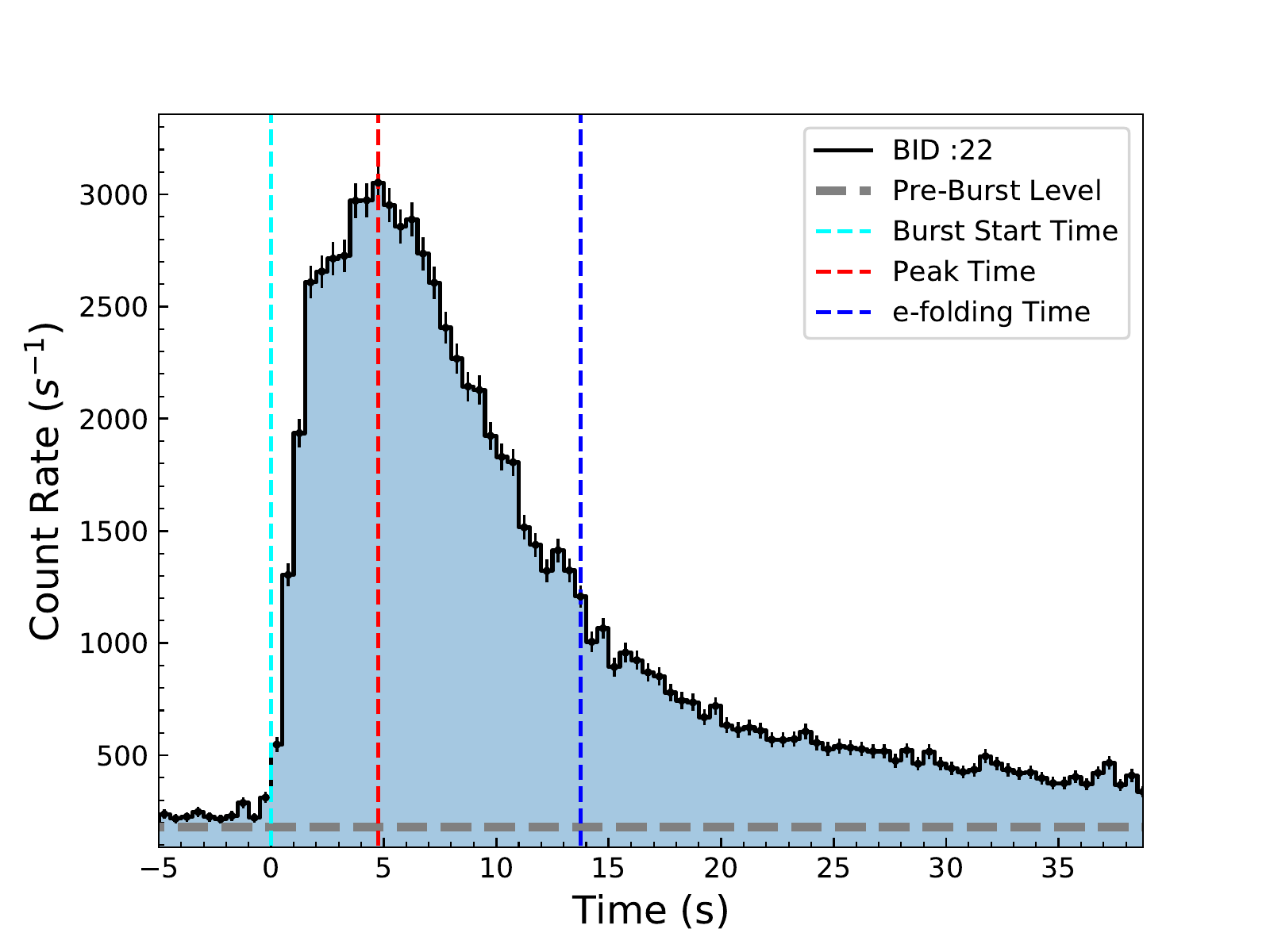}
    \caption{Same as \autoref{fig:burst_lc_1}.}
    \label{fig:burst_lc_3}
\end{figure*}


\section{Time evolution of spectral parameters for each burst} 
\label{app:sp_time_ev}
Figures of time resolved spectral evolution for all the bursts are shown here. 

\begin{figure*}
	\includegraphics[scale=0.5]{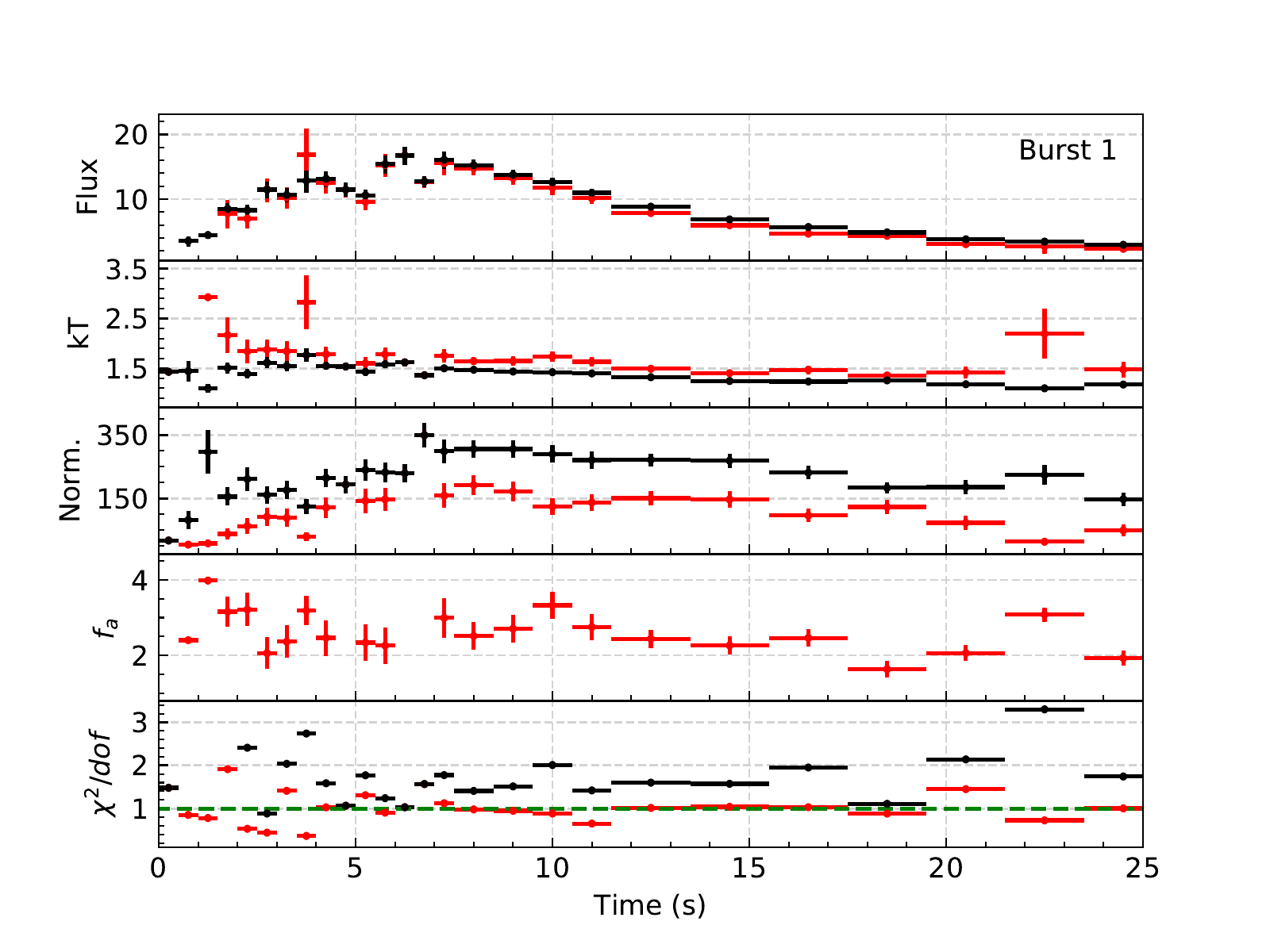}
	\includegraphics[scale=0.5]{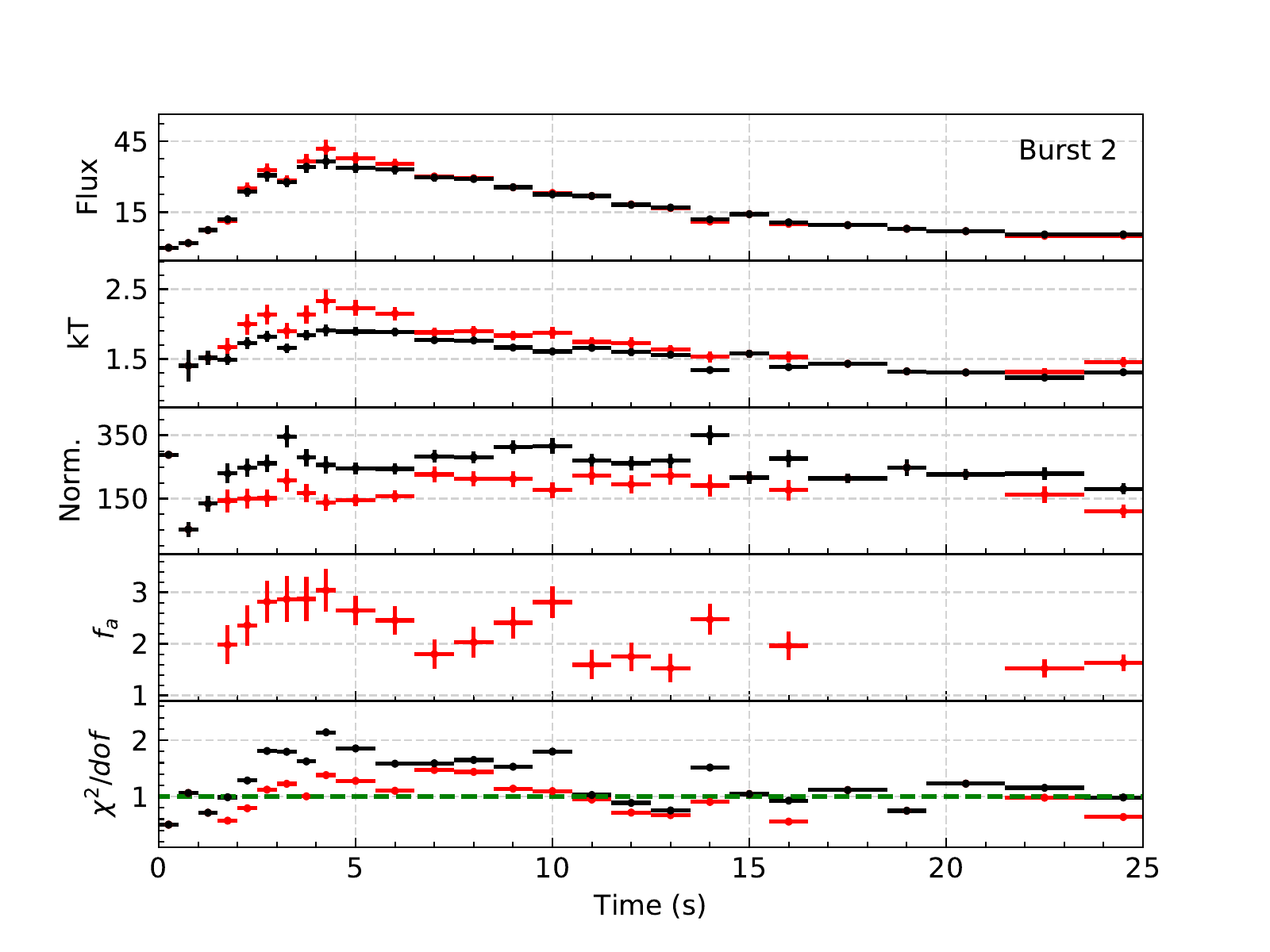}
	\includegraphics[scale=0.5]{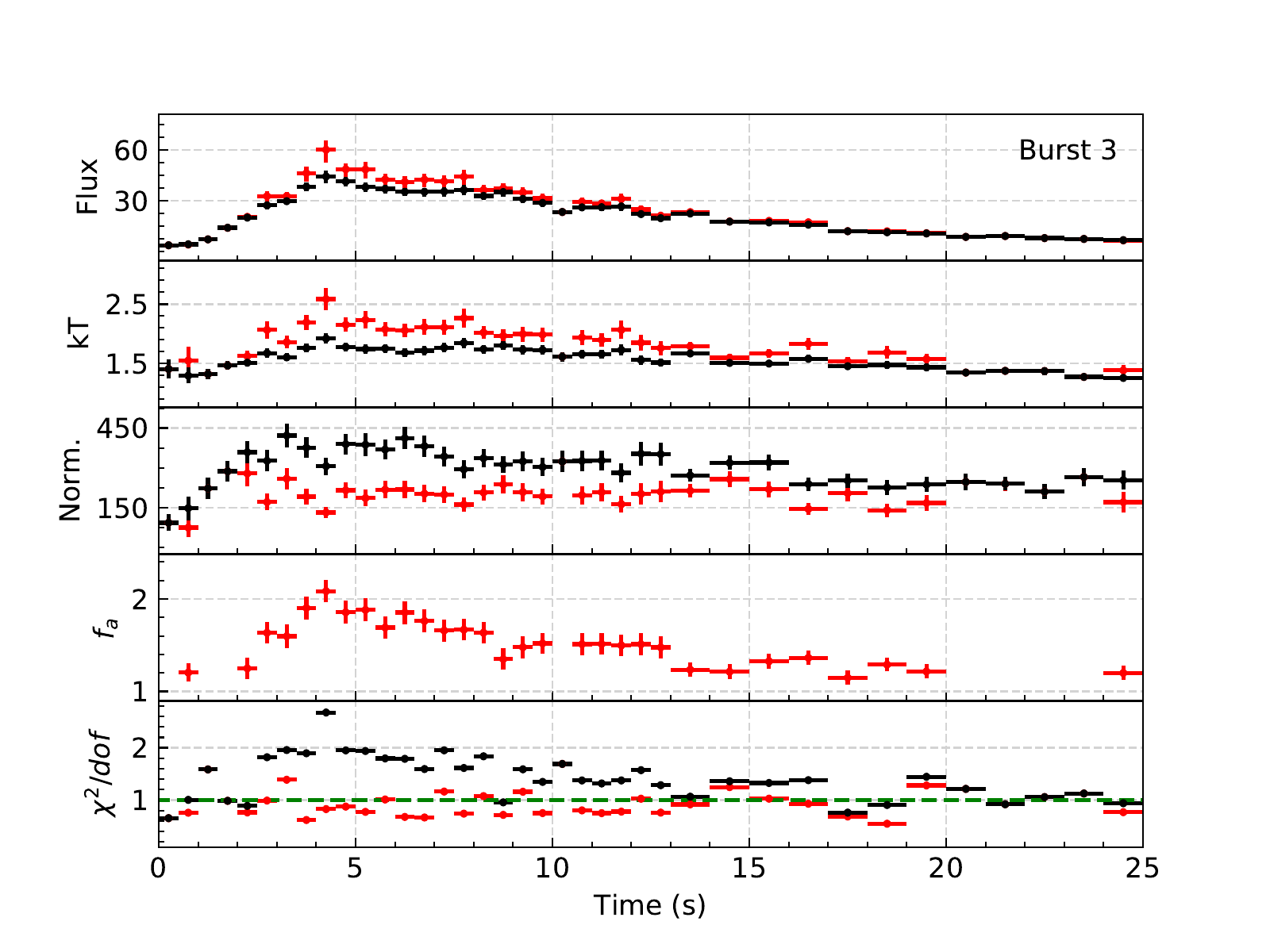}
	\includegraphics[scale=0.5]{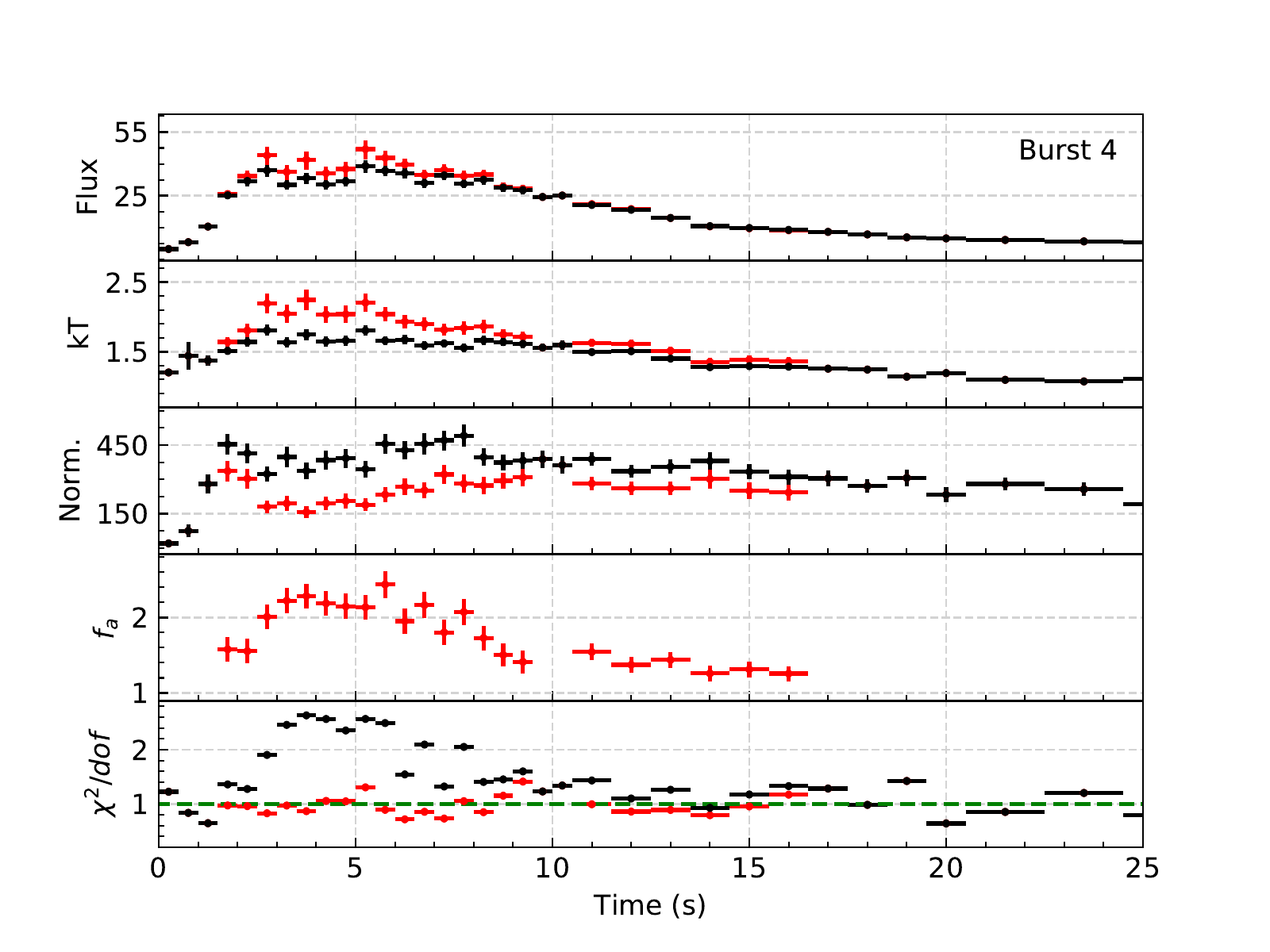}
	\includegraphics[scale=0.5]{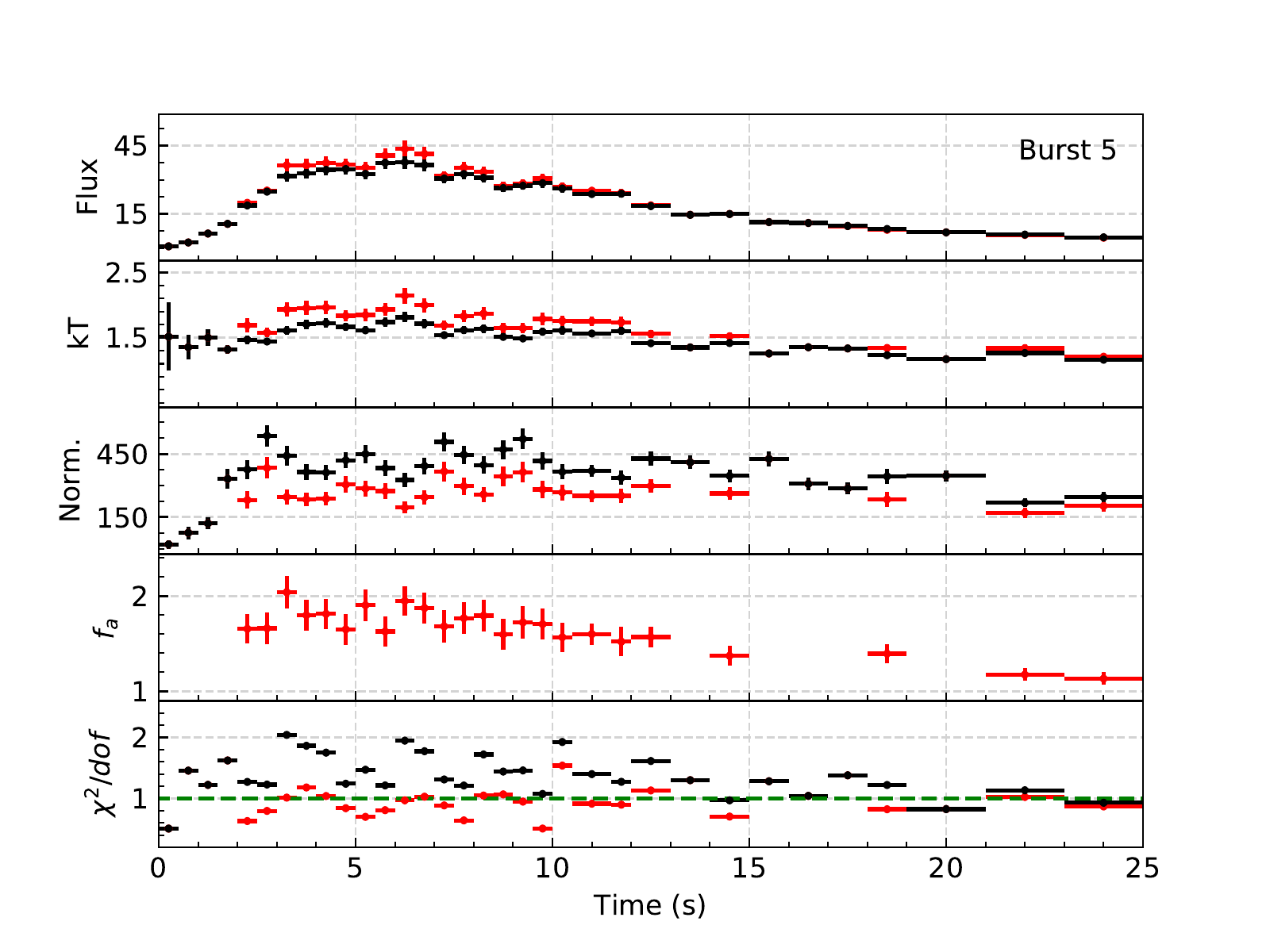}
	\includegraphics[scale=0.5]{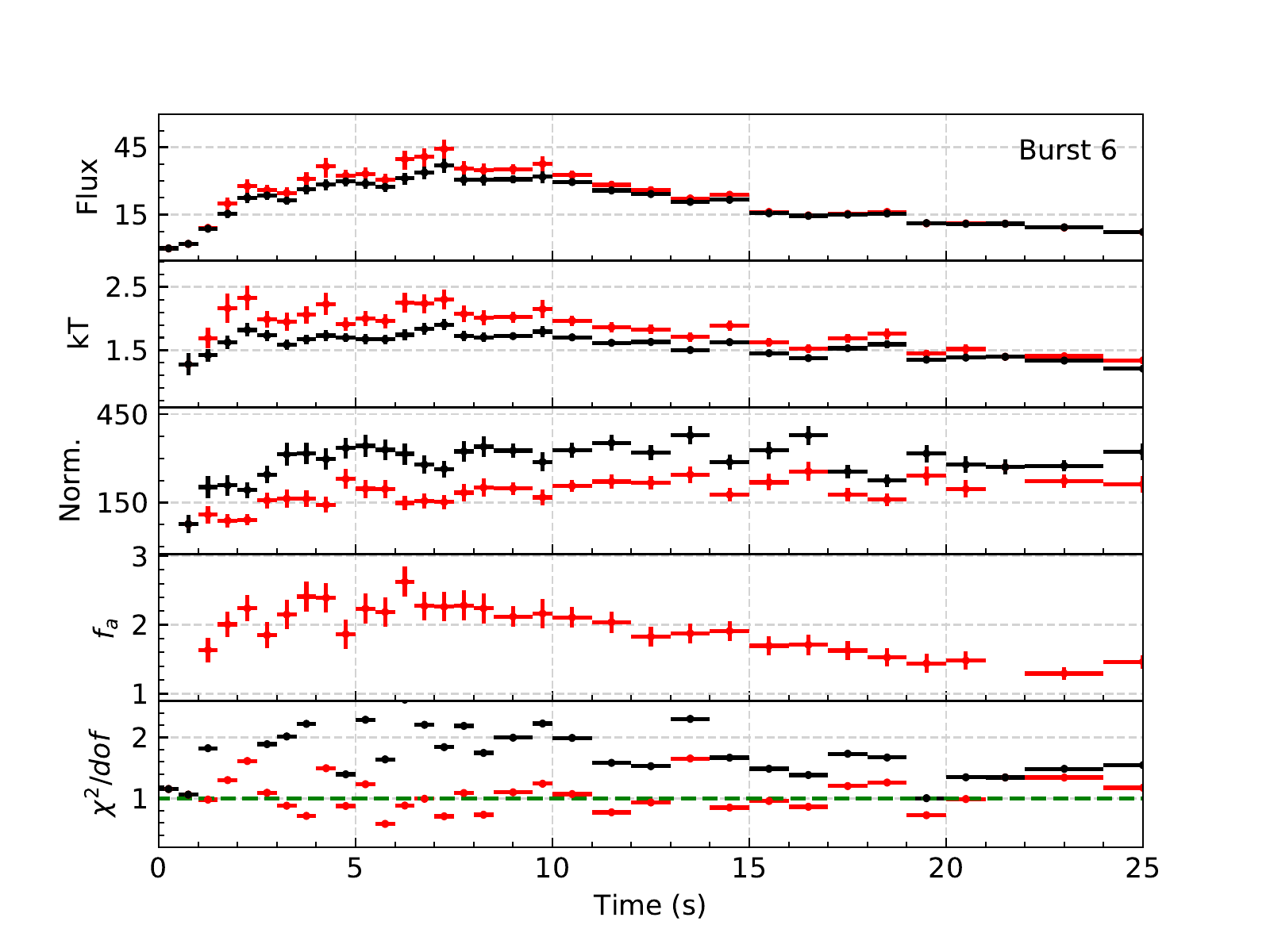}
    \caption{Time evolution of spectral parameters are shown. Red  symbols show the results of $f_a$ method and black symbols show the results for constant background case. We show, from top to bottom: the 0.5--10~keV X-ray flux, the temperature, the blackbody normalization, $f_a$ and finally, the fit statistic. In all panels fluxes are bolometric and in units of $\times10^{-9}$~\fluxcgs~. The temperature and blackbody normalization are in units of keV and $R^2_{km}/D^2_{10\rm{kpc}}$, respectively.}
    \label{fig:burst_plots_1}
\end{figure*}

\begin{figure*}
    \includegraphics[scale=0.5]{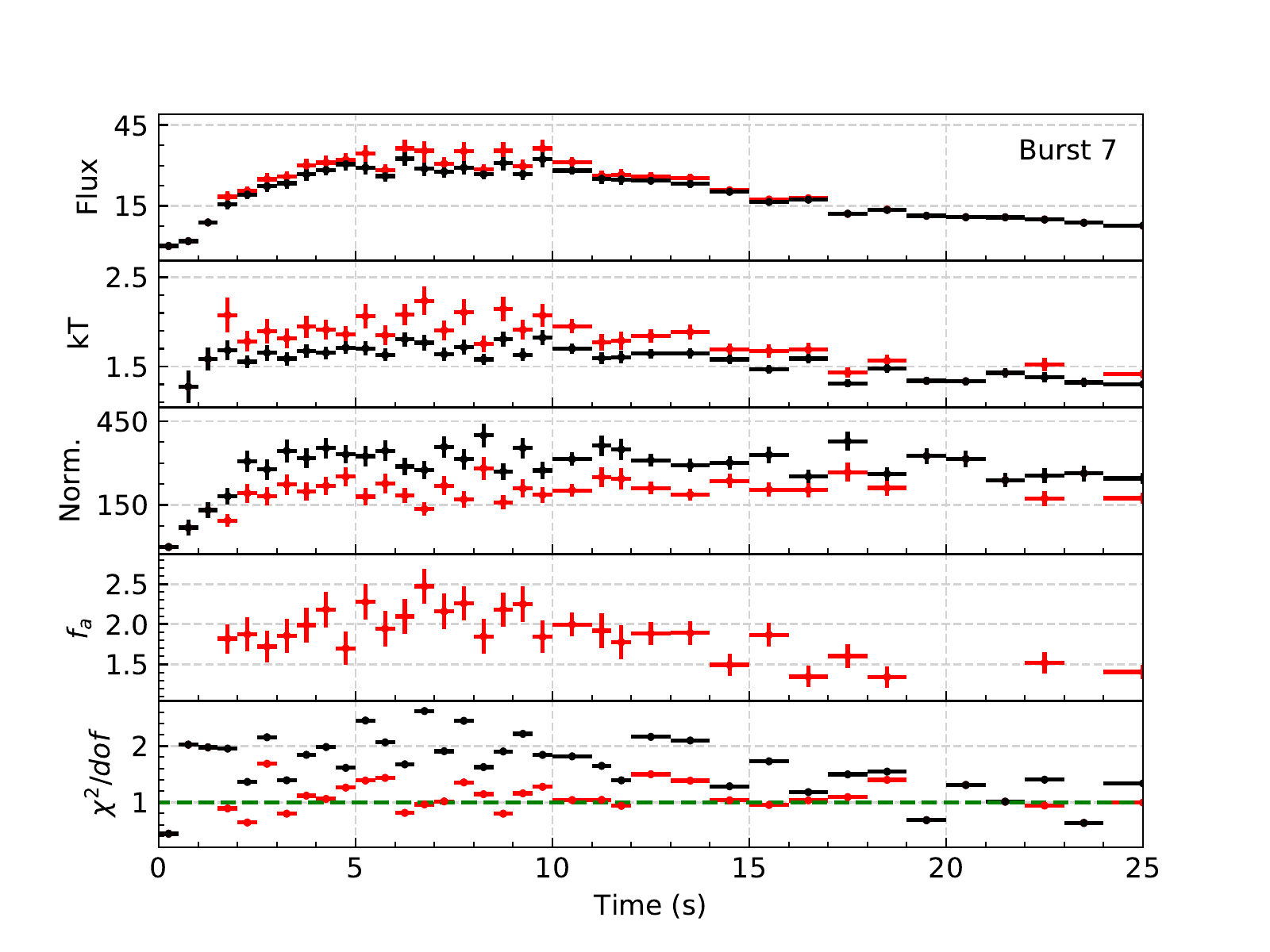}
	\includegraphics[scale=0.5]{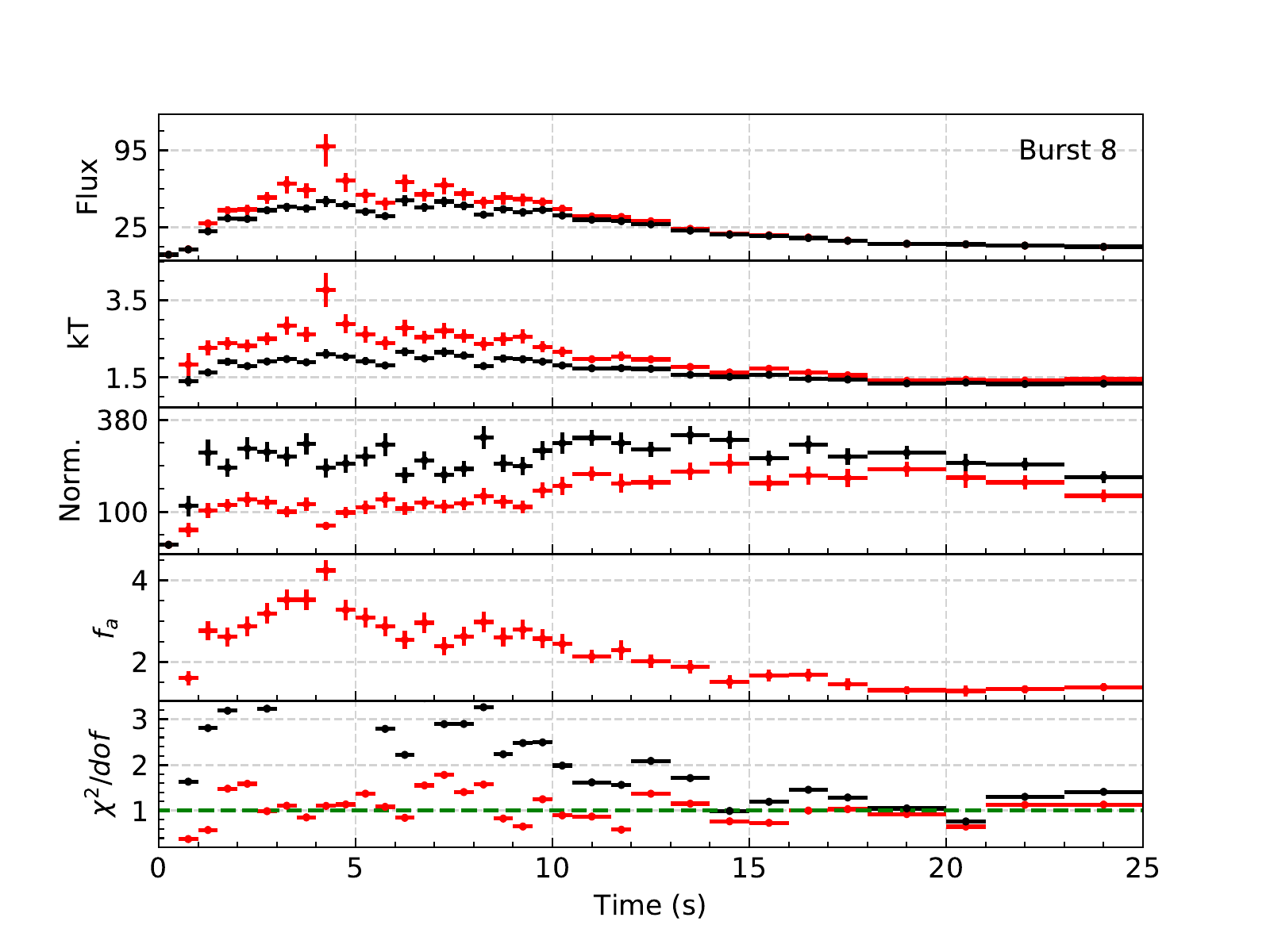}
	\includegraphics[scale=0.5]{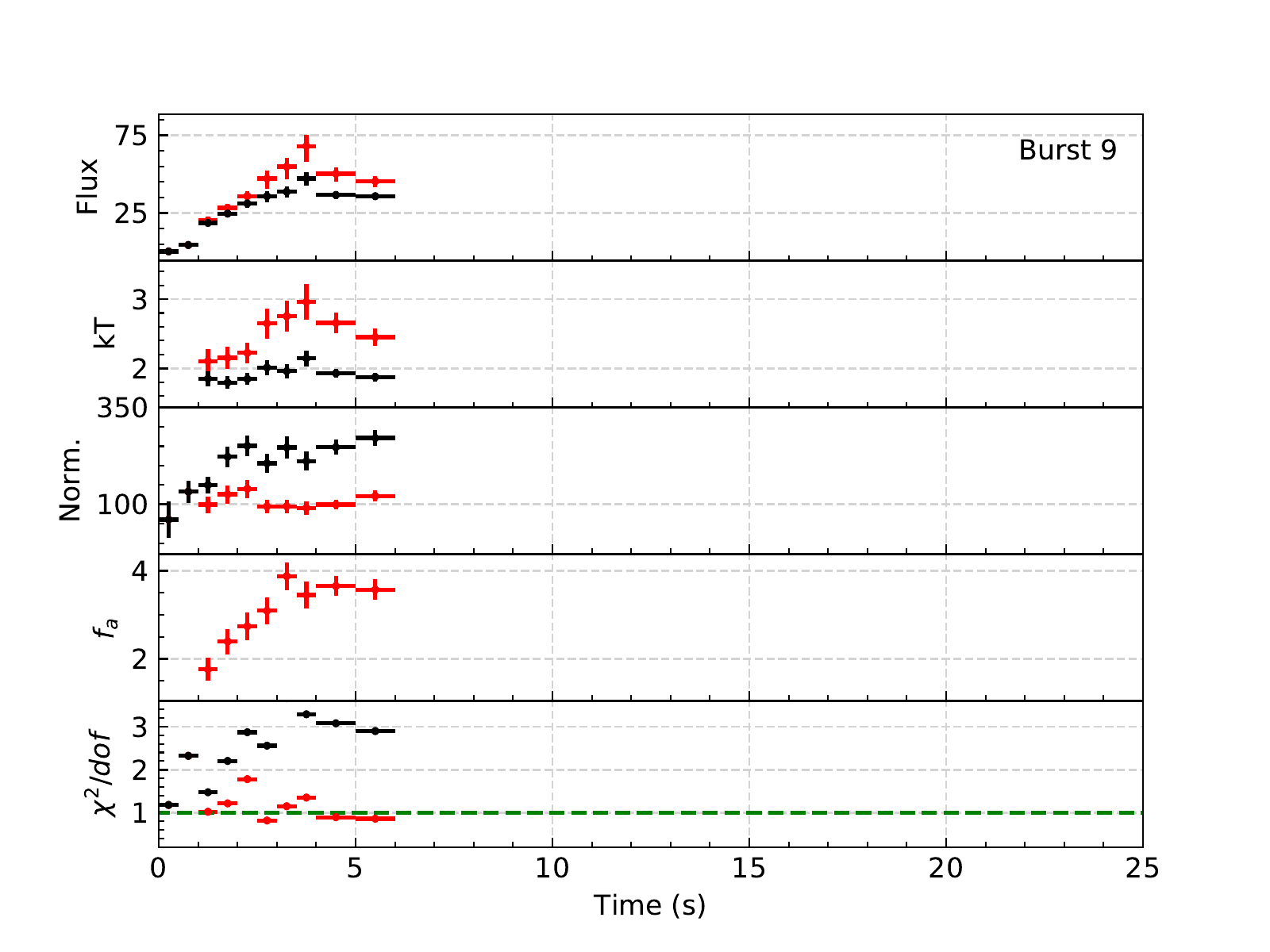}
	\includegraphics[scale=0.5]{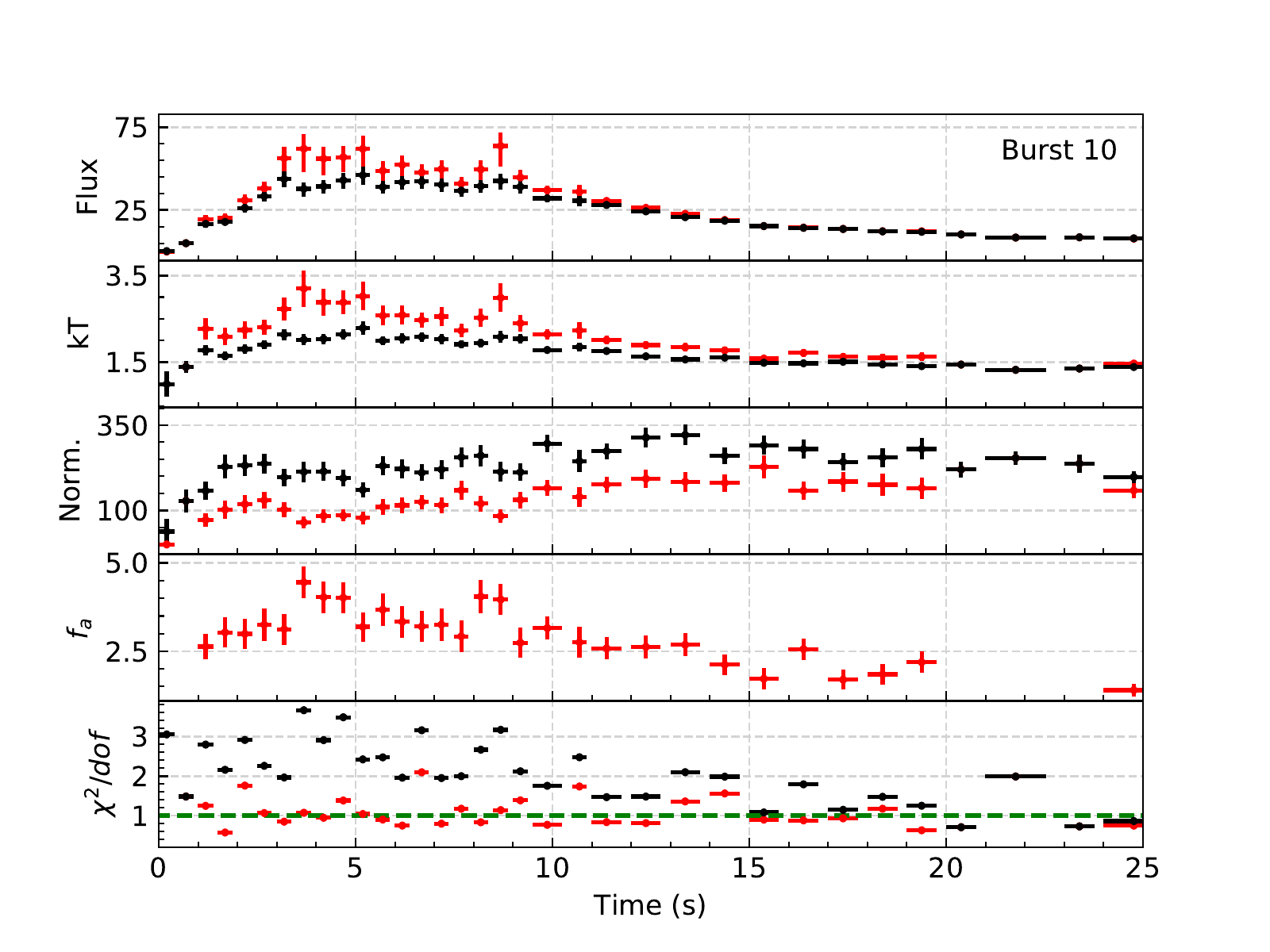}
	\includegraphics[scale=0.5]{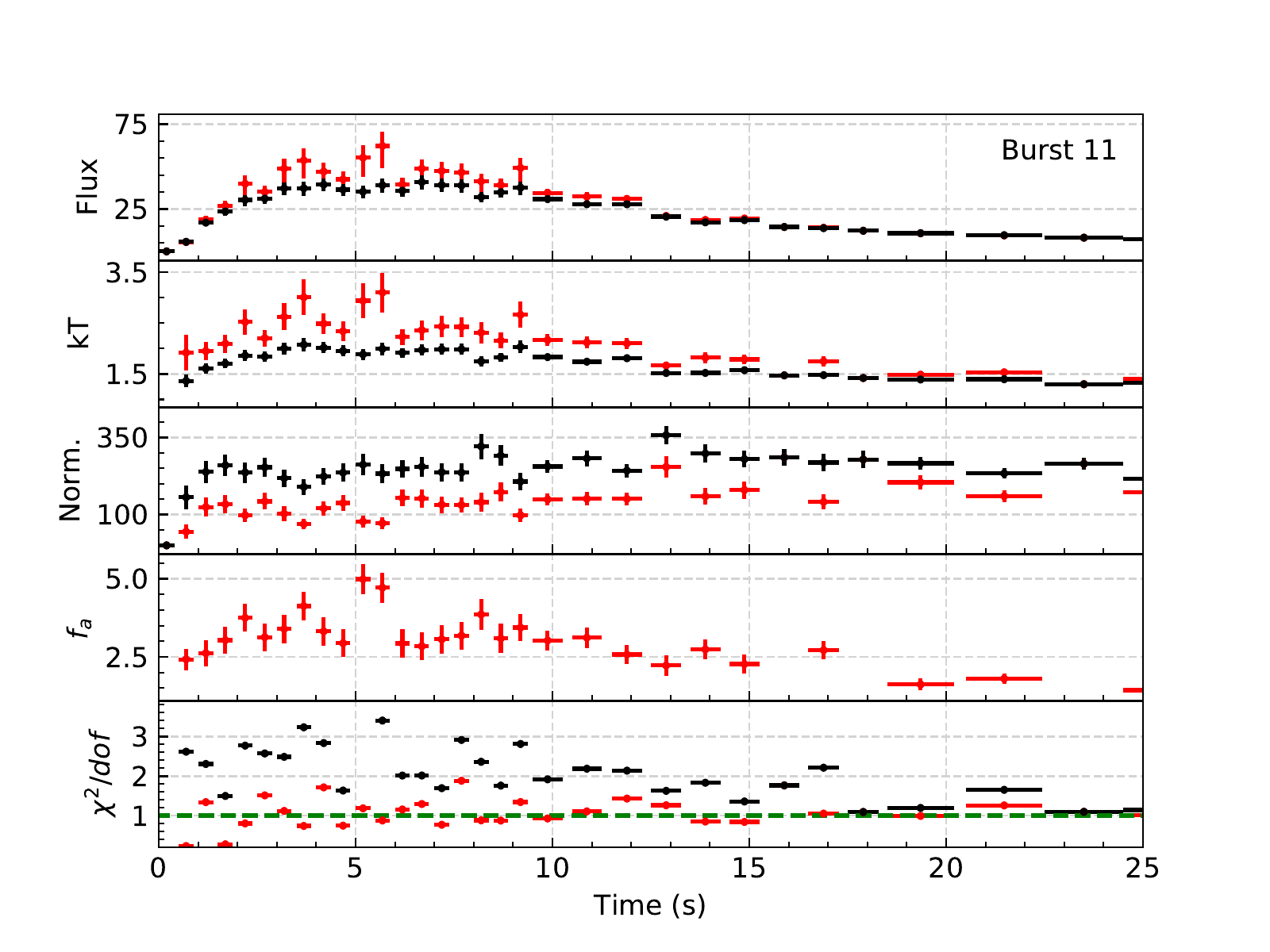}
	\includegraphics[scale=0.5]{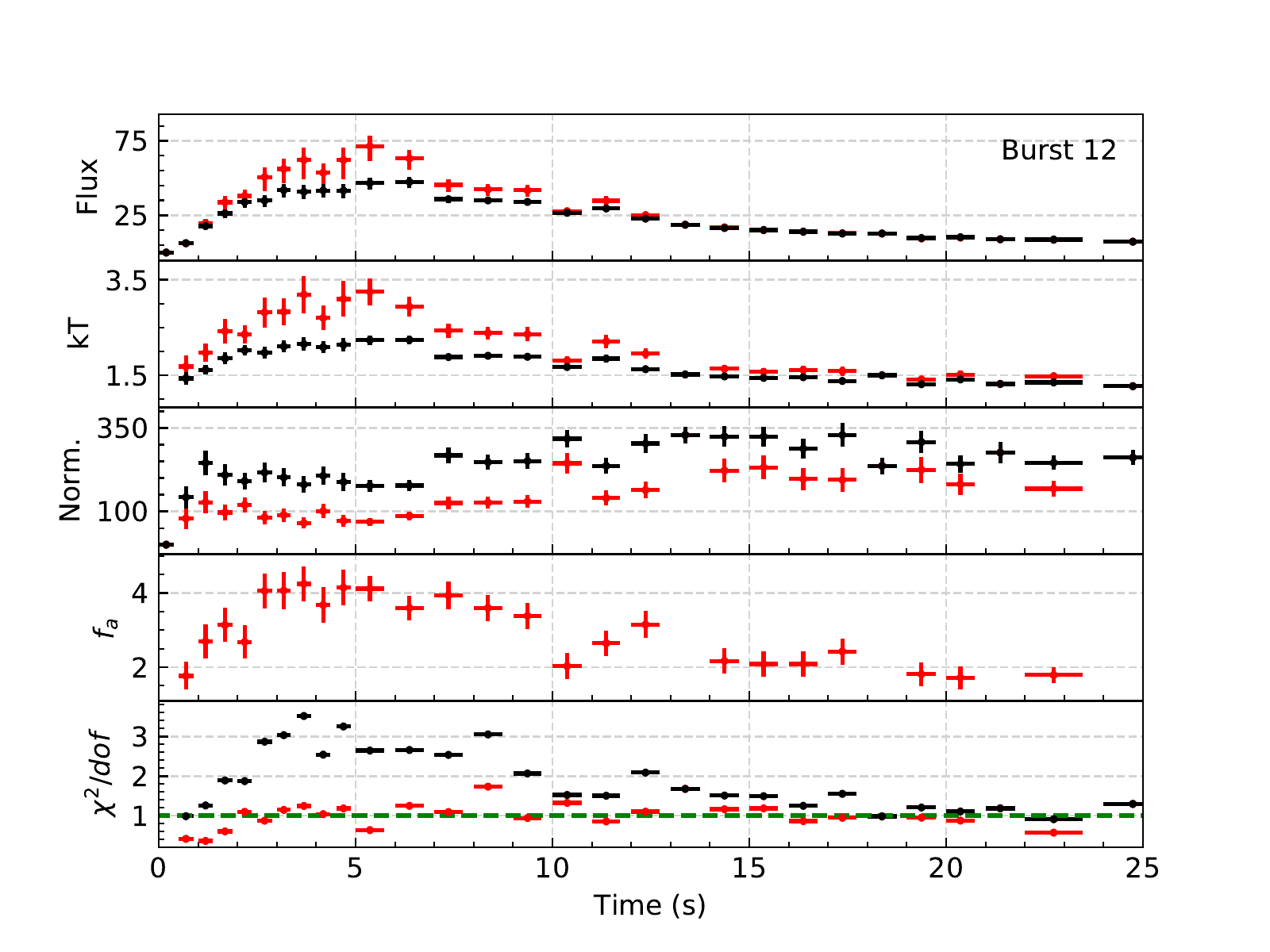}
    \caption{Same as Figure \ref{fig:burst_plots_1}.}
    \label{fig:burst_plots_2}
\end{figure*}

\begin{figure*}
	\includegraphics[scale=0.5]{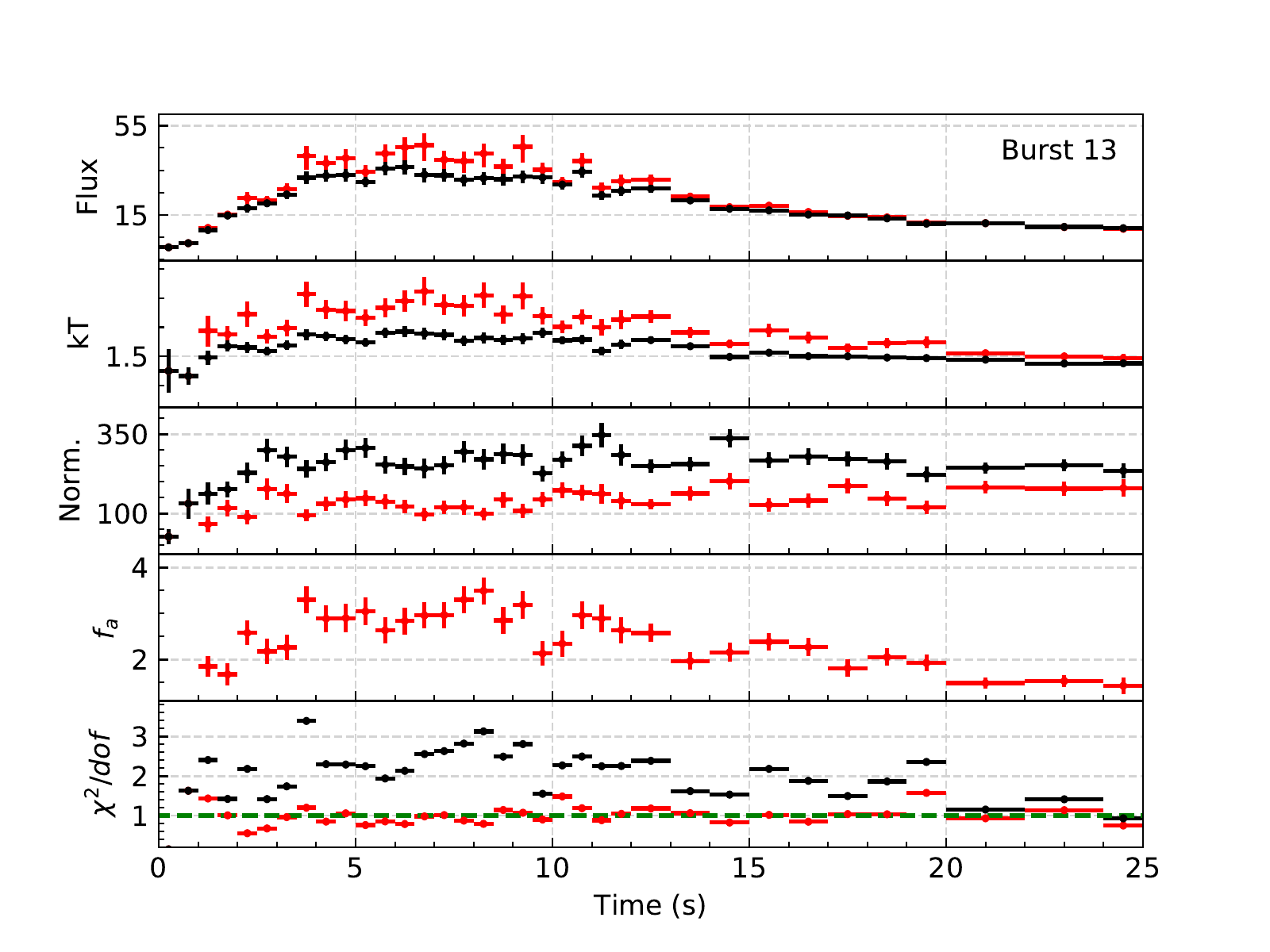}
	\includegraphics[scale=0.5]{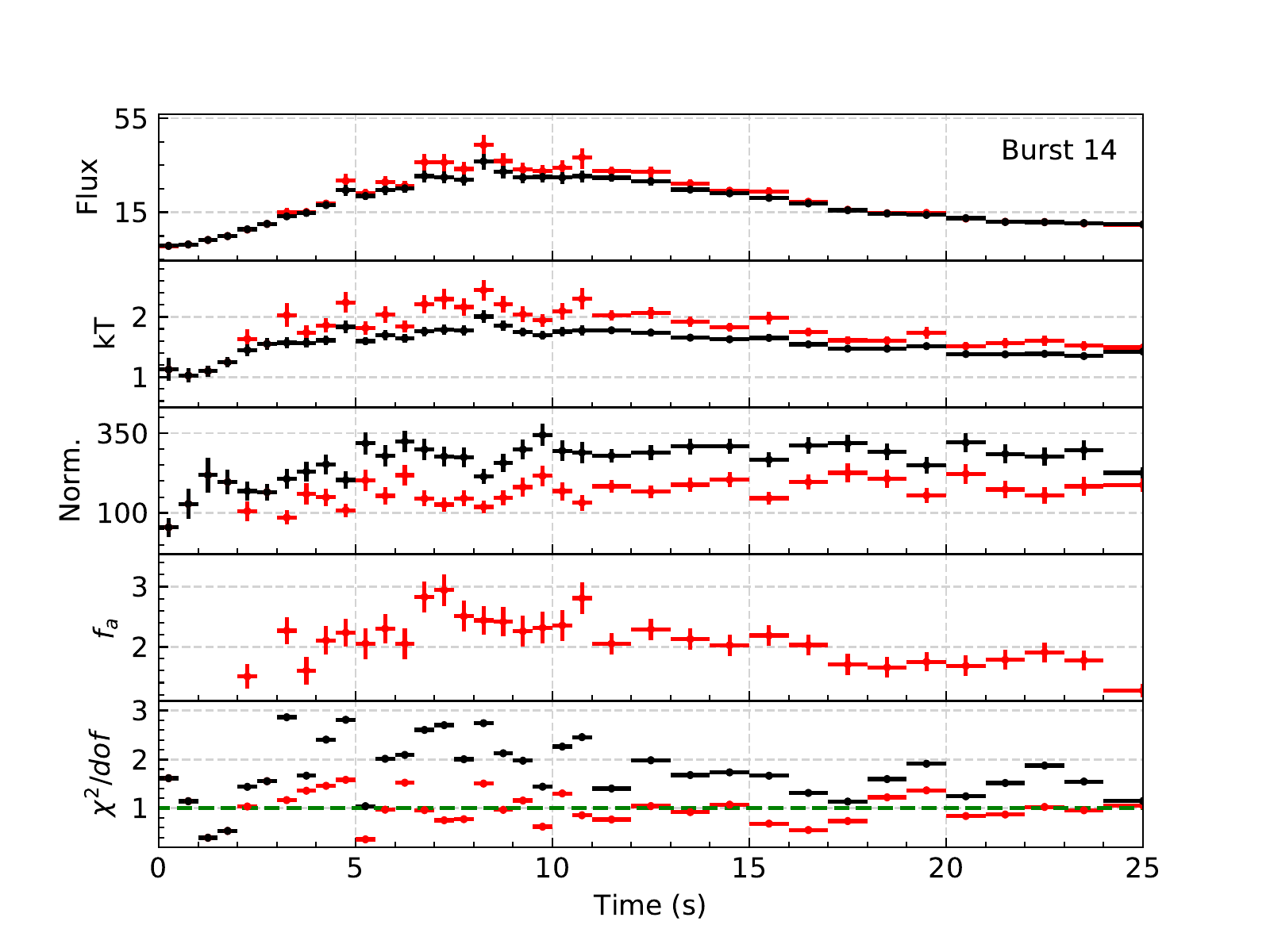}
	\includegraphics[scale=0.5]{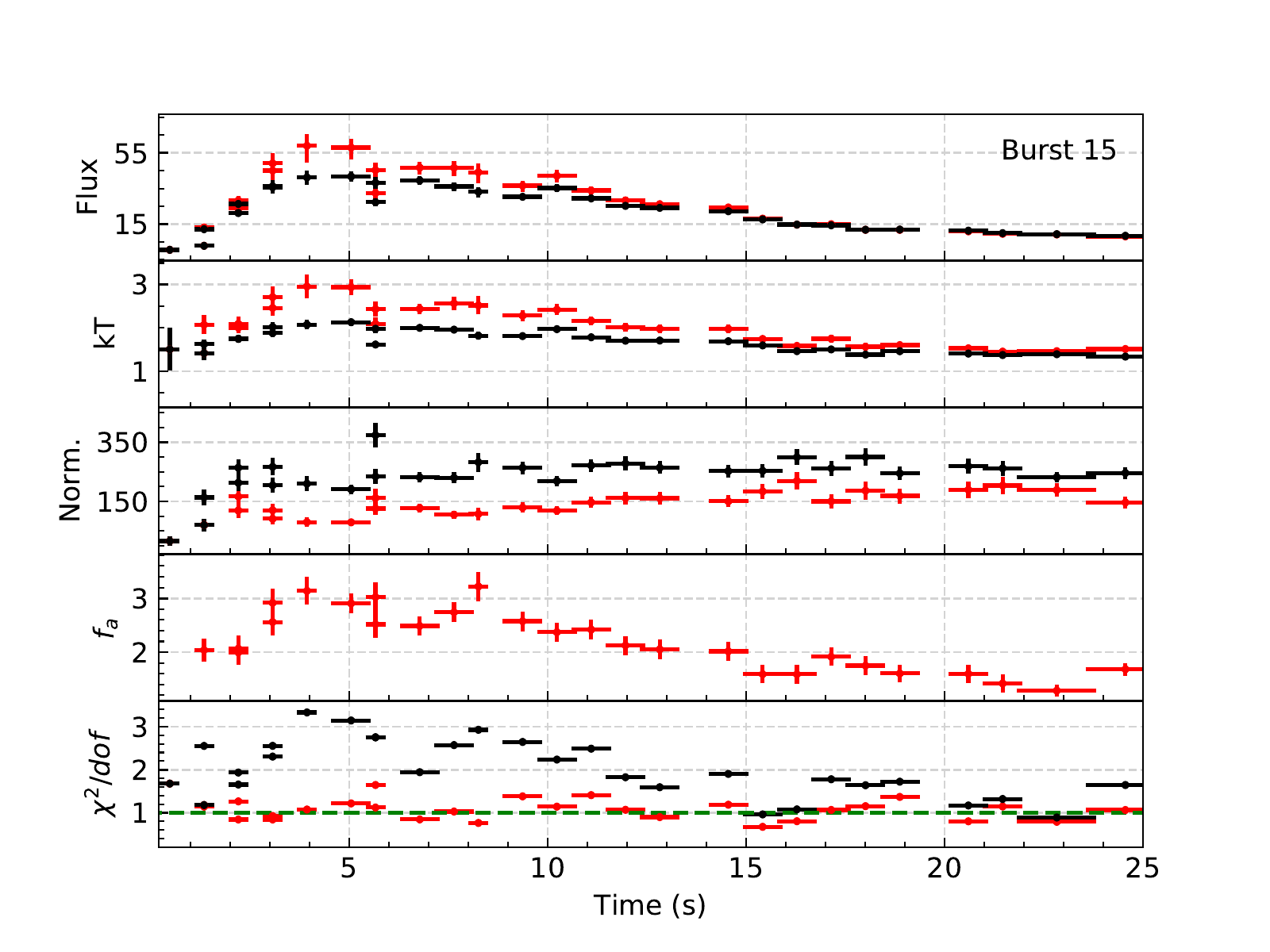}
	\includegraphics[scale=0.5]{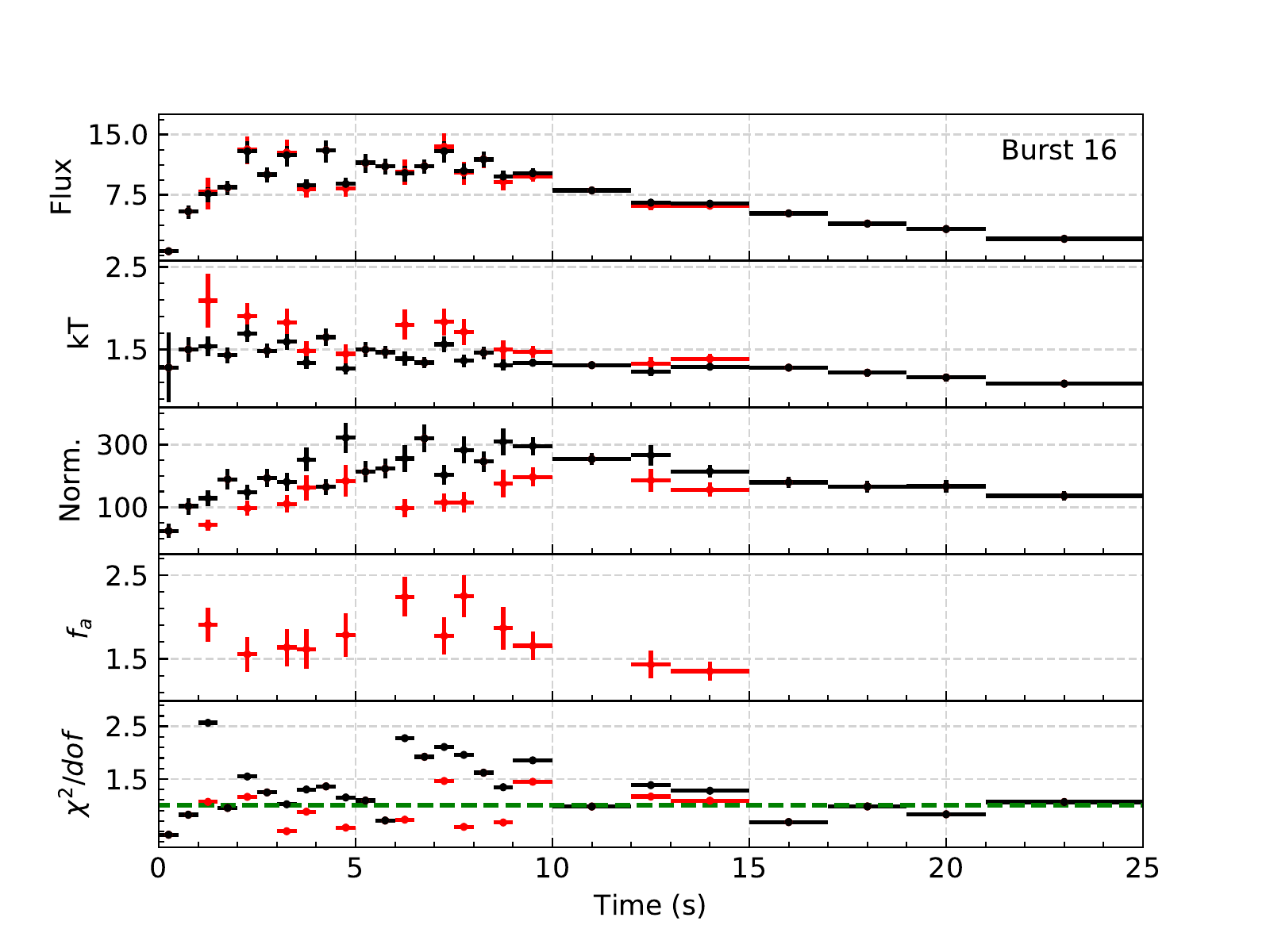}
	\includegraphics[scale=0.5]{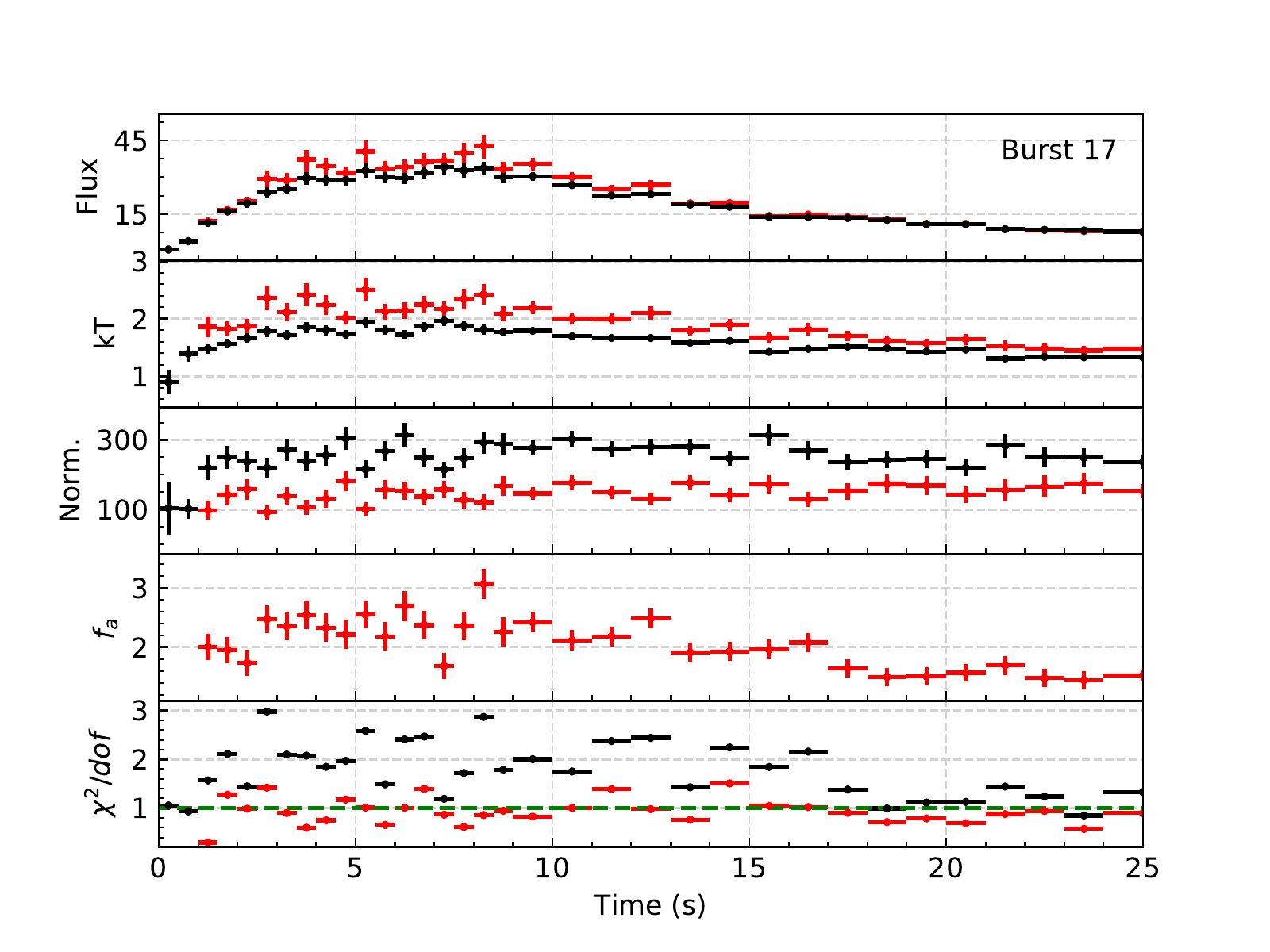}
	\includegraphics[scale=0.5]{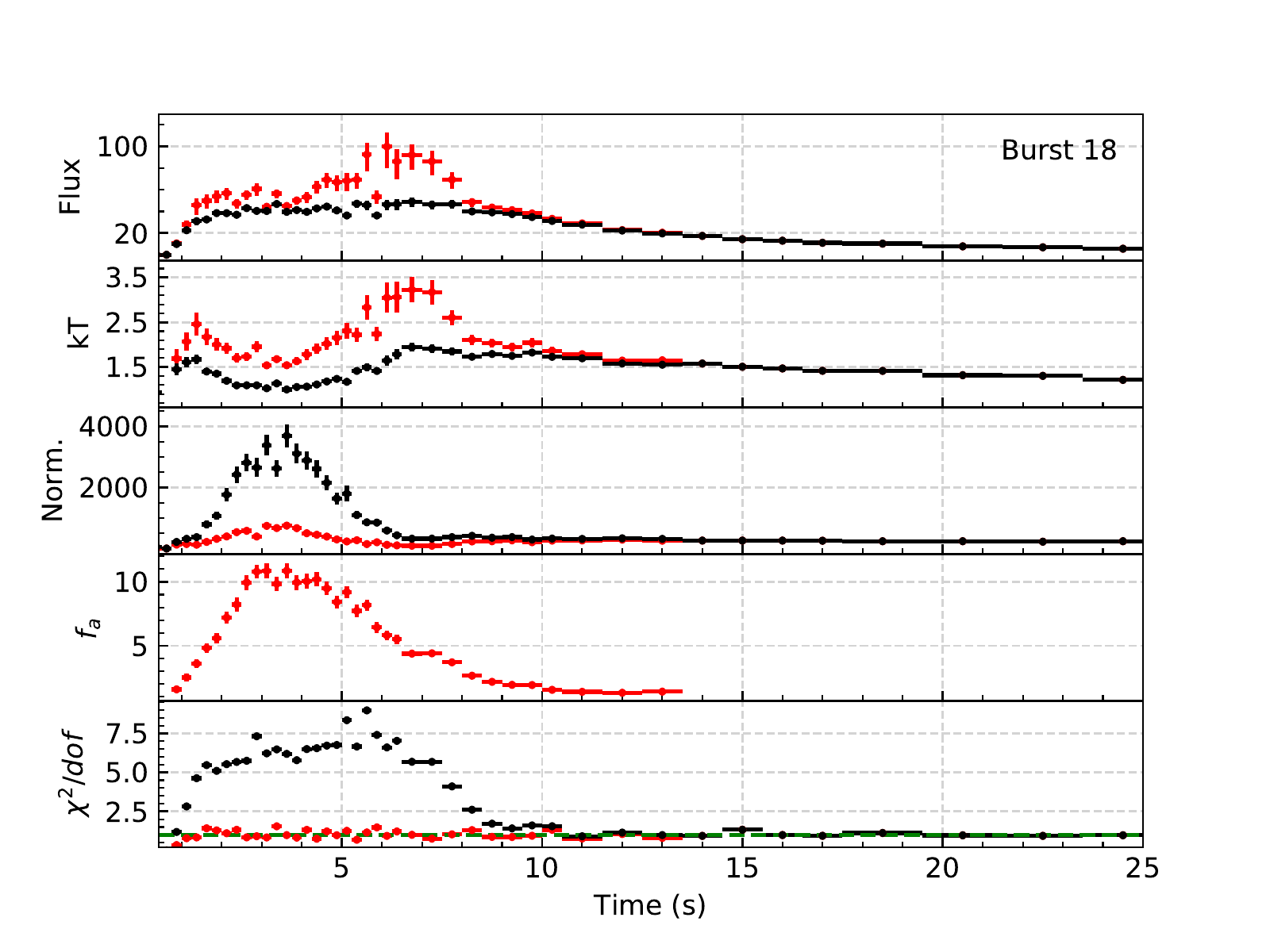}
    \caption{Same as Figure \ref{fig:burst_plots_2}.}
    \label{fig:burst_plots_3}
\end{figure*}

\begin{figure*}
	\includegraphics[scale=0.5]{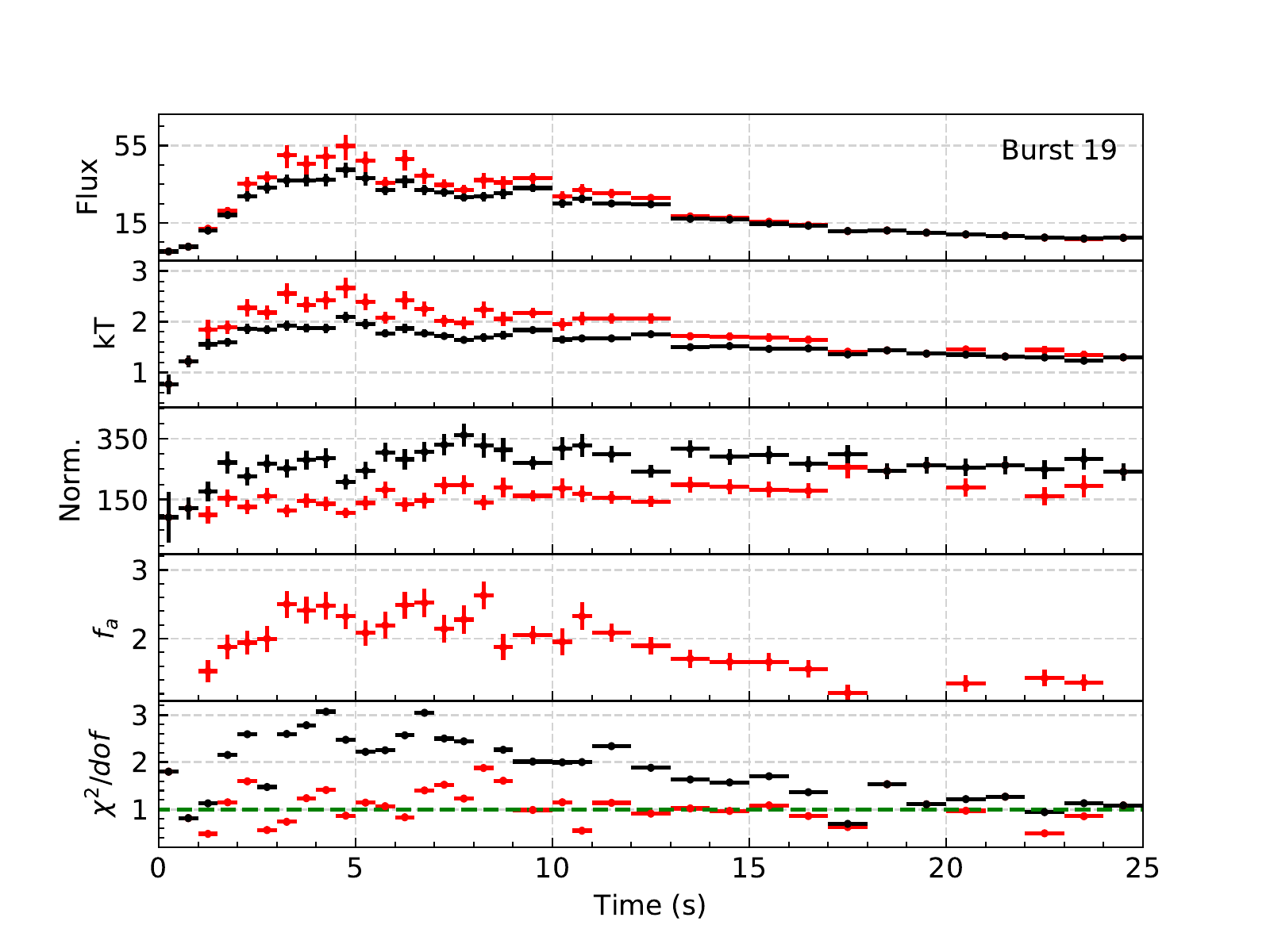}	\includegraphics[scale=0.5]{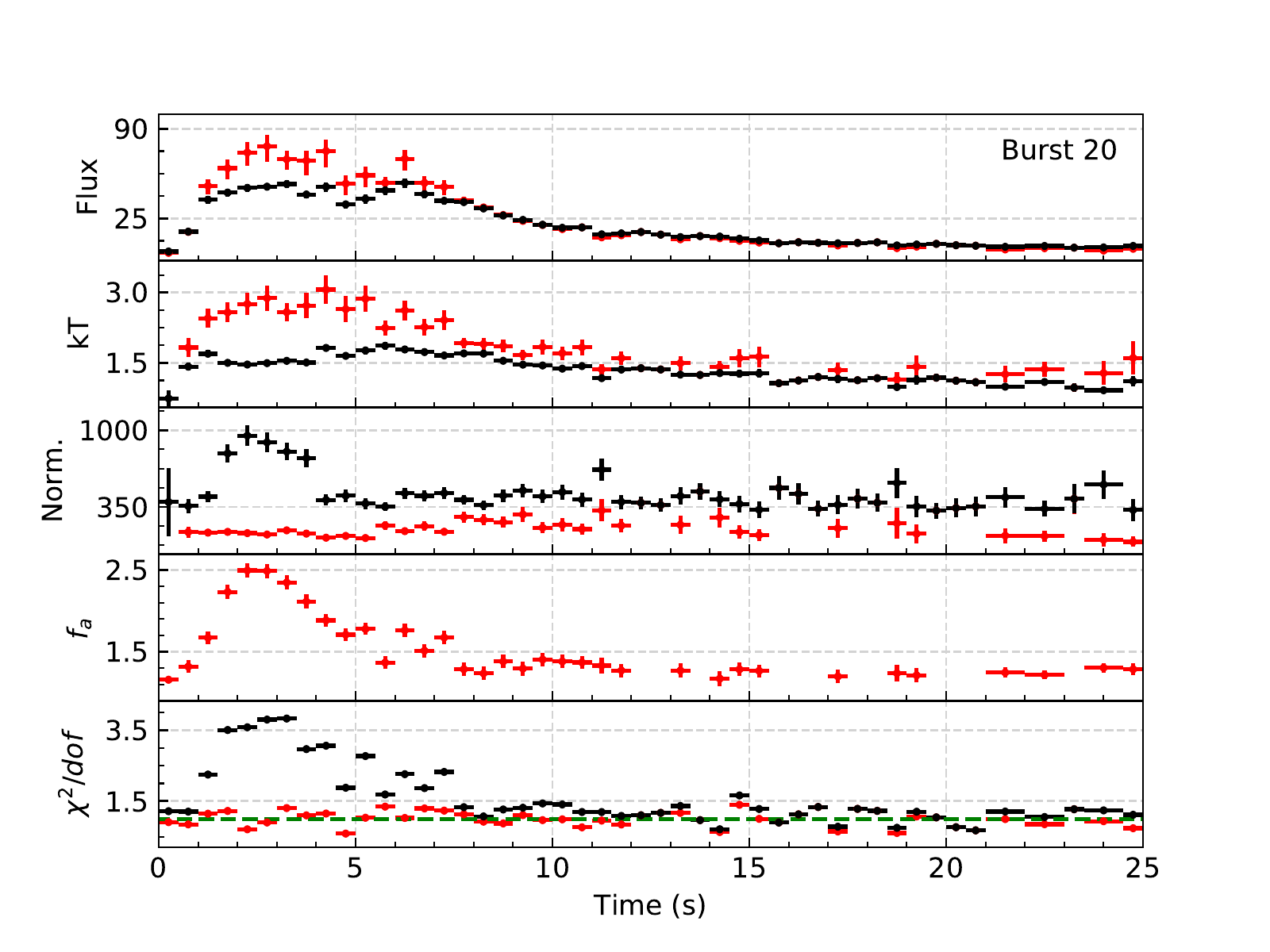}
	\includegraphics[scale=0.5]{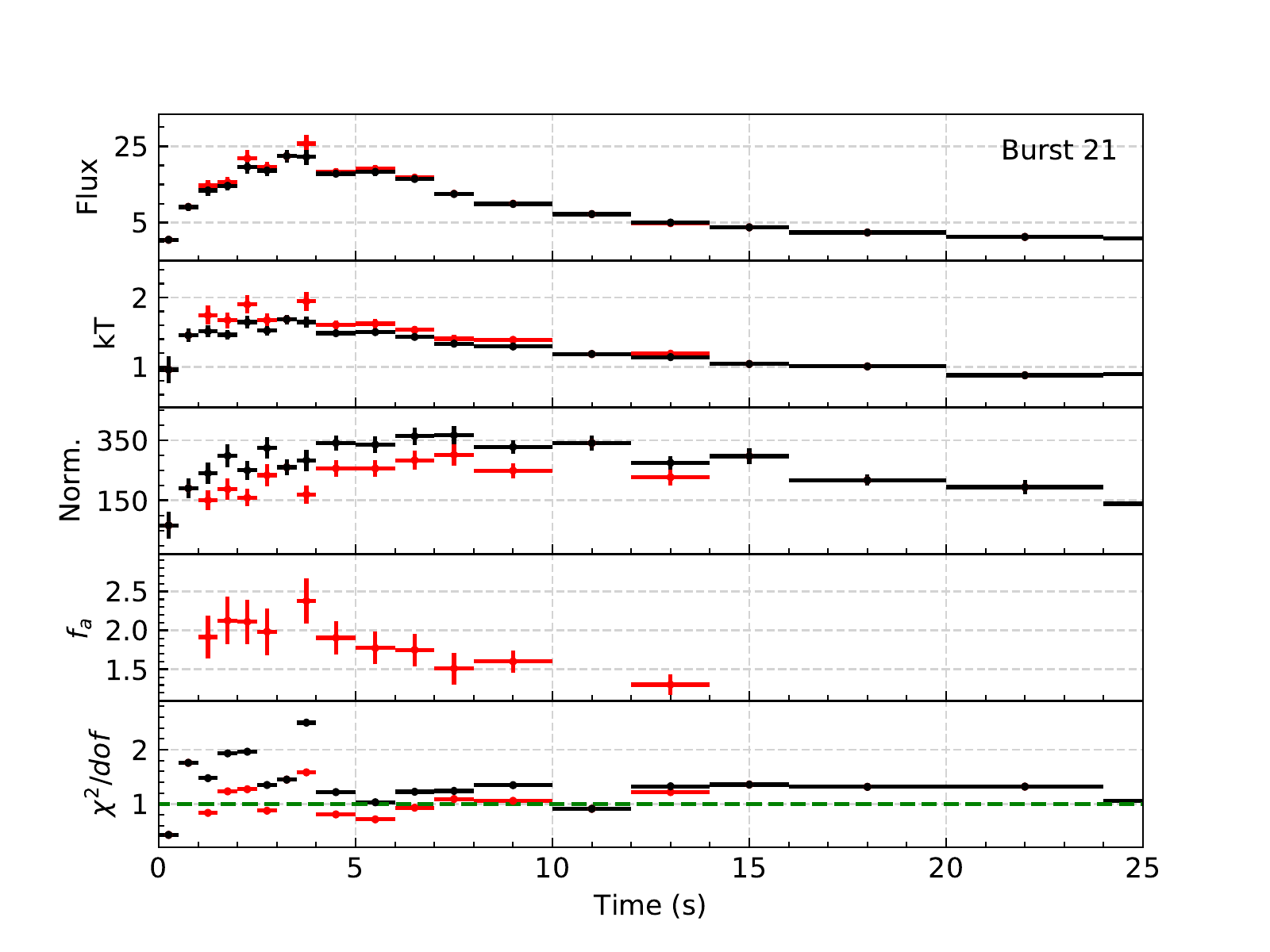}
	\includegraphics[scale=0.5]{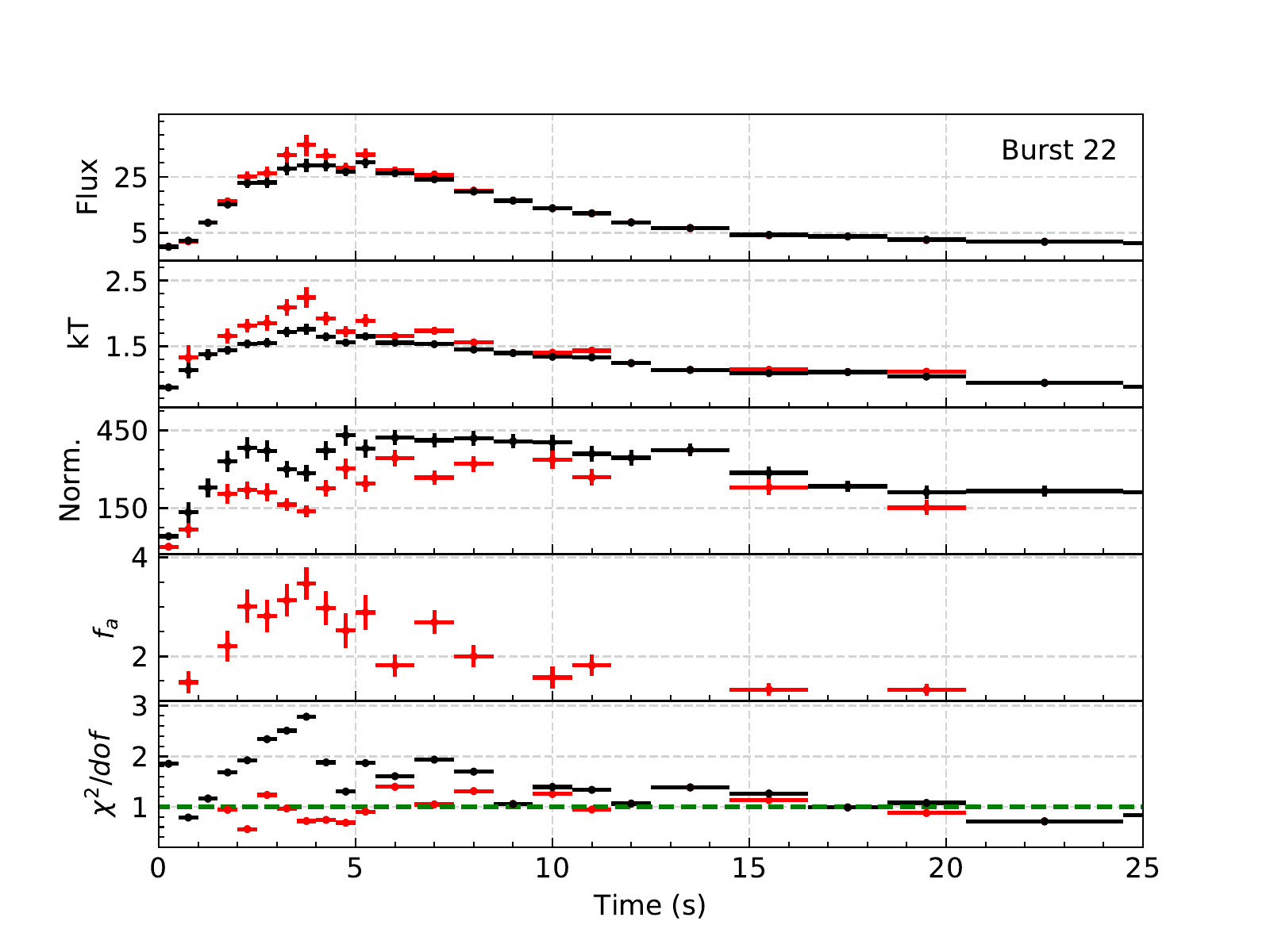}
    \caption{Same as Figure \ref{fig:burst_plots_3}.}
    \label{fig:burst_plots_4}
\end{figure*}

\bsp	
\label{lastpage}
\end{document}